\documentclass[a4paper,clock]{article}
\usepackage[dvips]{epsfig}
\usepackage{amssymb}
\usepackage{bm}
\usepackage{dcolumn}

\catcode`\@=11
\def\marginnote#1{}
\newcount\hour
\newcount\minute
\newtoks\amorpm
\hour=\time\divide\hour by60
\minute=\time{\multiply\hour by60 \global\advance\minute
by-\hour}\edef\standardtime{{\ifnum\hour<12
\global\amorpm={am}%
        \else\global\amorpm={pm}\advance\hour by-12 \fi
        \ifnum\hour=0 \hour=12 \fi
        \number\hour:\ifnum\minute<10
0\fi\number\minute\the\amorpm}}
\edef\militarytime{\number\hour:\ifnum\minute<10
0\fi\number\minute}

\def\draftlabel#1{{\@bsphack\if@filesw {\let\thepage\relax
   \xdef\@gtempa{\write\@auxout{\string
      \newlabel{#1}{{\@currentlabel}{\thepage}}}}}\@gtempa
   \if@nobreak \ifvmode\nobreak\fi\fi\fi\@esphack}
        \gdef\@eqnlabel{#1}}
\def\@eqnlabel{}
\def\@vacuum{}
\def\draftmarginnote#1{\marginpar{\raggedright\scriptsize\tt#1}}
\def\draft{\oddsidemargin -.5truein
        \def\@oddfoot{\sl preliminary draft \hfil
        \rm\thepage\hfil\sl\today\quad\militarytime}
        \let\@evenfoot\@oddfoot \overfullrule 3pt
        \let\label=\draftlabel
        \let\marginnote=\draftmarginnote

\def\@eqnnum{(\theequation)\rlap{\kern\marginparsep\tt\@eqnlabel}%
\global\let\@eqnlabel\@vacuum}  }

\def\numberbysection{\@addtoreset{equation}{section}
        \def\theequation{\thesection.\arabic{equation}}}

\def\underline#1{\relax\ifmmode\@@underline#1\else
 $\@@underline{\hbox{#1}}$\relax\fi}

\catcode`@=12
\relax

\numberbysection

\advance \topmargin by -\headheight
\advance \topmargin by -\headsep

\evensidemargin \oddsidemargin
\def\br{\begin{eqnarray}}
\def\er{\end{eqnarray}}
\def\be{\begin{equation}}
\def\ee{\end{equation}}

\def\({\left(}
\def\){\right)}

\relax

\def\a{\alpha}

\def\b{\beta}

\def\d{\delta}
\def\D{\Delta}

\def\g{\gamma}

\def\L{\Lambda}

\def\O{\Omega}

\def\pa{\partial}

\def\tp0{\Theta_{+}^{(0)}}
\def\tm0{\Theta_{-}^{(0)}}

\def\vp{\varphi}

\def\f#1#2#3 {f^{#1#2}_{#3}}

\def\win1{{\sf w_{1+\infty}}}

\def\Win1{{\sf W_{1+\infty}}}

\def\rlx{\relax\leavevmode}
\def\inbar{\vrule height1.5ex width.4pt depth0pt}
\def\IZ{\rlx\hbox{\sf Z\kern-.4em Z}}
\def\IR{\rlx\hbox{\rm I\kern-.18em R}}
\def\IC{\rlx\hbox{\,$\inbar\kern-.3em{\rm C}$}}
\def\IN{\rlx\hbox{\rm I\kern-.18em N}}
\def\IO{\rlx\hbox{\,$\inbar\kern-.3em{\rm O}$}}
\def\IP{\rlx\hbox{\rm I\kern-.18em P}}
\def\IQ{\rlx\hbox{\,$\inbar\kern-.3em{\rm Q}$}}
\def\IF{\rlx\hbox{\rm I\kern-.18em F}}
\def\IG{\rlx\hbox{\,$\inbar\kern-.3em{\rm G}$}}
\def\IH{\rlx\hbox{\rm I\kern-.18em H}}
\def\II{\rlx\hbox{\rm I\kern-.18em I}}
\def\IK{\rlx\hbox{\rm I\kern-.18em K}}
\def\IL{\rlx\hbox{\rm I\kern-.18em L}}
\def\one{\hbox{{1}\kern-.25em\hbox{l}}}
\def\0#1{\relax\ifmmode\mathaccent"7017{#1}%
B        \else\accent23#1\relax\fi}

\def\PRA#1#2#3{{\sl Phys. Rev.} {\bf A#1} (#2) #3}

\def\JMP#1#2#3{{\sl J. Math. Phys.} {\bf #1} (#2) #3}

\def\PR#1#2#3{{\sl Phys. Reports} {\bf #1} (#2) #3}

\def\JPAMT#1#2#3{{\sl J. Physics A: Math. Theor.} {\bf A#1} (#2) #3}

\def\PHSD#1#2#3{{\sl Physica} {\bf D#1} (#2) #3}

\def\JHEP#1#2#3{{\sl JHEP} {\bf #1} (#2) #3}

\def\JCP#1#2#3{{\sl Journal of Computational Physics} {\bf #1} (#2) #3}

\def\MAA#1#2#3{{\sl Methods and  Applications of Analysis} {\bf #1} (#2) #3}
\def\OL#1#2#3{{\sl Optics Letters} {\bf #1} (#2) #3}
\def\Nonl#1#2#3{{\sl Nonlinearity} {\bf #1} (#2) #3}
\def\CJP#1#2#3{{\sl Canadian Journal of Physics} {\bf #1} (#2) #3}
\def\OC#1#2#3{{\sl Optics Communications} {\bf #1} (#2) #3}

\begin{document}

\begin{titlepage}

\vspace{.2in}
\begin{center}
{\large\bf Quasi-integrable 
 non-linear Schr\"odinger models, infinite towers of exactly conserved charges and bright solitons}
\end{center}

\vspace{.2in}

\begin{center}

H. Blas, A.C.R. do Bonfim and A.M. Vilela

\par \vskip .2in \noindent

 Instituto de F\'{\i}sica\\
Universidade Federal de Mato Grosso\\
Av. Fernando Correa, $N^{0}$ \, 2367\\
Bairro Boa Esperan\c ca, Cep 78060-900, Cuiab\'a - MT - Brazil  
\normalsize
\end{center}

\vspace{.3in}

\begin{abstract}
\vspace{.3in}

Deformations of the focusing non-linear Schr\"odinger model (NLS) are considered in the context of the quasi-integrability concept. We strengthen the results of JHEP09(2012)103 for bright soliton collisions. We addressed the focusing NLS as a complement to the one in JHEP03(2016)005, in which the modified defocusing NLS models with dark solitons were shown to exhibit an infinite tower of exactly conserved charges. We show, by means of analytical and numerical methods, that for certain two-bright-soliton solutions, in which the modulus and phase of the complex modified NLS field exhibit even parities under a space-reflection symmetry, the first four and the sequence of even order charges are exactly conserved during the scattering process of the solitons. We perform extensive numerical simulations and consider the bright solitons with deformed potential $ V = \frac{ 2\eta}{2+ \epsilon} \( |\psi|^2\)^{2 + \epsilon}, \epsilon \in \IR, \eta<0$. However, for two-soliton field components without definite parity we also show numerically the vanishing of the first non-trivial anomaly and the exact conservation of the relevant charge. So, the parity symmetry seems to be a sufficient but not a necessary condition for the existence of the infinite  tower of conserved charges.  The model supports elastic scattering of solitons for a wide range of values of the amplitudes and velocities and the set $\{\eta, \epsilon\}$. Since the NLS equation is ubiquitous, our results may find potential applications in several areas of non-linear science.
\end{abstract}

\end{titlepage}

\section{Introduction}

The quasi-integrability concept has been introduced in the context of certain deformations of the integrable models \cite{jhep1, jhep2}. There have been shown that many non-integrable theories possess solitary wave solutions resembling to true solitons, i.e. the scattering of such solitons preserve their shapes and velocities. So, certain deformations of the sine-Gordon (SG) and
non-linear Schr\"odinger (NLS) models were considered to be  quasi-integrable  theories possessing infinite number of charges that are asymptotically
conserved. Recently, these results have been tested in the collective coordinate approach to the scattering
of solitons  \cite{jhep5}. In addition, the deformed SG models have been shown to posses a subset of infinite number of exactly conserved charges, provided that the two-soliton field configurations are eigenstates of the space-reflection parity symmetry \cite{jhep4}. Similar results have been put forward for the  deformed defocusing NLS model with dark solitons \cite{jhep3}. A particular deformation of the  NLS model, depending on the sign $\pm$ of the undeformed cubic self-interaction term, can be dubbed as focusing or defocusing, respectively. The focusing NLS supports bright solitons with vanishing boundary conditions, whereas the defocusing one supports dark solitary waves with non-vanishing boundary conditions.

In this paper we strengthen the results of \cite{jhep2} and  build on some results of our
earlier work \cite{jhep3}, relevant to bright solitons,  by deriving infinite towers of exactly conserved charges for the collision of special  two-bright solitons and numerically simulating a representative charge and vanishing anomaly. The deformed focusing NLS with bright soliton solutions and the structures responsible for the phenomenon of quasi-integrability have been discussed in \cite{jhep2}. It has been shown that this model possesses an infinite number of asymptotically conserved charges. An explanation found so far  for this behaviour of the charges is that some special soliton type solutions are eigenstates of a
space-time parity transformation.  As we will show here, there are some soliton-like solutions which present a special space-reflection symmetry at any time, such that the sequence of the even order charges are exactly conserved. As in the defocusing NLS case \cite{jhep3}, the demonstration of these results involve an interplay between the space-reflection parity and internal transformations in the affine Kac-Moody algebra underlying the anomalous Lax equation. 

However, it seems to be that such space-reflection parity is a sufficient but not a necessary condition in order to have the tower of exactly conserved charges. In fact, as we will show by numerical simulations of the first non-trivial quasi-conservation law, there are certain two-soliton like configurations without this symmetry which also exhibit such conserved charges. So, these properties constitute the distinguishing new features associated to the deformed focusing NLS with bright soliton solutions, as compared to the previous quasi-integrable focusing NLS model \cite{jhep2}. Thus, our results turn out to be complemantary to the ones in \cite{jhep3} and  strengthen the ones in \cite{jhep2}, and since the NLS equation is ubiquitous, with potential applications in several areas of non-linear science, we believe that they deserve to be available to the general community.  

In order to simulate the time dependence of field configurations for computing bright soliton quantities we have used an efficient and accurate numerical method, the so-called time-splitting cosine pseudo-spectral finite difference method (TSCP) \cite{bao}, in order to control the highly oscillatory phase background. This method allowed us to improve in several orders of magnitude the accuracy in the computations of the charges and anomalies presented in \cite{jhep2}.  In fact, the charge $Q^{(4)}$, which has been regarded in \cite{jhep2} as asymptotically conserved, is indeed an exactly conserved charge, as we will show by simulating the vanishing of the corresponding anomaly $\beta^{(4)}$ for general two-bright soliton configuration and several values of the deformation parameter $\epsilon$.

The paper is organized as follows. In the next section we introduce the deformed focusing NLS model. In section \ref{quasi}, we discuss the concept of quasi-integrability  for deformed focusing NLS following \cite{jhep2}. In  \ref{stp} we discuss the relationships between the space-time parity and asymptotically conserved charges. In \ref{srs} the space-reflection parity and the exactly conserved charges are discussed following \cite{jhep3}. In section \ref{charges} we list  the first four conserved charges and the first non-trivial quasi-conservation law and provide the first anomaly $\beta^{(4)}$.  In section \ref{bright} we discuss the space-time and space-reflection symmetries of bright solitons.   In section \ref{simul} we present the results of our numerical simulations of bright soliton scattering of the model (\ref{nlsd})-(\ref{pot1}) for several values of the deformation parameter $\epsilon$. In section \ref{conclu} we present some conclusions and discussions. The appendix presents relevant expressions of the quasi-conservation laws (\ref{cons1}) \cite{jhep2}.

\section{Deformations of focusing NLS}

\label{defor}
We will consider non-relativistic models   in $(1+1)-$dimensions with  equation of motion given by
\br
\label{nlsd}
i \frac{\pa}{\pa t} \psi(x,t) + \frac{\pa^2}{\pa x^2} \psi(x,t) -  \frac{\pa V[|\psi(x,t)|^2]}{\pa |\psi(x,t)|^2} \psi(x,t) =  0,\er 
where $\psi$ is a complex scalar field and $V: \IR_{+} \rightarrow \IR$.

The model (\ref{nlsd}) defines the deformed NLS model and it supports bright and dark soliton type solutions in analytical form for some special functions $V[I]$, $I\equiv |\psi|^2$. The potential  $V[I] = \eta I^2,\,(\eta<0)$, corresponds to the integrable focusing NLS model and supports N-bright soliton solutions. The potential  $V[I] = \eta  I^2-\epsilon I^3/6$ defines the non-integrable cubic-quintic NLS model (CQNLS) which possesses analytical bright and dark type solitons \cite{cowan, crosta}. In \cite{sombra, cowan} the bright solitary waves of the cubic-quintic focusing NLS have been regarded as quasi-solitons presenting partially inelastic collisions in certain region of parameter space. Among the models with saturable non-linearities \cite{kivshar}, the case $V[I] = \frac{1}{2} \rho_s (I+\frac{\rho_s^2}{I+ \rho_s})$ also exhibits analytical dark solitons \cite{kroli}. The deformed NLS model with $V'[I] = 2 \eta I - \epsilon \frac{I^q}{1+ I^q}$ ($ V'[I]  \equiv  \frac{d V[I]}{d I} $)  passes  the Painlev\'e test for arbitrary positive integers  $q\in \IZ_{+}$ and $\epsilon =1$ \, \cite{enns}. However, its Lax pair formulation and analytical solutions for a general set $\{ \eta, \epsilon, q\}$, to the best of our knowledge, are lacking.

Among the possible deformations of the NLS model the case 
\br
\label{pot1}
 V = \frac{ 2\eta}{2+ \epsilon} \( |\psi|^2\)^{2+\epsilon},\,\,\, \epsilon \in \IR, \,\,\,\, \eta<0,
\er 
has recently been considered in \cite{jhep2} in order to study  the concept of quasi-integrability for bright soliton collisions. An analytical solitary wave solution with vanishing boundary condition (bright soliton) for this potential is well known in the literature (see for example the equation (4.6) of \cite{jhep2})
\br 
\label{solitary}
\psi(x,t) = \Big[\frac{2 + \epsilon}{2} \frac{\rho^2}{|\eta|} \frac{1}{\cosh^2{[(1+ \epsilon) \rho (x - v t -x_0)]}}\Big]^{\frac{1}{2(1+ \epsilon)}} \,\, e^{i [(\rho^2-\frac{v^2}{4})t + \frac{v}{2} x]}.
\er
In this paper we will study analytically and numerically  some deformations of the NLS model of the type (\ref{nlsd}), such that in the limit $\epsilon \rightarrow 0 $ we recover the usual focusing NLS ($\eta<0$). In our numerical simulations we will consider the potential of type (\ref{pot1}).

\section{Quasi-integrability of deformed NLS}
\label{quasi} 

We follow the developments and notations of \cite{jhep2, jhep3} on quasi-integrability in deformed NLS models. Let us consider an anomalous zero curvature representation of the deformed NLS model (\ref{nlsd}) with the connection given by  
\br
\label{con1}
A_{x} &=&- i\, T_{3}^{1} + \bar{\g} \bar{\psi} \,T_{+}^{0}+\g \psi \,T_{-}^{0},\\
\nonumber
A_{t} &=& i \,T^{2}_{3} + i \frac{\d V}{\d |\psi|^2} \,T_{3}^{0} -( \bar{\g} \bar{\psi} \,T_{+}^{1}+\g \psi \,T_{-}^{1}) -i (\bar{\g} \pa_x \bar{\psi} \,T_{+}^{0}-\g \pa_x \psi \,T_{-}^{0}),
\er  
where the above Lax potentials are based on a $sl(2)$ loop algebra (see more details in \cite{jhep2}). It follows that the curvature of the connection (\ref{con1}) is given by
\br
\label{zero0}
F_{xt} & \equiv &  \pa_{t}A_x - \pa_{x} A_t + [A_x\,,A_t] \\
 &=&X T_{3}^{0}+ i \bar{\g} \Big[-i \pa_{t} \bar{\psi} + \pa_x^2 \bar{\psi} - \bar{\psi} \frac{\d V}{\d |\psi|^2} \Big] T_{+}^{0}-i \g \Big[i \pa_{t} \psi + \pa_x^2 \psi - \psi \frac{\d V}{\d |\psi|^2} \Big] T_{-}^{0}\label{zero1}
\er
with
\br
\label{an0}
X \equiv -i \pa_x \(\frac{\d V}{\d |\psi|^2} -2 \bar{\g} \g |\psi|^2 \).
\er

Notice that when the equation of motion (\ref{nlsd}) and its complex conjugate  are satisfied the terms proportional to the Lie algebra generators $T_{\pm}^0$ vanish.  In addition, the quantity $X$ vanishes for the usual non-linear Schr\"{o}dinger potential
\br
\label{pot0}
V(|\psi|^2) = \eta (|\psi|^2)^2,\,\,\,\,\eta \equiv   \bar{\g} \g.   
\er
Then, the curvature vanishes for the usual NLS model, rendering this theory an integrable field theory. For the deformations of the potential (\ref{pot0}), the above curvature is non-vanishing and the model is regarded as non-integrable. 

Let us denote
\br
\label{exp1}
\psi = \sqrt{R} e^{i \frac{\vp}{2}}
\er
and parametrize $\g = i \sqrt{\eta} e^{i \a},\,\,\g = -i \sqrt{\eta} e^{-i \a},\,\,\eta \equiv   \bar{\g} \g$. Substituting the parametrization  (\ref{exp1}) into (\ref{nlsd}) one gets the system of eqs. of motion
\br\label{eq111}
 \pa_t R + \pa_x \( R \pa_x \vp \) &=& 0\\
 \pa_t \vp + \frac{1}{2} (\pa_x \vp)^2 - \frac{\pa_x^2 R}{R} + \frac{1}{2} (\frac{\pa_x R}{R})^2+2 \, \frac{\d V}{\d R } &=&0\label{eq222} 
\er 

We specialize the parametrizations for the case of the focusing NLS, i.e. the case $\eta < 0$. In this case the usual NLS model admits  a bright soliton solution with  the boundary condition 
\br
\label{vbc}
R(x,t)|_{ |x| \rightarrow \infty} = 0,\,\,\,\,\,\frac{1}{2}\vp(x, t)|_{ |x| \rightarrow \infty} =  v x - \frac{1}{2}(v^2-a^2) t + \theta_0.
\er 
 For a deformed NLS we can apply the abelianization procedure in order to gauge transform the curvature (\ref{zero0}) and  get the quasi-conserved charges (see more details in \cite{jhep2})   
\br
\label{cons1}
\pa_{t} a_x^{(3, -n)} - \pa_{x} a_{t}^{(3, -n)} = X \a^{(3, -n)};\,\,\,\,n=0,1,2,... 
\er
In the appendix we present the first quantities $a_x^{(3, -n)}$\, and  $\a^{(3, -n)}$.
 
So, in the focusing NLS case with bright soliton solutions and vanishing boundary condition (\ref{vbc}) one has that the $a_t$ component of the connection satisfies a boundary condition such that $a_{t}^{(3, -n)}(x=+\infty) = a_{t}^{(3, -n)}(x=-\infty)$. Then from  (\ref{cons1}) we have the anomalous conservation laws
\br
\label{ano1}
\frac{d Q^{(n)}}{dt} = \beta^{(n)},\,\,\,\,\,\mbox{where}\,\,\,\,\,Q^{(n)}= -i \int_{-\infty}^{\infty} dx \, a_x^{(3, -n)}\,\,\,\,\,\,\mbox{and}\,\,\,\,\,\,\b^{(n)} = -i \int_{-\infty}^{\infty} dx \,X \, \a^{(3, -n)} .
\er
Thus, the non-vanishing of the quantity $X$ given in (\ref{an0}) and the anomalies $\b_n$ above imply the non-conservation of the charges.  Therefore, the charges and anomalies in (\ref{ano1}) are valid for the deformed NLS model with vanishing boundary condition (\ref{vbc}) and bright soliton solutions.

\subsection{Space-time parity and asymptotically conserved charges}
\label{stp}

For a subset of solutions of the deformed model (\ref{nlsd}) the charges $Q^{(n)}$ satisfy a {\sl mirror type symmetry} \cite{jhep2}. In fact, for every solution belonging to this subset one can find a  point  $(x_{\rho},t_{\rho})$ in space-time    
such that the fields $R$ and $\vp$ transform as
\br
\label{pari1}
R \rightarrow R,\,\,\,\,\vp \rightarrow -\vp + const., 
\er
under the  parity transformation 
\br
\label{pari0}
P: (\widetilde{x},\,\widetilde{t}) \rightarrow  (-\widetilde{x},\,-\widetilde{t}),\,\,\,\,\widetilde{x}\equiv x-x_{\rho},\,\widetilde{t}\equiv t-t_{\rho}.       
\er

Let us summarize the main results of \cite{jhep2}. 

1. If one has a two-bright-soliton-like solution of (\ref{nlsd}), transforming under the space-time parity (\ref{pari0}) as in (\ref{pari1}), i.e.
\br
\label{summ1}
P(R) = R,\,\,\,\,P(\vp)= - \vp + \mbox{constant,}
\er

and

2. If the potential $V(R)$ in (\ref{nlsd}) evaluated on such a solution is even under the parity $P$, i.e. 
\br
\label{summ2}
P(V) = V
\er
such that 
\br
\label{summ20}
P(X) = - X,
\er  

3. Then, one has an infinite set of asymptotically conserved charges, i.e.
\br
\label{summ3}
  Q^{(n)} (t=+\infty) =  Q^{(n)} (t=-\infty),\,\,\,\, n=0,1,2,3,...
\er

Therefore, the values of the charges in the remote past, before the collision of the solitons,
are the same as in the far future, after the collision. In particular,  the deformed  NLS model (\ref{nlsd}) with potential (\ref{pot1}) can be shown to satisfy the requirements (\ref{summ1}) and (\ref{summ2}) for some field configurations, and then it belongs to the class of quasi-integrable theories.

\subsection{Space-reflection symmetry and conserved charges}
\label{srs}

There are some bright-soliton-like solutions which present a  space-reflection symmetry at any time, such that the sequence of the even order charges are exactly conserved. The demonstration of these results follows the same steps as in  \cite{jhep3}, where the defocusing NLS case  has been considered, and it involves an interplay between the space-reflection parity and the internal transformations in the affine Kac-Moody algebra underlying the anomalous Lax equation. So, for the details of the demonstration we refer to \cite{jhep3} and in the present work we will specialize to the case with $\eta < 0$ (focusing) and the vanishing boundary condition (\ref{vbc}) for bright solitons.  

So, we consider some special solutions which  exhibit  a space-reflection symmetry
\br
\label{spari11}
{\cal P}_x: (\widetilde{x},\,\widetilde{t}) \rightarrow  (-\widetilde{x},\,\widetilde{t}),\,\,\,\,\widetilde{x}\equiv x-x_{\rho},\,\widetilde{t}\equiv t-t_{\rho},       
\er
such that the fields $R$ and $\vp$ transform as
\br
\label{xsym} 
R(-\widetilde{x}, \widetilde{t}) \rightarrow R(\widetilde{x}, \widetilde{t}),\,\,\,\,\vp(-\widetilde{x}, \widetilde{t}) \rightarrow \vp(\widetilde{x}, \widetilde{t}). 
\er

As we will show below, the special case of 2-bright-solitons moving in opposite directions and  equal velocities, such that they undergo a head-on collision, exhibits a space-reflection symmetry.

The implication of this additional symmetry of the fields under space-reflection (for any shifted time), i.e. $R$ and $\vp$ being even fields, on the behaviour of the quantities $\a^{(3,-n)}$ deserves a further analysis.  The main results of \cite{jhep3} are summarized as follows.
  
1. The $\a^{(3, -n)}$'s with $n=0,2,4,...$ are even under ${\cal P}_x$. On the other hand, since $R$ is even under  ${\cal P}_x$ and that $X$ given in (\ref{an0}), is a $x-$derivative of a functional of $R$, one can show  that $X$ is odd under ${\cal P}_x$, i.e. ${\cal P}_x(X) = - X$ and so    
\br \nonumber
\b^{(n)} &=& -i \int_{-\infty}^{+\infty} dx \, X \a^{(3, -n)}\\
 &=& 0,\,\,\,\,\,\,\,\,\,\,\,\,\,\,\,\,\,\,\,\,\,\,\,\,\,\,n=0,2,4,...\label{qcon}
\er
 
2. Then, the anomalous conservation laws (\ref{ano1}) imply 
\br
\label{charge11}
\frac{d Q^{(n)}}{dt} =0,\,\,\,\,\,\,\,\,\,\,n=0,2,4,...
\er 

Consequently, the even order charges are exactly conserved. Notice that this property holds provided that the  potential $V$ depends only on the modulus $|\psi|$, and that the field components satisfy (\ref{xsym}). In \cite{jhep3} it has been performed a construction in perturbation theory around solutions of the usual NLS model, by expanding the equations of motion and the solutions into power series in $\epsilon$. The results  in \cite{jhep3} did show that the zero'th order approximation symmetries hold at higher orders and
 at all times, and the charges are exactly conserved.

Let us summarize the main results so far. The deformed NLS model presents an infinite number of asymptotically  conserved charges as in (\ref{summ3}) for solitons satisfying the space-time symmetry (\ref{pari0})-(\ref{pari1}) (main result of \cite{jhep2}) . In addition, for bright solitons satisfying the same space-time symmetry, as well as the space-reflection symmetry (\ref{spari11})-(\ref{xsym}) one can say even more. In this case,  the sequence of the even order charges become indeed exactly conserved (\ref{charge11}). So, the model supports infinite towers  of alternating conserved and asymptotically conserved charges provided that the solutions  satisfy the both, space-reflection and space-time symmetries, respectively.  
   
\section{Charges and anomalies}  
\label{charges}

Following \cite{jhep2, jhep3} let us write the first five charges and anomalies in (\ref{ano1}). Since we will consider the vanishing boundary condition it is no necessary to renormalize the quasi-conservation laws as in \cite{jhep2}, when the non-vanishing boundary conditions have been imposed at $x \rightarrow \pm \infty$. So, next we list the first charges and anomalies corresponding  to  the vanishing boundary conditions (\ref{vbc}). See the appendix for the relevant  expressions of $a_{x}^{(3,-n)}$ for $n=0,1,2,3, 4$ of the quasi-conservation laws (\ref{cons1}).
  
1. The $Q^{(0)}$ charge

One has 
\br 
Q^{(0)} &=& \int dx \,  [\frac{1}{2} \pa_x \vp ]\\
&=& -i \int dx \, \(  \frac{\bar{\psi} \pa_x \psi - \psi \pa_x \bar{\psi}}{2 |\psi|^2} \).
\er
The anomaly vanishes  
\br
\b^{(0)} = -i \int_{-\infty}^{+\infty} dx \, X = 0.
\er
The charge $Q^{(0)}$ is associated to the phase difference (or the phase jump) of the solutions.

2. The $Q^{(1)}$ charge 

We have 
\br
Q^{(1)} = 2  \eta \int_{-\infty}^{+\infty} dx \,|\psi|^2  .
\er 

This charge $Q^{(-1)}$ defines the normalization of the solution $\psi$ and it is related to the $U(1)$ internal symmetry of the model: $\psi \rightarrow e^{i \a} \psi,\,\,\a=\mbox{const}.$ The anomaly $\b^{(1)}$ vanishes identically.

3. The $Q^{(2)}$ charge

It can be shown that the anomaly $\b_{2}$ vanishes 
\br
\b^{(2)} = 2 \eta \int_{-\infty}^{ +\infty} dx \, \pa_x \Big[ V - \eta R^2\Big] = 0.
\er
So, one has 
\br
Q^{(2)} &=& 2  \eta \int_{-\infty}^{ +\infty} dx\, \Big[ R \pa_x \vp   \Big].\\
&=& 2  \eta \int_{-\infty}^{+\infty} dx\, \Big[\bar{\psi} \pa_x \psi- \psi \pa_x \bar{\psi}   \Big]. 
\er
This charge is related to the space translation symmetry of the model.  In the last line  the expression (\ref{exp1}) has been used. 

4. The $Q^{(3)}$ charge

The expression for $\b^{(3)}$  becomes
\br
\b^{(3)} &=& - 2  \eta \int_{-\infty}^{ +\infty} dx\, \pa_x \Big[ \frac{\d V}{\d R} - 2 \eta R\Big] R \pa_x \vp\\
&=& - 2  \eta \int_{-\infty}^{ +\infty} dx\, \pa_t \( V- \eta R^2\),
\er
where we have used the eq. of motion for $R$ given in (\ref{eq111}). By adding this total time derivative to the expression  in the l.h.s of  (\ref{ano1}), and discarding `surface' terms,  one can define the charge as \cite{jhep3} 
\br
Q^{(3)} = \int_{-\infty}^{ +\infty} dx\,\Big[ |\pa_x \psi|^2 + V\Big],
\er
such that  $\frac{d}{dt} Q^{(-3)}_{r} =0$. It is just the energy of the system related to time translations.

5. The $Q^{(4)}$ charge: the first asymptotically-conserved charge
 
We have that the charge and anomaly, respectively,  are
\br
\label{ch4}
Q^{(4)} &=& \frac{\eta}{4}   \int_{-\infty}^{ +\infty} dx\, \Big[ 12 \eta R^2 \pa_x \vp + 3  \pa_x \vp \frac{(\pa_x R)^2}{R}+  R \( (\pa_x \vp)^3 - 4 \pa_x^3 \vp\) \Big]\\
\b^{(4)}&=&- \eta  \int_{-\infty}^{ +\infty} dx\,\pa_x \( \frac{\d V}{\d R} - 2 \eta R\) \Big[ 6 \eta R^2 + \frac{3}{2} R (\pa_x \vp)^2 - 2 \pa^2_x R+ \frac{3}{2} \frac{(\pa_x R)^2}{R} \Big]\label{ano44}.
\er 
So, consider the  quasi-conservation law
\br
\label{qcon11}
\frac{dQ^{(4)}}{dt} = \b^{(4)}
\er
with 
\br
\b^{(4)} &\equiv & - \eta  \int_{-\infty}^{+\infty} dx \(V''[R] -2\eta \) \Big\{   \frac{3}{4} \pa_x R^2 (\pa_x \vp)^2 -\pa_x (\pa_x R)^2 + \frac{3}{2} \frac{(\pa_x R)^3}{R}\Big\}
\label{beta11},\\
&\equiv& \int_{-\infty}^{+\infty} \, dx\,  \g(x,t) \label{func11},
\er
where we have discarded a ``surface" term in (\ref{ano44}) and, for later purposes, defined the anomaly density  $\g(x,t)$.  The time integrated anomaly  becomes  
\br 
\label{tint} 
\int_{-\widetilde{t}_0}^{+\widetilde{t}_0}\b^{(4)}_r(t') dt' = \int_{-\widetilde{t}_0}^{+\widetilde{t}_0} \int_{-\infty}^{+\infty} \, dx \, dt'\, \g(x,t).
\er    
This is the first non-trivial non-vanishing anomaly. One notices that in the limit $\epsilon \rightarrow 0$ the anomaly $\b^{(4)}$ vanishes identically since $V''[R] \rightarrow 2 \eta$ in this limit. However, by inspecting the form of the anomaly density in (\ref{beta11}) one can have vanishing anomaly $\b^{(4)}$, associated to an exactly conserved charge $Q^{(4)} $, for solutions satisfying the symmetry (\ref{xsym}). Moreover, this anomaly vanishes for the free field continuous wave background, $\psi \sim e^{i(kx - wt)}$, i.e. $R = \mbox{const.},\,\,\frac{\vp}{2} = k x -w t$. In addition,  we can show, following a similar approach to the one in \cite{jhep3},  that this anomaly  vanishes for a general travelling solitary wave solution of the deformed NLS.  We will compute numerically the anomaly $\b^{(4)}$ for certain two-bright soliton configurations in the deformed NLS model (\ref{nlsd}) with potential (\ref{pot1}).

\section{Symmetries of focusing NLS and bright solitons}

\label{bright}
Next we discuss the both space-time and space-reflection symmetries in the general 2-bright soliton solutions of the integrable focusing NLS model. So,  consider the focusing NLS equation 
\br
\label{nls1}
i \pa_{t} \psi +  \pa_{xx} \psi - 2 \eta |\psi|^{2} \psi=0,\,\,\,\, \eta < 0.
\er
In the following we  refer to the relevant solitons of the integrable NLS model (\ref{nls1}) as  N-soliton (1-soliton, 2-soliton, etc.) solutions.
   
\subsection{Space-time parity transformation}
\label{spacetime}
Let us consider the 2-bright soliton solution \cite{jhep2}
\br
\label{sol2}
\psi_{o}(x,t) = \frac{e^{-\D/2}}{\sqrt{|\eta|}} e^{-i \O_{+}} \frac{\hat{\cal N}}{\hat{\cal D}},
\er
where
\br
\hat{\cal D} = \cosh{z_{+}} + e^{−\D} \cosh{z_{-}} - 16 \frac{|\rho_1 | |\rho_2|}{\L_{-}} \cos{[2 (\O_{-} + c)]}
\er
and
\br
\hat{\cal N} = e^{-\frac{z_{+}}{2}} e^{i \d_{-}} \Big[ |\rho_1| e^{-i (\O_{-} + c + \d_{+})} e^{\frac{z_{-}}{2}}
 + |\rho_2| e^{i (\O_{-} + c + \d_{+})} e^{-\frac{z_{-}}{2}} \Big] + \\
 e^{\frac{z_{+}}{2}} e^{-i \d_{-}} \Big[ |\rho_1| e^{-i (\O_{-} + c - \d_{+})} e^{-\frac{z_{-}}{2}}
 + |\rho_2| e^{i (\O_{-} + c - \d_{+})} e^{\frac{z_{-}}{2}} \Big].
\er
The parameter $\D$ is defined by 
\br
\label{delta}
e^{\Delta} = \frac{\L_{-}}{\L_{+}} =  \frac{(v_1-v_2)^2+ 4 (\rho_1-\rho_2)^2}{(v_1-v_2)^2+ 4 (\rho_1+\rho_2)^2},
\er
and the coordinates
\br
z_{+} = X_1 + X_2 + \D,\,\,\,z_{-} = X_1 - X_2,
\er
such that 
\br
\label{x12}
X_{i} = \rho_i (x - v_i t - x_i^{(0)}),\,\,\,\,\O_{i} = (\frac{v_i^2}{4}- \rho_i^2) t - \frac{v_i}{2}  x  + \theta_i + \zeta_i ,\,\,\,\,\, i = 1,2,
\er
where
\br
\d_{\pm} = \arctan{\frac{2(\rho_1\pm \rho_2)}{v1-v2}}.
\label{sol2parameters}
\er
The $\O_{\pm}$ exhibit homogeneous dependences on $z_{\pm}$ as $\O_{\pm}= \b_{\pm}^{+} z_{+}+\b_{\pm}^{-} z_{-}$, with $\b_{\pm}^{-}$ being some constants. These are defined as  
\br
\label{omegas}
\frac{\O_{1}-\O_{2}}{2} - \d_{+} = \O_{-} + c,\,\,\,\,\,\frac{\O_{1}+\O_{2}}{2} + \d_{-} = \O_{+} + d.
\er

The 2-soliton (\ref{sol2}) depends on eight free real parameters, i.e. $v_i,\,\rho_i,\,x_{i}^{(0)} (i=1,2), c\,\, \mbox{and}\,\, d $. Consider the space-time parity transformation
\br
\label{stsymm}
P:   (\widetilde{x}, \widetilde{t})\rightarrow (-\widetilde{x}, -\widetilde{t}),\,\,\,\,\, \widetilde{x}= x - x_{\D},\,\,\,\, \widetilde{t}= t - t_{\D},
\er
where $x_{\D}, t_{\D}$ are some constant parameters depending on  $v_i,\,\rho_i,\,x_{i}^{(0)} (i=1,2)$. In \cite{jhep2} it has been shown that the solution (\ref{sol2}) possesses the space-time parity symmetry  (\ref{stsymm}) provided that the parameter $c$ satisfies
\br
\label{cc}
c= n \frac{\pi}{2},\,\,\,\, n \in \IZ .
\er 
In terms of the components $R_0$ and $\vp_0$, namely for $\psi_0 =\sqrt{R_0} e^{i \frac{\vp_0}{2}}$, this symmetry property can be written as 
\br
\label{symm2br}
P: R_0 \rightarrow R_0 ;\,\,\,\,\,\,\,\,\,\,\vp_0 \rightarrow -\vp_0+ 2 \pi n. 
\er

The Fig. 1 shows these functions for $\rho_1 =2, \rho_2= 4, v_1 = -5, v_2 = 5$ for three successive times, before $t_i$ (green), during  $t_c$ (blue) and after collision $t_f$(red).

In \cite{jhep2} the initial conditions for the simulations have been used the integer $n=0$ ($c=0$ in (\ref{cc})), as well as other values for $c$ corresponding to non-integer values of $n$, in order to examine the behaviour of the integrated anomalies associated to asymptotically conserved charges for the collision of two bright solitons in the deformed focusing NLS model. The $c \neq  n \frac{\pi}{2}\,  (n\in \IZ)$ cases have shown to exhibit non-vanishing integrated anomalies for the soliton collisions. However, as we will show below, an accurate numerical computation reveals that the $\beta^{(4)}$ anomaly indeed vanishes for any two-bright solitons, regardless of the existence of the symmetry (\ref{symm2br}).

\begin{figure}
\centering
\label{fig1}
	\includegraphics[width=12cm,scale=3, angle=0,height=4.5cm]{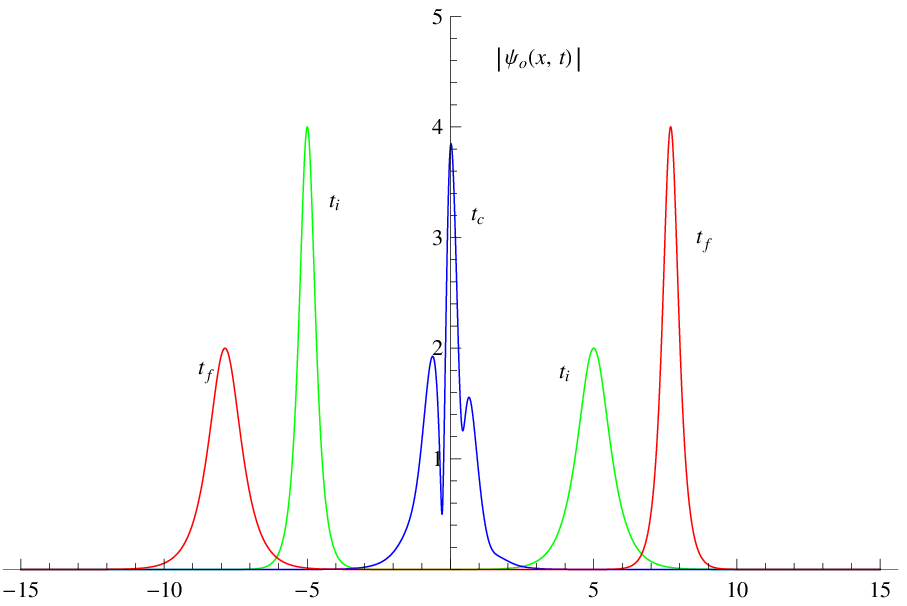}
\includegraphics[width=12cm,scale=3, angle=0,height=4.5cm]{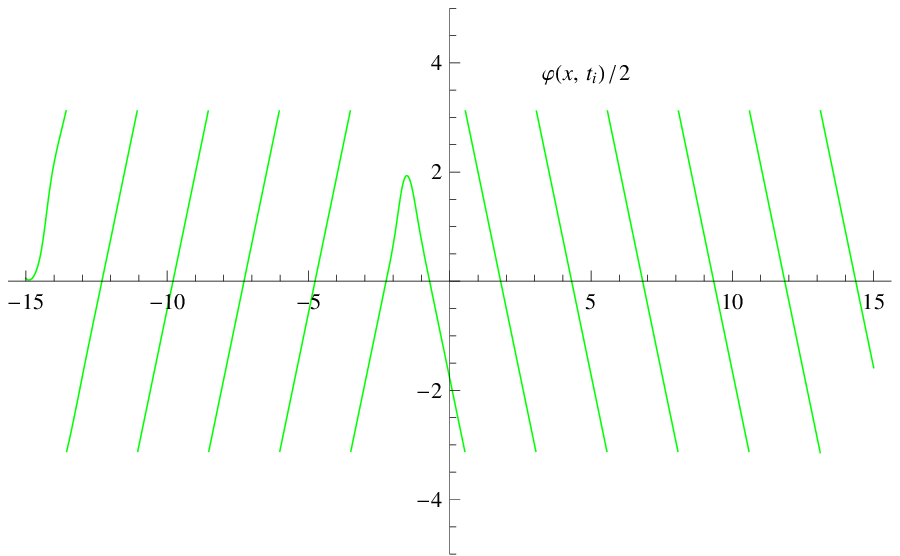}
\includegraphics[width=12cm,scale=3, angle=0,height=4.5cm]{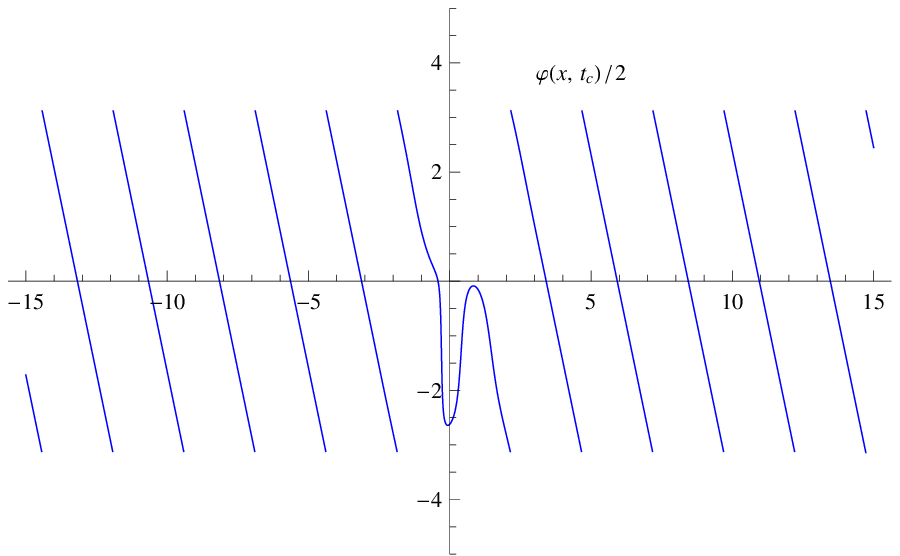}
\includegraphics[width=12cm,scale=3, angle=0,height=4.5cm]{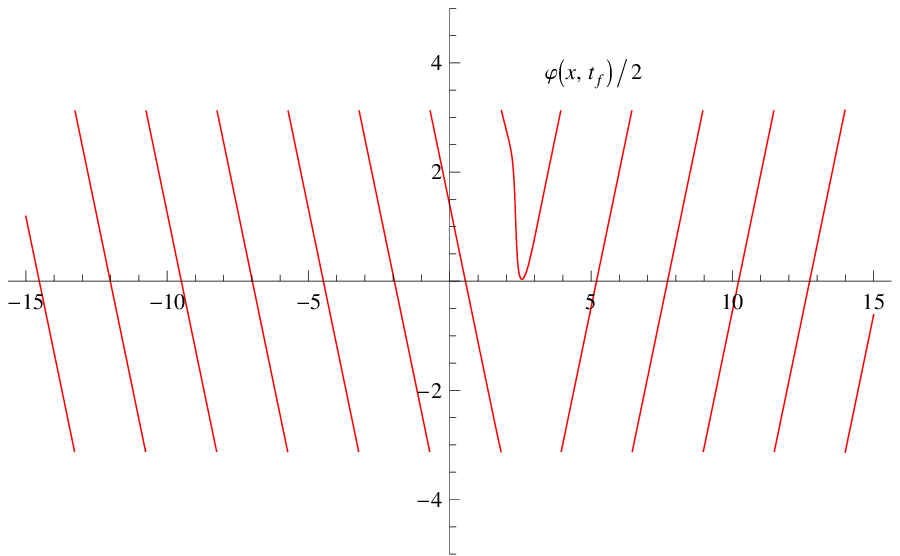}
\parbox{6in}{\caption{(color online) The asymmetric amplitude $|\psi(x,t)|$ vs $x$ (top left)  and  phase $\frac{\vp}{2}(x, t)$ vs $x$ (top right and bottom line) for two bright solitons, with $v_1=-v_2=-5$,\,amplitude$_{(1)} = 2$,\,amplitude$_{(2)} = 4$, for three successive times, initial $t_i$ (green), collision  $t_c$ 
(blue) and final $t_f$ (red), respectively. At $t_c$ the solitons form an asymmetric collision pattern, and after collision  each soliton emerges with its characteristic velocity and shape.}}
\end{figure}

The general form of the 2-bright soliton solutions as presented in (\ref{sol2}) have been derived in \cite{jhep2} using the Hirota method. For an earlier calculation of the N-soliton, in the context of the dressing transformation and tau function methods of the AKNS model, see e.g.  \cite{solvint}.  

\subsection{Space-reflection parity transformation}
\label{space1}

Next, let us discuss the behaviour of the fields $R$ and $\vp$ for the 2-soliton solutions such that   
\br
\label{srsymm}
P_x:   (\widetilde{x}, t)&\rightarrow& (-\widetilde{x}, t),\,\,\,\,\,\,\,\,\,\,\, \widetilde{x}= x - x_{\rho},\\
P_x :  R_0(-\widetilde{x}, t) &\rightarrow & R_0(\widetilde{x}, t) ;\,\,\,\,\,\,\,\,\,\,\vp_0(-\widetilde{x}, t) \rightarrow \vp_0(\widetilde{x}, t), \label{srsymm1}
\er
for any time $   t $. The real parameter $x_{\rho}$ depends on $v_i,\,\rho_i,\,x_{i}^{(0)} (i=1,2), c$\, and $d$.

Next we show that the 2-bright-soliton solution (\ref{sol2}) contains a subset of solutions which exhibit the above space-reflection symmetry. Consider the head-on collision of 2-bright solitons with equal amplitudes and opposite equal velocities. Accordingly, let us take the  following set of special parameters
\br
\label{srparam1} 
\rho_1 = \rho_2 = \rho,\,\,\,\,\,v_1=-v_2 = -v.
\er
So, from the equations (\ref{delta})-(\ref{omegas}) one can get
\br
z_{+} &=& 2 \rho x-\rho_1 x_{1}^{(0)}-\rho_2 x_{2}^{(0)}  +\D,\,\,\,\,\,z_{-} = 2 v t+\rho_2 x_{2}^{(0)}-\rho_1 x_{1}^{(0)},\\
 d+ \O_{+} & = & (\frac{v^2}{4}-\rho^2) t + \d_{-} + \frac{\theta_1+\zeta_1+\theta_2+\zeta_{2}}{2},\\
 c+ \O_{-} & = & \frac{v}{2} x - \d_{+} + \frac{\theta_1+\zeta_1-\theta_2-\zeta_{2}}{2},\\
 \d_{+} &=& - \arctan{\frac{2 \rho}{v}},\,\,\,\,\,\,\,\d_{-} = 0,\,\,\,\,\,\,\, e^{\D} = \frac{\L_-}{\L_+} = \frac{v^2}{v^2+4 \rho^2}.
\er
Without loss of generality, let us set the above parameters such that  $x_{\rho} =0$ which amounts to impose 
\br
\D &= &\rho [ x^{(0)}_1 + x^{(0)}_2 ],\nonumber \\
{\hat c} & \equiv & \d_{+} -\frac{(\theta_1+ \zeta_1)-(\theta_2+ \zeta_2)}{2}\nonumber \\
 & = &  \frac{\pi}{2}  n , \,\,\,\,\,n \in 2\IZ, \label{hc}
 \er 
where the new parameter ${\hat c}$ has been defined in terms of the even integers $n \in 2\IZ$; this is a subset of the integers $n\in \IZ$ appearing in the definition of the parameter $c$ in (\ref{cc}).  Then the above relationships imply 
\br
\nonumber
R_0(x,t) =\frac{e^{-\D}}{|\eta|} \frac{|\hat{\cal N}|^2}{{\cal D}^2} &=& \frac{e^{-\D}}{|\eta|} \frac{2 \rho^2}{\hat{\cal D}^2} \Big[ \cosh{(2 v t-2 \rho x)}+\cosh{(2 v t+2 \rho x)}+ \\&& 2 \cos{(2 \d_{+})} + e^{2 \rho x} \cos{(v x - 2\d_{+})} +
 e^{-2 \rho x} \cos{(v x + 2\d_{+})} + \nonumber \\
&& 2 \cosh{(2 v t)} \cos{(v x)}\Big] \label{r0}\\ 
\hat{\cal D} &=&  \cosh{(2 \rho x)} + e^{-\D} \cosh{(2 v t)}- 16 (\frac{\rho}{v})^2 \cos{(v x)}\\
\label{vp0}
\frac{\vp_0(x,t)}{2} &=& \arctan{\Big\{\(\frac{1-e^{2 v t}}{1+e^{2 v t}}\) \frac{\sin{(\frac{v x}{2} + \d_{+})} - e^{2 \rho x} \sin{(\frac{v x}{2} - \d_{+})}}{\cos{(\frac{v x}{2} + \d_{+})} + e^{2 \rho x} \cos{(\frac{v x}{2} - \d_{+})}}\Big\}} - (\frac{v^2}{4}-\rho^2) t.
\er
Clearly, the above components $R_0$  and $\vp_0$ of the 2-bright soliton $\psi_0$  are even parity eigenstates of the space-reflection  operator $P_x$ (\ref{srsymm})-(\ref{srsymm1}) provided that $x_{\rho} =0$. The Fig. 2 shows these functions for $\rho = 2$ and $v=5$ and for three successive times, before   $t_i$ (green), during $t_c$ (blue) and after collision $t_f$(red).

\begin{figure}
\centering
\label{fig2}
	\includegraphics[width=12cm,scale=3, angle=0,height=4.5cm]{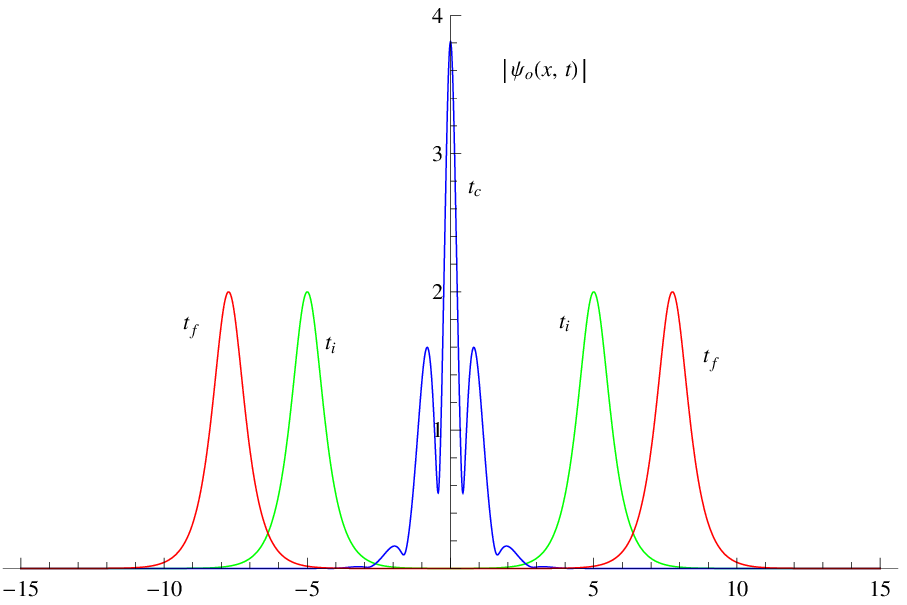}
\includegraphics[width=12cm,scale=3, angle=0,height=4.5cm]{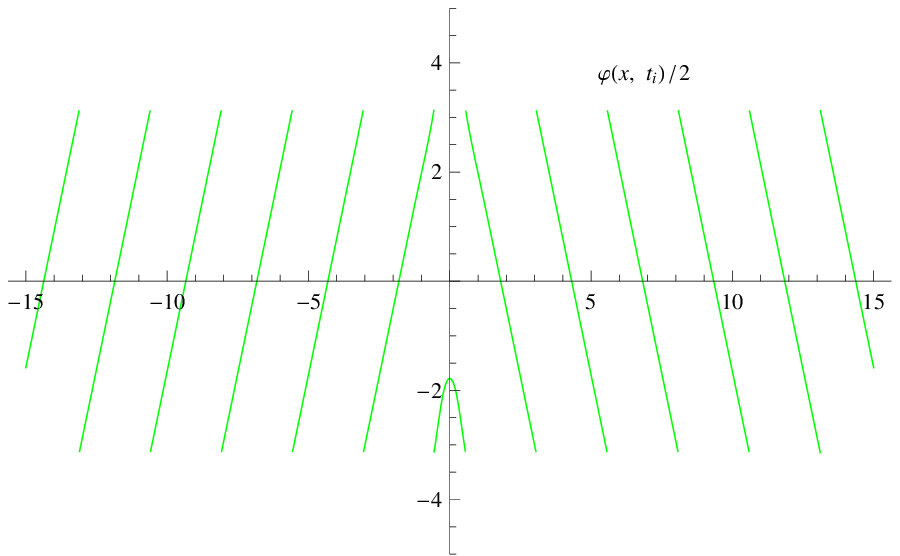}
\includegraphics[width=12cm,scale=3, angle=0,height=4.5cm]{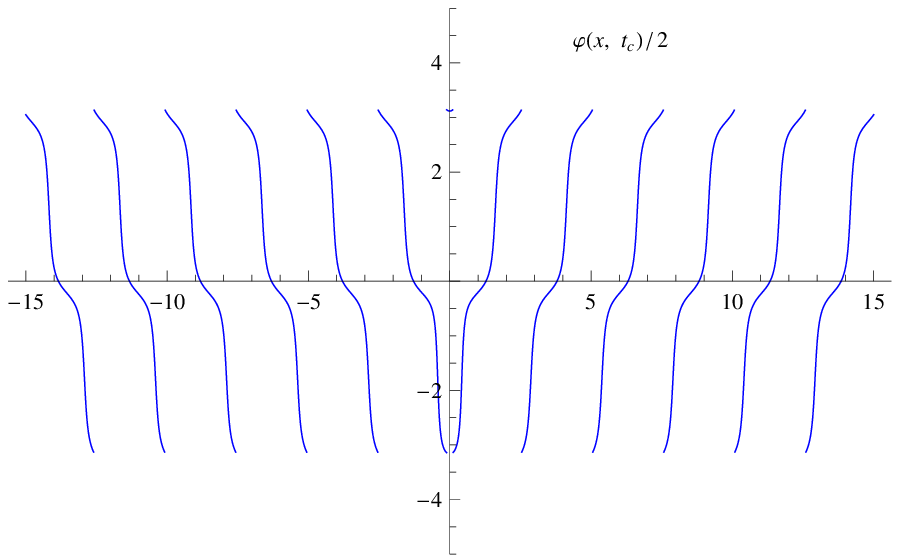}
\includegraphics[width=12cm,scale=3, angle=0,height=4.5cm]{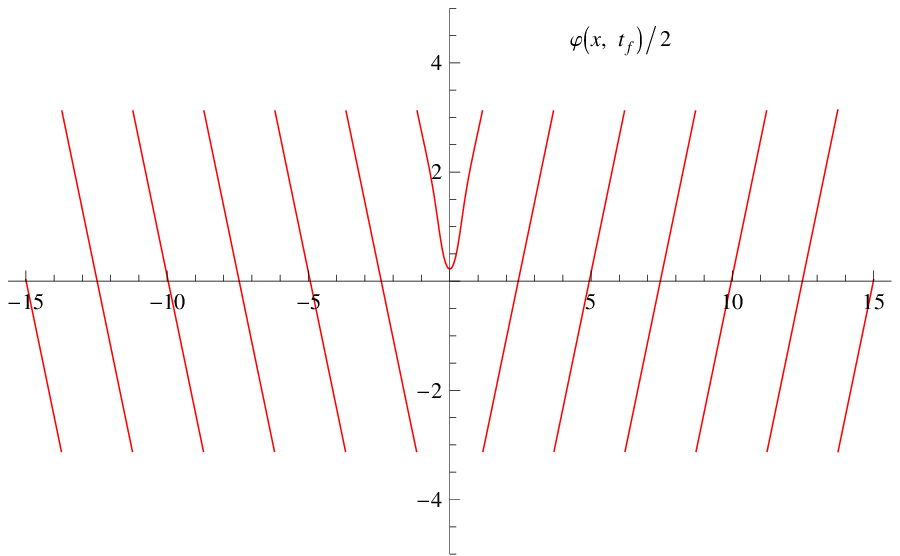}
\parbox{6in}{\caption{(color online) Amplitude $|\psi(x,t)|$ vs $x$ (top left) and phase $\frac{\vp}{2}(x, t)$ vs $x$ (top right and bottom line) for two bright solitons with space-reflection symmetries at any time,  for 
$v_1=-v_2=-5$, amplitude$_{(1)}=$amplitude$_{(2)}=2$, for three successive times, initial $t_i$ (green), collision  $t_c$ 
(blue) and final $t_f$ (red) times, respectively.}}
\end{figure}

For later purposes let us  perform the asymptotic expansion of the above 2-bright soliton solution (\ref{sol2})-(\ref{sol2parameters})  and decompose it as a sum of individual 1-solitons. The quantity $X_{i},\,i=1,2$ in (\ref{x12}) characterize  each of  the individual 1-solitons. In fact, the  asymptotic form of  (\ref{sol2}) as $X_2 \rightarrow - \infty$ becomes
\br
\label{asymp1}
\psi_0 \sim \frac{1}{\sqrt{|\eta|}} i  |\rho_1| e^{-i \Big[(\frac{v_1^2}{2} - \rho_1^2) t - \frac{v_1}{2} x +\theta_1+\zeta_1 \Big]} \mbox{sech} [\rho_1 (x - v_1 t - x_{1}^{(0)})].
\er
Actuallly, this is a 1-bright soliton of the NLS model with four free parameters $\rho_1, v_1, \theta_1, x_{1}^{(0)}$.

Similarly, the asymptotic form of (\ref{sol2}) as $X_1 \rightarrow + \infty$ becomes
\br
\label{asymp2}
\psi_0 \sim \frac{1}{\sqrt{|\eta|}} i   |\rho_2| e^{-i \Big[(\frac{v_2^2}{2} - \rho_2^2) t - \frac{v_2}{2} x +\theta_2+\zeta_2+2(\delta_{+}+\delta_{-})\Big]} \mbox{sech} [\rho_2 (x - v_2 t - x_{2}^{(0)} + \frac{\Delta}{\rho_2})].
\er
In fact, this is the second 1-bright soliton   with four free parameters $\rho_2, v_2, \theta_2, x_{2}^{(0)}$. 

Let us define the constant phases, respectively, as
\br
\phi_2\equiv \theta_2 + \zeta_2 + 2(\delta_{+}+\delta_{-}),\,\,\,\,\phi_1\, \equiv  \theta_1 + \zeta_1\label{nls2par}.
\er

Notice that the general 2-bright soliton  (\ref{sol2})-(\ref{sol2parameters}) exhibits eight free paramateres, i.e. the set  
\br
\{\rho_i, v_i, x_{i}^{(0)},  c,  d\},\,\,\,\, \mbox{for}\,\,\,  i=1,2;
\er
where the pair $\{\theta_{i},\,i=1,2\}$ has  been traded for the  parameters $c$ and $d$ in (\ref{omegas}); so,  without loss of generality we will  consider the conditions
\br
\label{inpos}
 x_{2}^{(0)} = \frac{\Delta}{\rho_2}- x_{1}^{(0)},\,\,\,\,\,\,\,d = 0. 
\er
Then, for 2-bright soliton solutions (\ref{sol2})-(\ref{sol2parameters}) satisfying the space-time parity symmetry (\ref{stsymm}) we can write the phase  parameters as
\br
\phi_1\, & \equiv &  c + \frac{\phi^{+}+\phi^{-}}{2};\,\,\,\phi_2\, \equiv  -c  + \frac{\phi^{+}-\phi^{-}}{2} \label{phases1},\,\,\,\,\,\,\,c=\frac{\pi}{2} n ,\,\,\,n \in \IZ\\
\phi^{+} &\equiv & \frac{1}{8 v_{r} \rho_1 \rho_2} \Big[ -8  x_1^{(0)} \rho_1 \rho_2 \(v_1 v_2 +2(\rho_1^2+ \rho_2^2)\) + 16 v_r  \rho_1 \rho_2 \arctan{(\frac{2 \rho_{+}}{v_{r}})} + \nonumber\\
&&\(\rho_{-} (v_1^2- v_2^2) + 2 v_1 v_ 2 \rho_{+} + 4 \rho_{+}  (\rho_1^2 +\rho_2^2)\)  \log{(1- \frac{16 \rho_1 \rho_2}{v_r^2+4 \rho_{+}^ 2})} \Big]\nonumber \\
\phi^{-} &\equiv &-\frac{1}{8 v_{r} \rho_1 \rho_2} \Big[ 16 v_{r} \rho_1 \rho_2 \arctan{(\frac{2 \rho_{-}}{v_{r}})} + \rho_{-} \(16 x_1^{(0)} \rho_1 \rho_2 \rho_{+} -(v_r^2+4 \rho_{+}^{ 2}) \log{(1- \frac{16 \rho_1 \rho_2}{v_r^2+4 \rho_{+}^ 2})}\)\Big]\nonumber\\
v_{r}\ &\equiv &  v_1 - v_2,\,\,\,\,\rho_{\pm} \equiv  \rho_1 \pm \rho_2,\nonumber
\er
where the parameter $c$, characterizing this type of symmetry,  has been defined in (\ref{cc}). 
So, once the individual 1-soliton parameters $\rho_i, v_i\,$ and $x_{i}^{(0)}$ ($i=1,2$) have been chosen, it is left the free parameter $c$, such that  for space-time parity symmetric solutions it must get the value (\ref{cc})  .
 
Therefore, for 2-bright soliton configuration with space-time parity symmetry,   the relative constant phase of its asymtotic expansion becomes
\br
\label{phi12}
\phi_1 - \phi_2 &=&2 c +  \phi^{-} \\
&=&\pi n + \phi^{-},\,\,\,\,\,n\in \IZ .\label{phi122}
\er
Similarly, for 2-bright soliton solutions (\ref{sol2})-(\ref{sol2parameters}) satisfying the space-reflection parity symmetry (\ref{srsymm}) we consider the relationships (\ref{srparam1}) between  the parameters; so,  since $\rho_1=\rho_2 \Rightarrow \rho_{-}=0$ one has  that $\phi^{-} = 0$ form  (\ref{phases1})-(\ref{phi12}), then  the relative constant phase for space-reflection symmetric configuration becomes 
\br
\label{phi12rs}
 \phi_1 - \phi_2 &=& {\hat{c}}\\
                    &=&\pi n,\,\,\,\,\,n\in \IZ,\label{phi12rs2}
\er   
where the parameter $\hat{c}$ has been defined in (\ref{hc}).

In our simulations below we will consider the above two asymptotic 1-solitons of the 2-bright soliton as a suitable initial condition and their relevant space-time symmetries will be encoded in the corresponding  relationships between the parameters.  
 
\section{Simulations}

\label{simul}
The deformed NLS model (\ref{nlsd}) with potential (\ref{pot1}) possesses the solitary wave (\ref{solitary}), and an analytic expression for a two-solitary wave is not  known. So, we would take two one-bright solitary waves some distance apart as the initial condition for our numerical simulations. However,  this approach presents a drawback,  we would not have  the relevant information to fix the phases of the initial individual solitons, and so  determine the relative phase, which must be provided as the initial condition.  Alternatively,  the collision of two-bright solitons in the deformed  NLS equation (\ref{nlsd}) can be simulated  numerically by considering the initial condition $\psi_0(x)$ defined as
\br
\label{nls2}
\psi_0(x) =  \rho_1\,e^{-i \phi_1}\, e^{i \frac{v_1 x}{2}}\,\mbox{sech}[\rho_1 (x - x_0)] +  \rho_2 \, e^{-i \phi_2}\, e^{i \frac{v_2}{2} x} \,\mbox{sech}[\rho_2 (x + x_0)], 
\er  
where two 1-bright soliton solutions of the usual NLS  model have been located some distance apart. 

The initial condition (i.c.) (\ref{nls2}) deserves some comments in the context of the asymptotic expansions   (\ref{asymp1}) - (\ref{asymp2})  of the general 2-bright soliton (\ref{sol2})-(\ref{sol2parameters}). First,  the two asymptotic wave forms in  (\ref{asymp1}) - (\ref{asymp2}) for $\eta = -1 $ and disregarding an overall phase factor $e^{i \pi/2}$, can be combined at $t=0$ by stitching them at the middle point $x=0$, in order to produce an uniformly valid solution, which is justified  by assuming  that the error terms  of the general 2-bright soliton solution (\ref{sol2})-(\ref{sol2parameters})  are exponentially very small.  Second, the above i. c.  is suitable for simulating the two-soliton interaction of the deformed model (\ref{nlsd})-(\ref{pot1}) such that  a general set of parameters $\{\rho_i, v_i, x_{0},  \phi_i \}$ for $i=1,2$ is assumed. The pair of parameters $\{c, d\}$ has been traded for the $\phi_i$'s, whereas  the pair  $\{x_{1}^{(0)}, x_{2}^{(0)}\}$ is traded for the single parameter $x_0$, provided that the two solitons are  equidistantly located around the origin $x=0$.  Third,  since (\ref{nls2}) is an approximate Ansatz, suitable as an initial condition for the simulation of two-bright soliton interaction of the deformed model, it is not expected to exhibit the relevant space-time and space parity symmetries of the full 2-bright soliton (\ref{sol2})-(\ref{sol2parameters}).  Fourth,  for space-time parity symmetric configurations  the i.c. must consider the relative phase (\ref{phi12})- (\ref{phi122}). In addition, for a space-reflection symmetric configuration we must have equal amplitudes $\rho_1=\rho_2$ and equal and opposite velocities $v_1 = - v_2$ and  its relative phase must satisfy  (\ref{phi12rs})-(\ref{phi12rs2}). In particular,  the i.c. (\ref{nls2}) possesses the space-reflection symmetry (\ref{srsymm1}) for $\phi_1=\phi_2\equiv \phi$, implying   the integer value  $n=0$ in (\ref{phi12rs2}).  Finally, since the deformed NLS (\ref{nlsd}) possesses the  internal symmetry $\psi \rightarrow e^{i \alpha} \psi,\,\,\,\alpha \equiv const.$, an  overall phase  $e^{-i \phi}$  in the i.c. can always be absorbed.

It is instructive  to consider two types of i.c.'s in (\ref{nls2}): the type I i.c. \,defined for $\phi_1=\phi_2$ with integer $n$ (type IA) and non-integer $n$ (type IB); and,  the type II\, i.c. \,defined for  $\phi_1\neq \phi_2$ with integer $n$ (type IIA) and non-integer $n$ (type IIB):
\br
\mbox{Type}\, I:   \phi_1=\phi_2   \left\{ \begin{array}{cc}
 IA  : & n \in \IZ \\
 IB:  & n =\mbox{non-integer} 
\end{array}  \right.   ;\,\,\,\,\,\,\,\,  
\mbox{Type} \, II:   \phi_1\neq \phi_2   \left\{ \begin{array}{cc}
 IIA  : & n \in \IZ \\
 IIB:  & n =\mbox{non-integer} 
\end{array} . \right.   
\er
Notice that the i.c.'s  IA and IIA are suitable for simulating the space-time symmetric configurations. On the other hand,  these IA and IIA i.c.'s, provided with the parameters relationships (\ref{srparam1}),  are suitable for the simulation of space-reflection symmetric configurations. According to our discussion above, an overall phase can be absorbed, so in the  case IA with space-reflection symmetry configuration, without loss of generality,  we will  consider the integer $n=0$ and vanishing phases $\phi_1=\phi_2\equiv 0$, which  satisfy  (\ref{phi12rs2}). 

The domain of the simulation is considered to be ${\cal D} =[-L,L] $ with $L=15$, mesh size $h = 0.017$ and time step $\tau = 0.00011$. The domain ${\cal D} $ is chosen such that the effect of the extreme regions near the points $x = \pm L$ do not interfere the dynamics of the solitons, i.e. the boundary condition (\ref{vbc}) is satisfied for each time step. In our numerical simulations we will  use the so-called time-splitting cosine pseudo-spectral finite difference (TSCP) method  \cite{bao}.

The two solitons are initially centered at $\pm x_0$ ($x_0>0$), the soliton centered initially at $-x_0$ ($t=0$) moves to the right with velocity $v_2 > 0$, whereas the soliton initially ($t=0$) centered  at $x_0$ travels to the left with velocity ($v_1 <0$). Notice that the direction of motion of each soliton is related to the sign of its phase slope. In addition, we will consider initially well-separated solitons, i.e. the parameter $2x_0$ is chosen to be  several times the width of the solitons ($\sim \frac{1}{\rho_i},\,i=1,2$)  and $2x_0 <  2 L$. So, the initial condition considers  two NLS bright solitons  which are stitched together at the middle point, and then we allow  the scattering of them, absorbing the radiation at the
edges of the grid. It amounts to maintain the vanishing boundary condition (\ref{vbc}) at the edges of the grid for each time step of the numerical simulation.

The Figs. 3-10 show the results of the simulations of two-bright soliton collisions of the model (\ref{nlsd})-(\ref{pot1}) with the types IA and IB  i.c.'s,  where  $\phi_1=\phi_2=0$  in (\ref{nls2}) is considered.  The two-soliton collisions with equal amplitudes, $\rho_1=\rho_2 $, and equal and opposite velocities, $v_1=-v_2=- 5$, corresponding to $n=0$ in (\ref{phi12rs}) (type IA) , and the relevant anomalies have been considered in Figs. 3-4 with parameters  $\epsilon = \pm 0.06$. These Figs. show the space-reflection symmetric functions $R(x,t)$ for initial (before collision), collision and final (after collision) times. The Fig. 3  also shows the symmetric  phase $\vp$ for initial and final times (middle figures).   

In the Figs. 5 and 6 we have simulated the collision process of two-solitons with different amplitudes and $\phi_1=\phi_2=0$  for $\epsilon = -0.03$ and  $0.06$, which correspond to non-integer parameters   (type IB i.c.)  $n=-1.18$ and $n=-3.734$, respectively,  in (\ref{phi12}).  The Figs. show the asymmetric function $R(x,t)$ (top left figures)  under space-reflection for three successive times. The Fig. 5 shows the asymmetric phase (middle figures) under space-reflection  before collision (green), collision (blue) and after collision (red) times, respectively. 

In the Figs. 3-6 (bottom figures) the relevant anomaly $\beta^{(4)}(t)$  (\ref{beta11}) and the integrated anomaly  $\int^t dt' \beta^{(4)}(t')$ (\ref{tint}) as functions of time, are plotted for several parameters. In Figs. 3-6 (top right) we plot  the anomaly density $\g(x,t)$ (\ref{func11}) as function of the space variable $x$ for three successive times in Figs 3-5 and at the collision time $t_c$ in Fig. 6.  One can see qualitatively the behavior of this function, for different parameters,  that would imply the vanishing of the anomalies. In fact, in Figs. 3-4 this density is an odd function of $x$, rendering a vanishing anomaly $\beta^{(4)}$ upon $x-$integration. There may be any other reasons for the  vanishing of the anomaly  $\beta^{(4)}$; in fact,  in Figs. 5-6 one notices the appearance of vanishing anomalies, within numerical accuracy, despite  the asymmetric anomaly densities.

In the Figs. 7-10 we also considered a variety of amplitudes and velocities with $\phi_1=\phi_2=0$  (type IB i.c.) , for fast and slow solitons. All of these figures show vanishing anomalies $\beta^{(4)}(t)$ and $t-$integrated anomalies $\int^t dt' \beta^{(4)}(t')$, within numerical accuracy.

The Figs. 11-17 show the results of the simulations of two-bright soliton collisions of the model (\ref{nlsd})-(\ref{pot1}) with the type IIA and IIB  i.c.'s,  where  $\phi_1\neq \phi_2$  in (\ref{nls2}) is considered.  In the Figs. 11 and 13  the relevant anomaly $\beta^{(4)}(t)$  (\ref{beta11}) (bottom left) and the integrated anomaly  $\int^t dt' \beta^{(4)}(t')$ (bottom right)  (\ref{tint}) as functions of time, are plotted for several parameters for type IIB; and in Fig. 12 for type IIA collision. In these Figs. (top right) we plot  the anomaly density $\g(x,t)$ (\ref{func11}) as function of the space variable $x$ for three successive times. In all of these Figs. one notices the vanishing of the anomalies, within numerical accuracy, despite  the asymmetric anomaly densities. 

In the Figs. 14-17 we have considered a variety of amplitudes and velocities of type IIA and IIB. All of these figures show vanishing anomalies $\beta^{(4)}(t)$ and $t-$integrated anomalies $\int^t dt' \beta^{(4)}(t')$, within numerical accuracy.

\begin{figure}
\centering
\label{fig3}
\includegraphics[width=2cm,scale=6, angle=0, height=4cm]{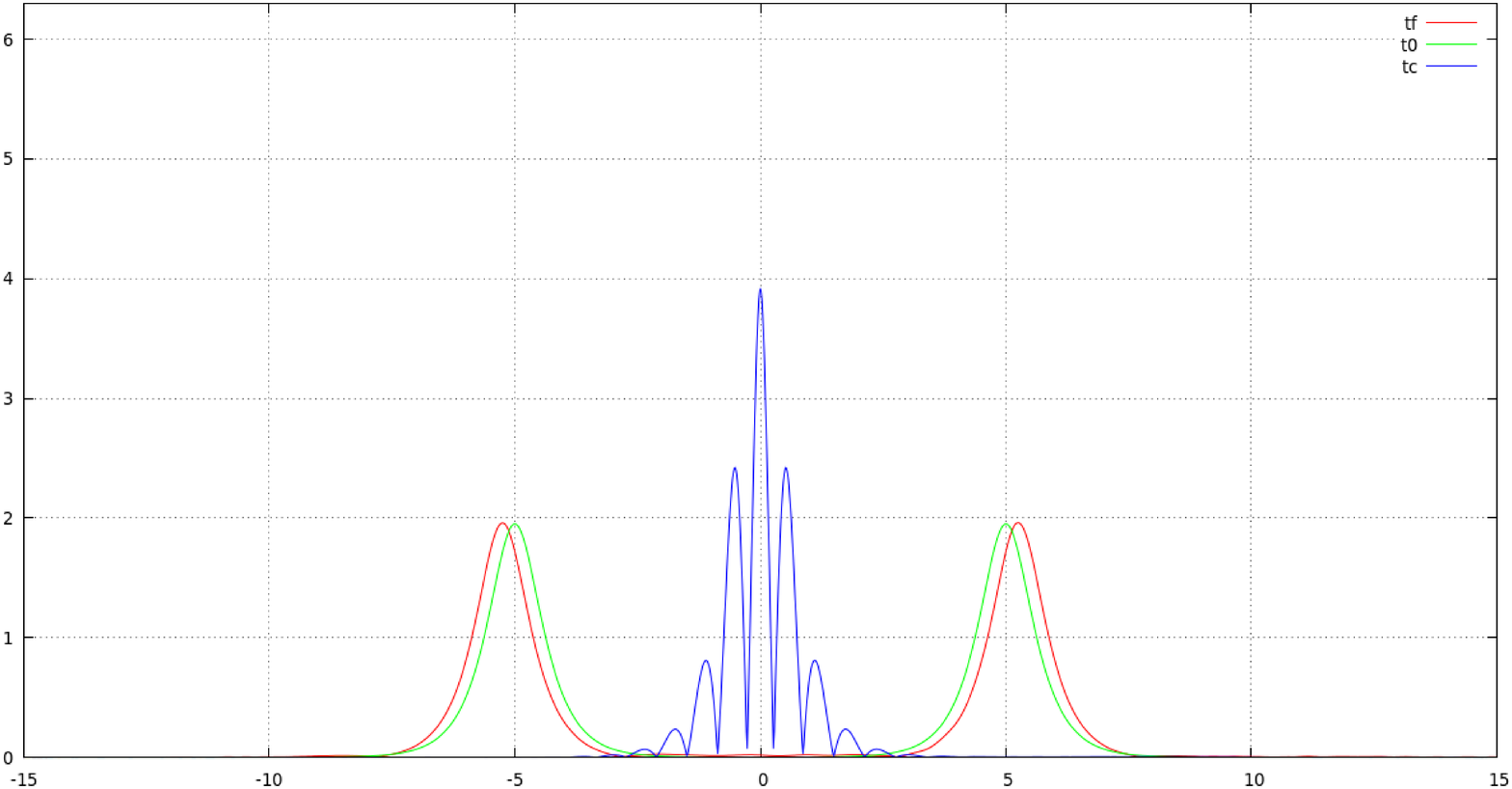} 
\includegraphics[width=2cm,scale=6, angle=0, height=4cm]{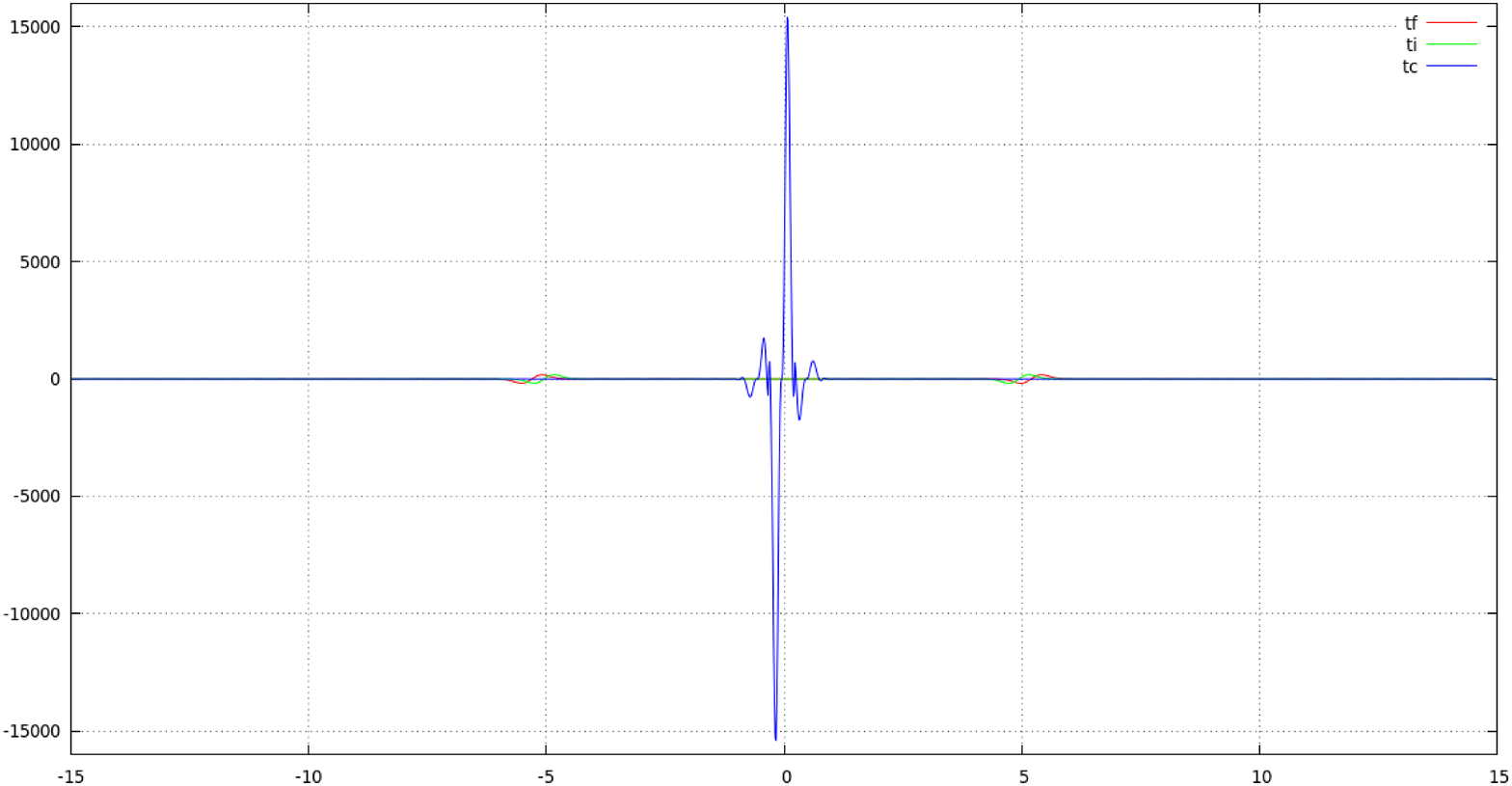}
\includegraphics[width=2cm,scale=6, angle=0, height=4cm]{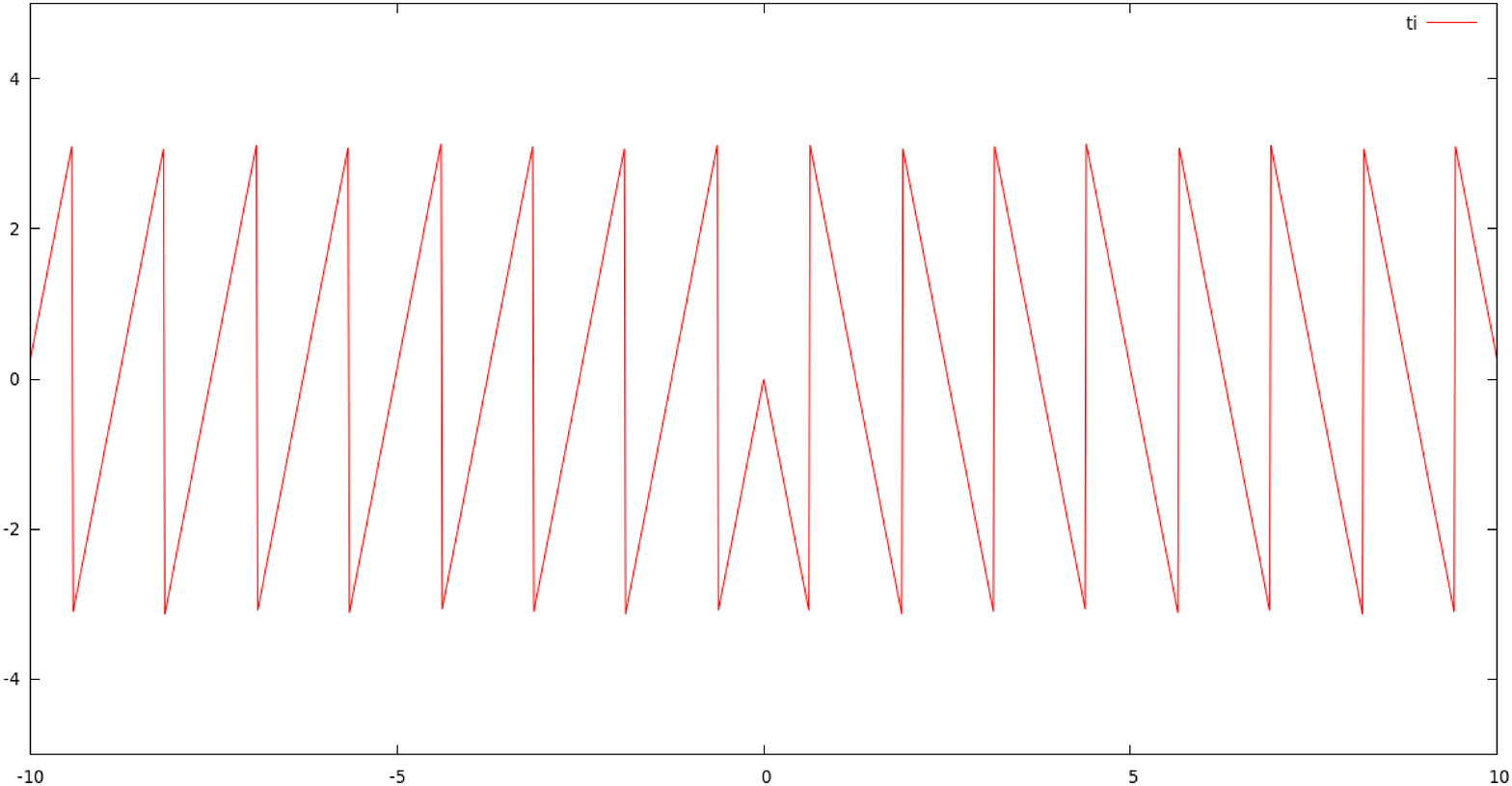} 
\includegraphics[width=2cm,scale=6, angle=0, height=4cm]{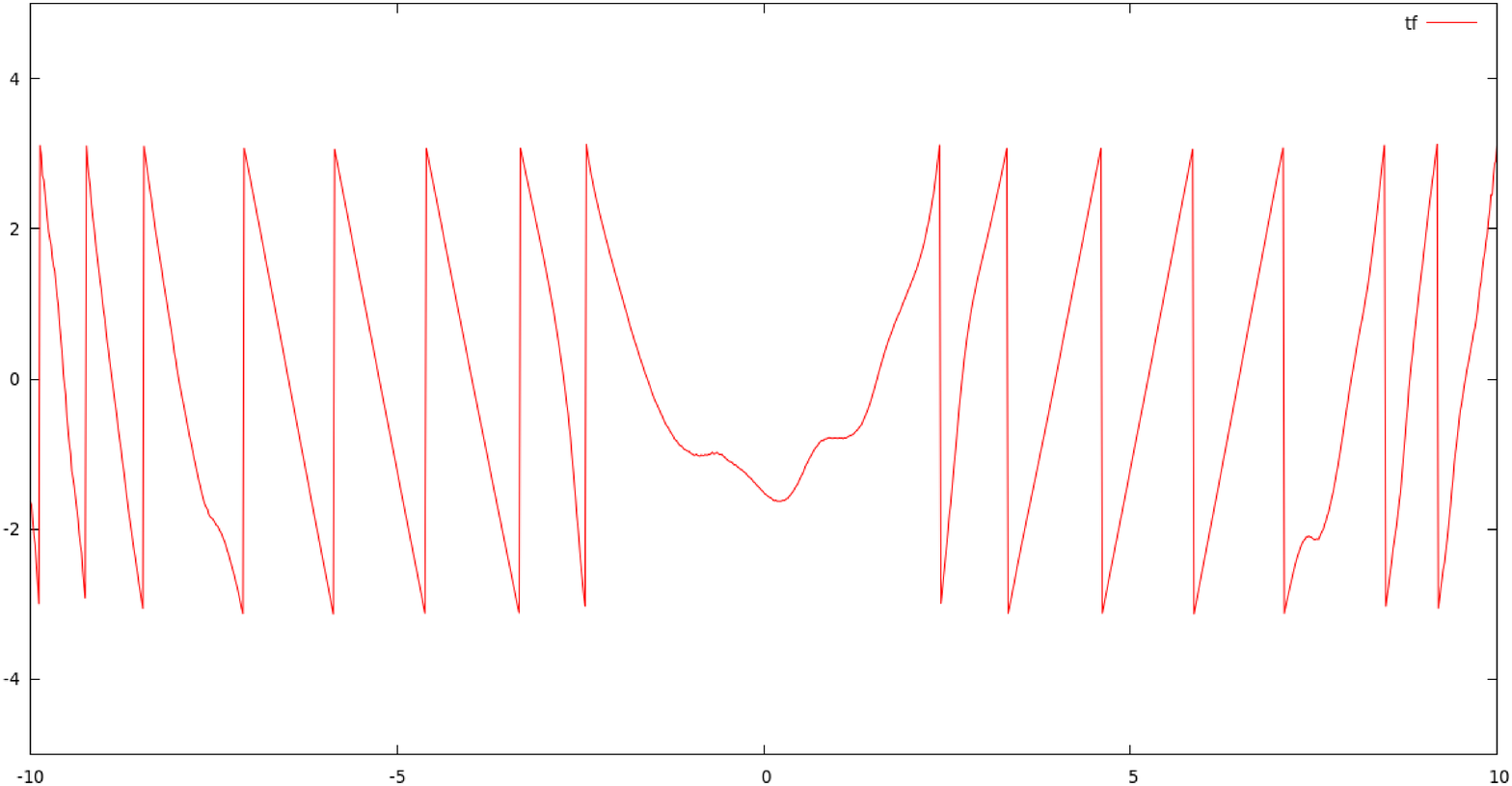}
\includegraphics[width=2cm,scale=6, angle=0, height=4cm]{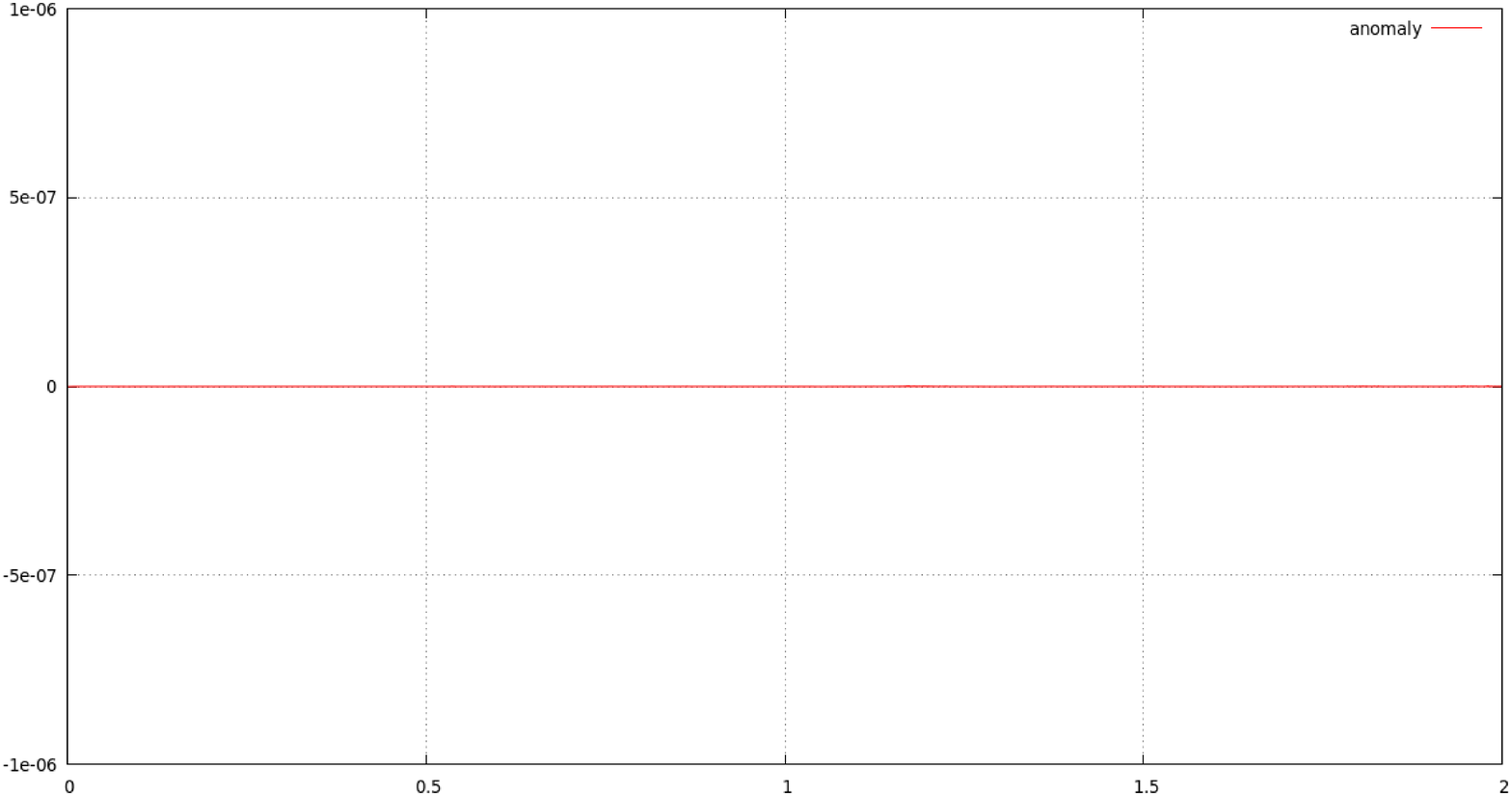}
\includegraphics[width=2cm,scale=6, angle=0, height=4cm]{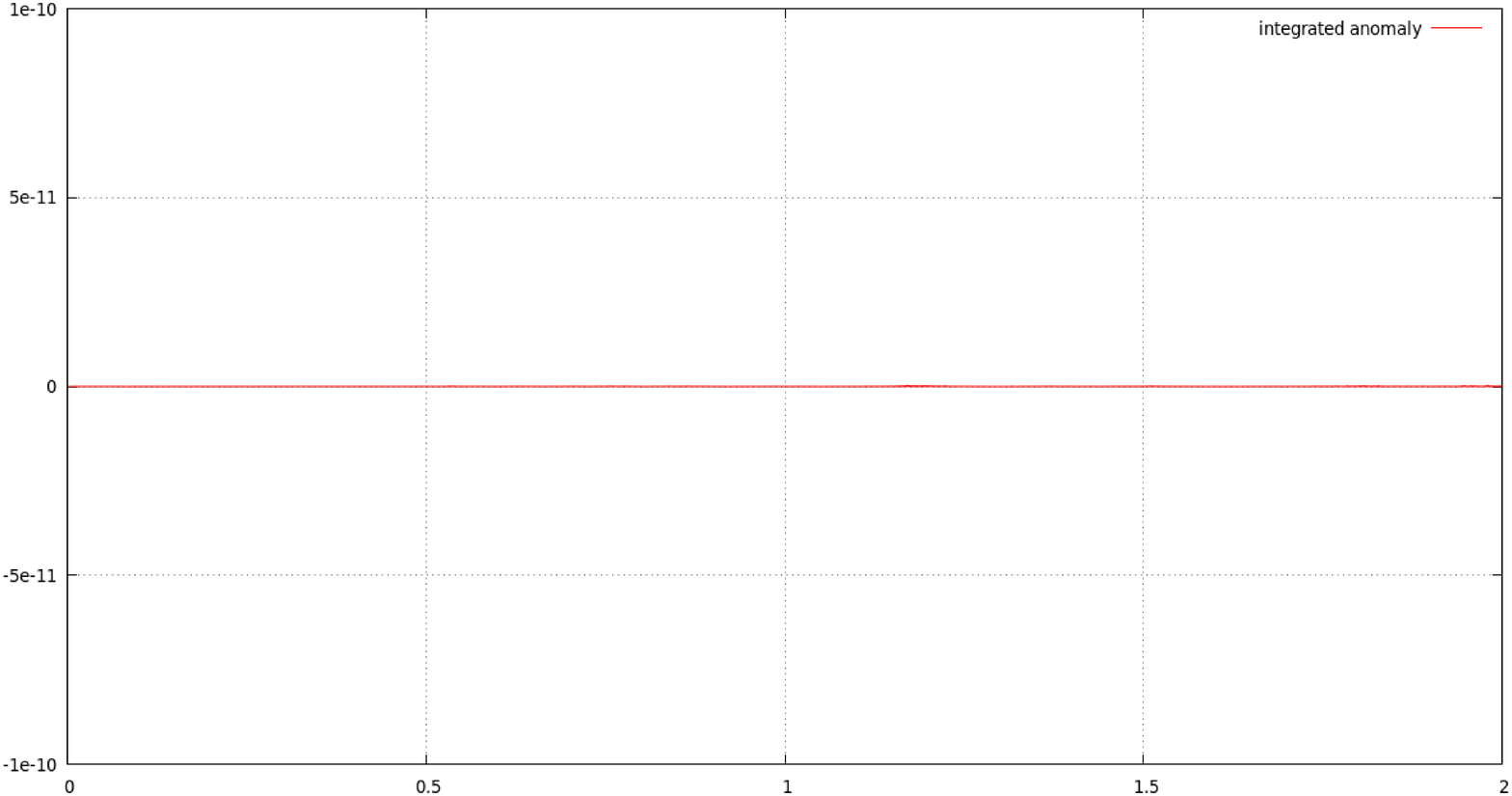}   
\parbox{6in}{\caption {(color online)Type IA.  Top left: the transmission of two bright solitons  of the deformed  NLS model (\ref{nlsd})-(\ref{pot1}) is plotted for $\epsilon=0.06$. The initial solitons $t_i$ (green) travel in opposite direction with velocity $v = 5$ and amplitude $1.98$ such that $n=0$ in (\ref{phi12rs}). They form a collision pattern $t_c$ (blue) in their closest approximation and then transmit to each other. The solitons after collision are plotted as red line ($t_f$). Top right: the anomaly density  $\g(x,t)$ plotted for three successive times $t_i$, $t_c$ and $t_f$. Middle: the phase $\vp$ plotted for an initial time $t_i$ (left) and final time $t_f$ (right).  Bottom: the anomaly $\beta^{(4)}(t)$ and time integrated anomaly $\int^t dt' \beta^{(4)}(t')$, respectively.}}
\end{figure}

\begin{figure}
\centering
\label{fig4}
\includegraphics[width=4cm,scale=6, angle=0, height=5cm]{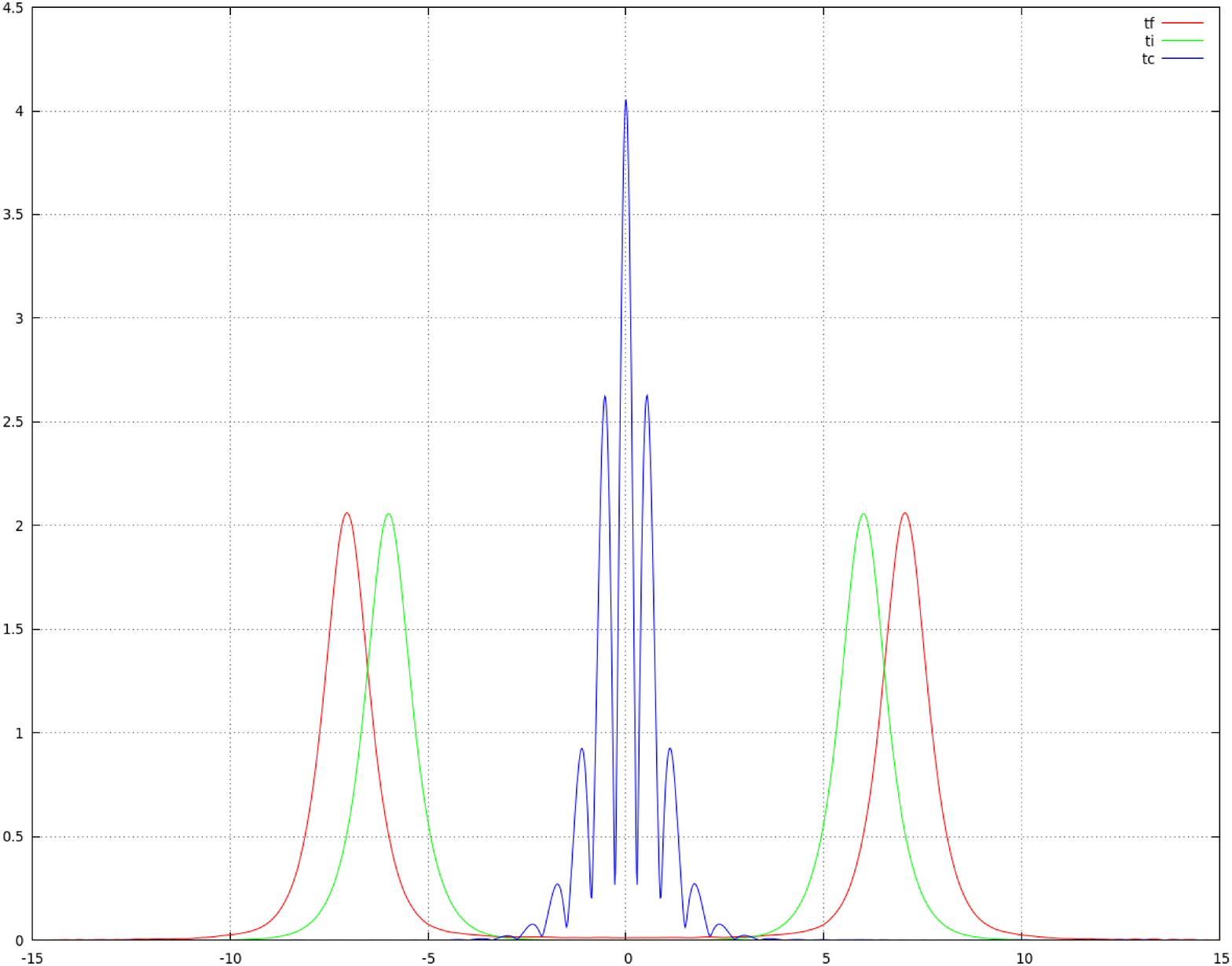} 
\includegraphics[width=4cm,scale=6, angle=0, height=5cm]{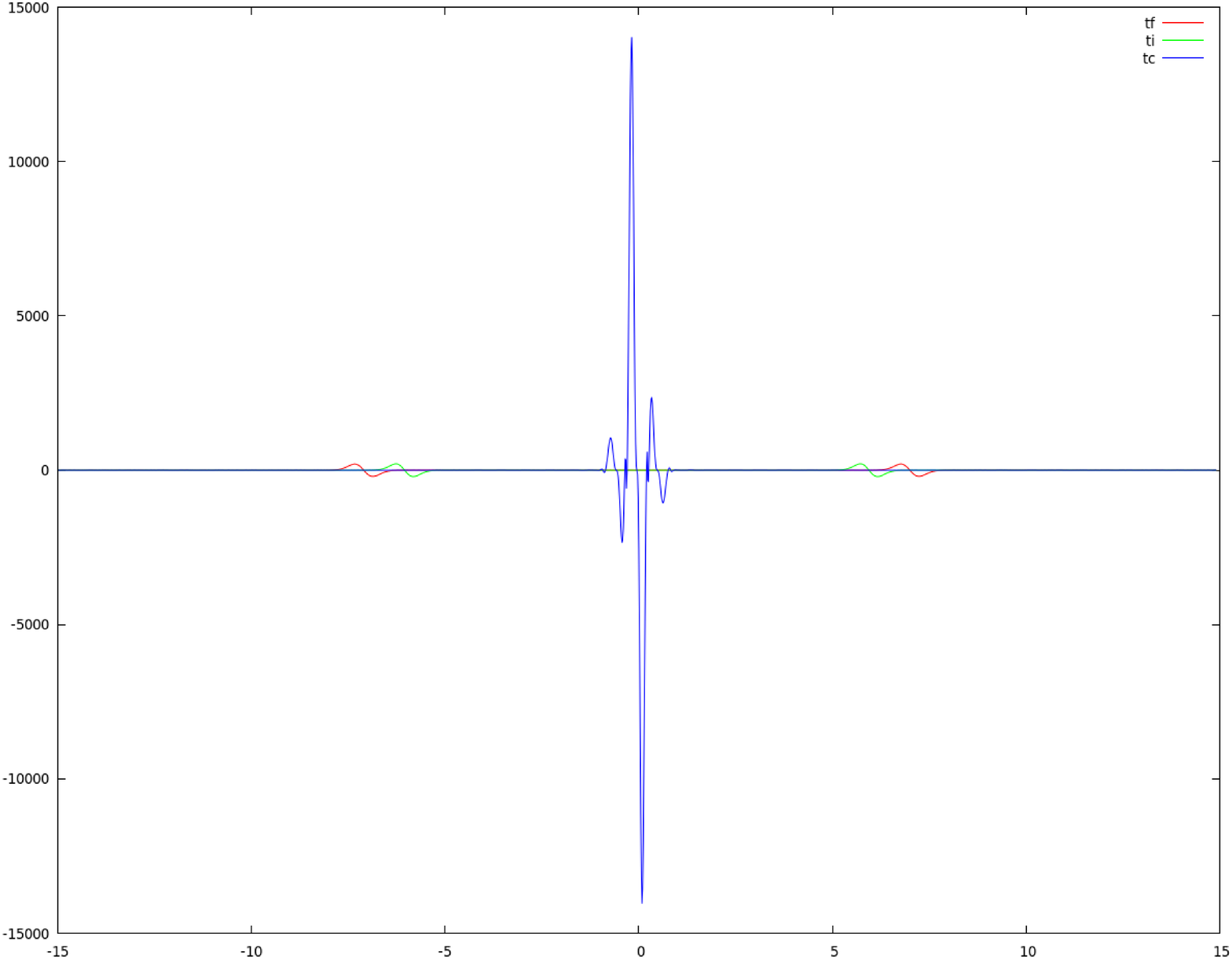}
\includegraphics[width=4cm,scale=6, angle=0, height=5cm]{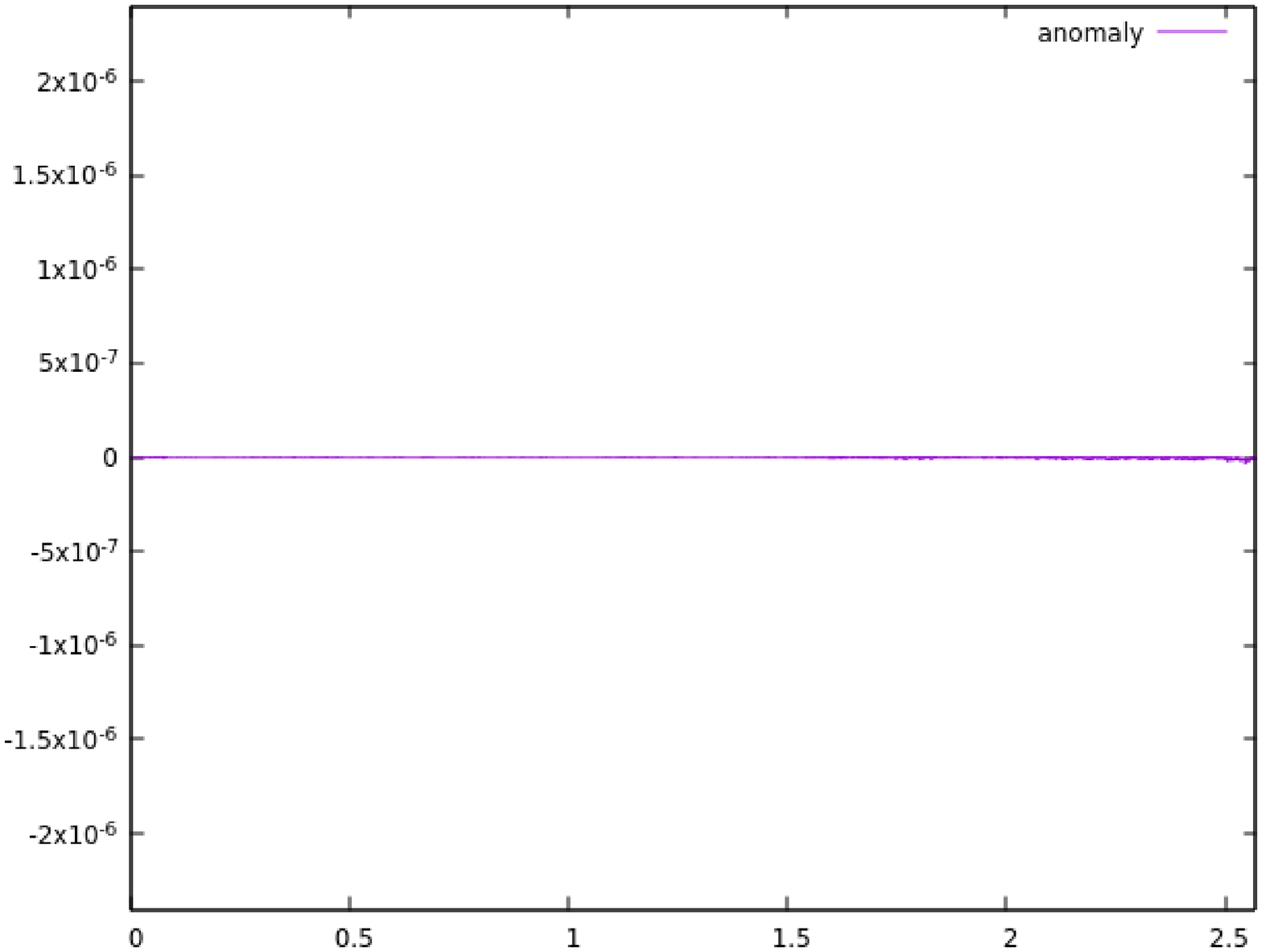}
\includegraphics[width=4cm,scale=6, angle=0, height=5cm]{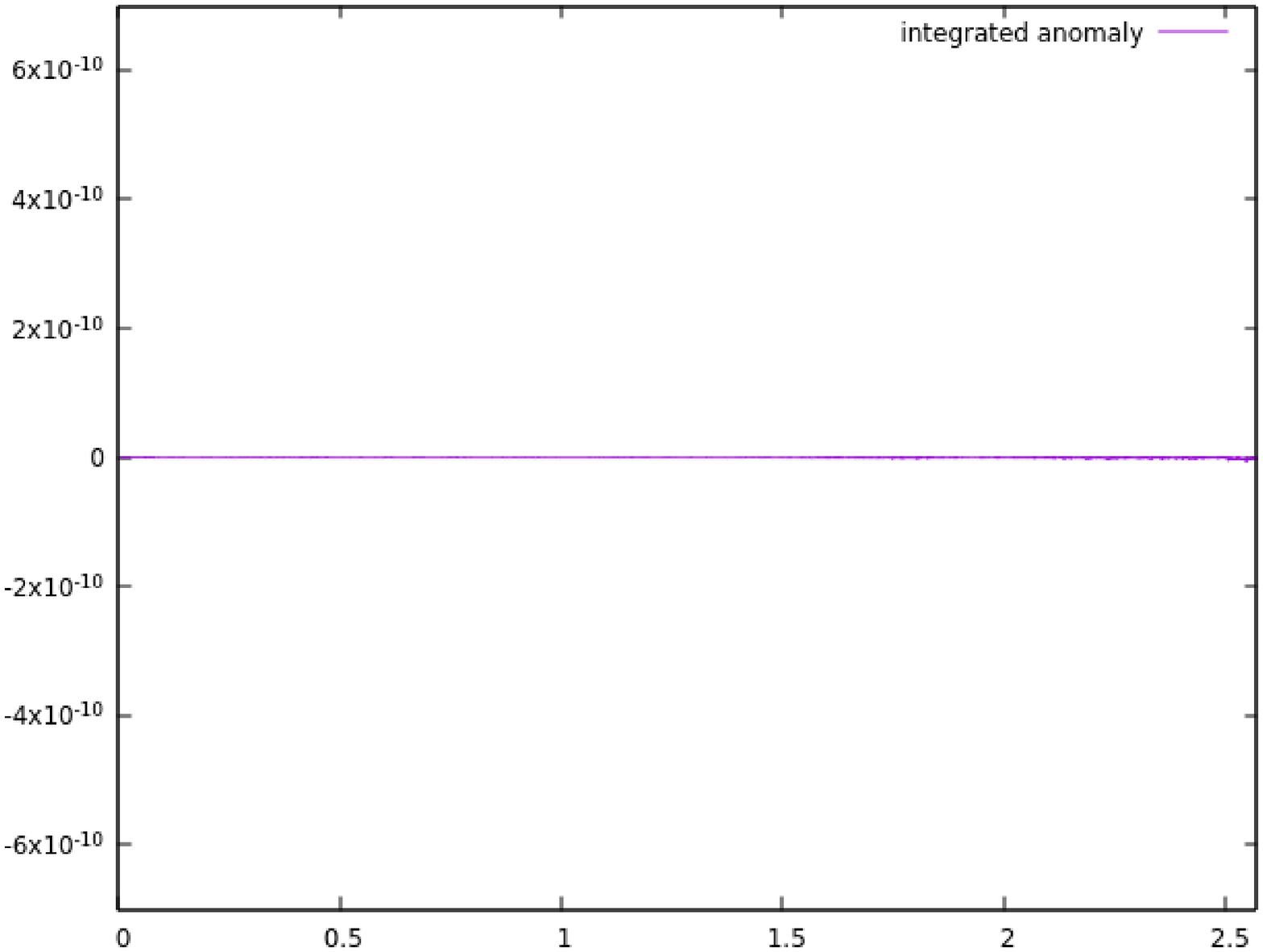}   
\parbox{6in}{\caption {(color online)Type IA.  Top left: the transmission of two bright solitons  of the deformed  NLS model (\ref{nlsd})-(\ref{pot1}) is plotted for $\epsilon=-0.06$. The initial solitons $t_i$ (green) travel in opposite direction with velocity $v = 5$ and amplitude $ 2.1$  such that $n=0$ in (\ref{phi12rs}). They form a collision pattern $t_c$ (blue) in their closest approximation and then transmit  to each other. The solitons after collision are plotted as a red line ($t_f$). Top right:  the anomaly density $\g(x,t)$  plotted for three successive times ($t_i$, $t_c$ and $t_f$). Bottom: the anomaly $\beta^{(4)}(t)$ and time integrated anomaly $\int^t dt' \beta^{(4)}(t')$, respectively.}}
\end{figure}
 
Since the two-bright soliton solutions admit the symmetry (\ref{pari1}), we expect the anomalies $\beta^{(n)}, \, n\geq 4,$ to vary during collision and the charges to be asymptotically conserved, as presented in section 3.1 of \cite{jhep2} through perturbation theory on the deformation parameter. However, for a special soliton solutions with space-reflection symmetry, the anomaly $\b^{(4)}$ belongs to the sequence of even order charges and it must vanish during the whole collision process. So, we can argue that the dynamics favours the soliton configurations with symmetry properties (\ref{xsym}) at all orders in perturbation theory, as presented in section 3.2 of \cite{jhep3}, for the collision of equal amplitude two-bright solitons with opposite and  equal velocities. Moreover, our numerical simulations show that this anomaly vanishes for a variety of soliton configurations and different values of the deformation parameter $\epsilon$.    

Notice that the well separated individual solitons provide vanishing anomalies. Remarkably, for two-bright soliton collisions with different velocities and amplitudes one notices that in Figs. 5-11 the anomalies vanish  during the  whole collision process for $\epsilon = \pm 0.06, -0.03$. The vanishing of the anomaly occurs also for a variety of parameters as shown in the Figs. 12-17, regardless of the value of the parameter $n$ to  be integer or non-integer. Even though the incident and outgoing solitons do not form a two-soliton solution with definite even parity under space-reflection ${\cal P}_x$, the collision dynamics reproduces the vanishing of $\beta^{(4)}$ in (\ref{qcon11}). Therefore, the charges must be exactly  conserved, within numerical accuracy. So, our numerical simulations suggest that the space-reflection ${\cal P}_x$ symmetry is a sufficient but not a necessary condition for the vanishing of the anomaly $\beta^{(4)}$.

\begin{figure}
\centering
\label{fig5}
\includegraphics[width=2cm,scale=6, angle=0, height=4cm]{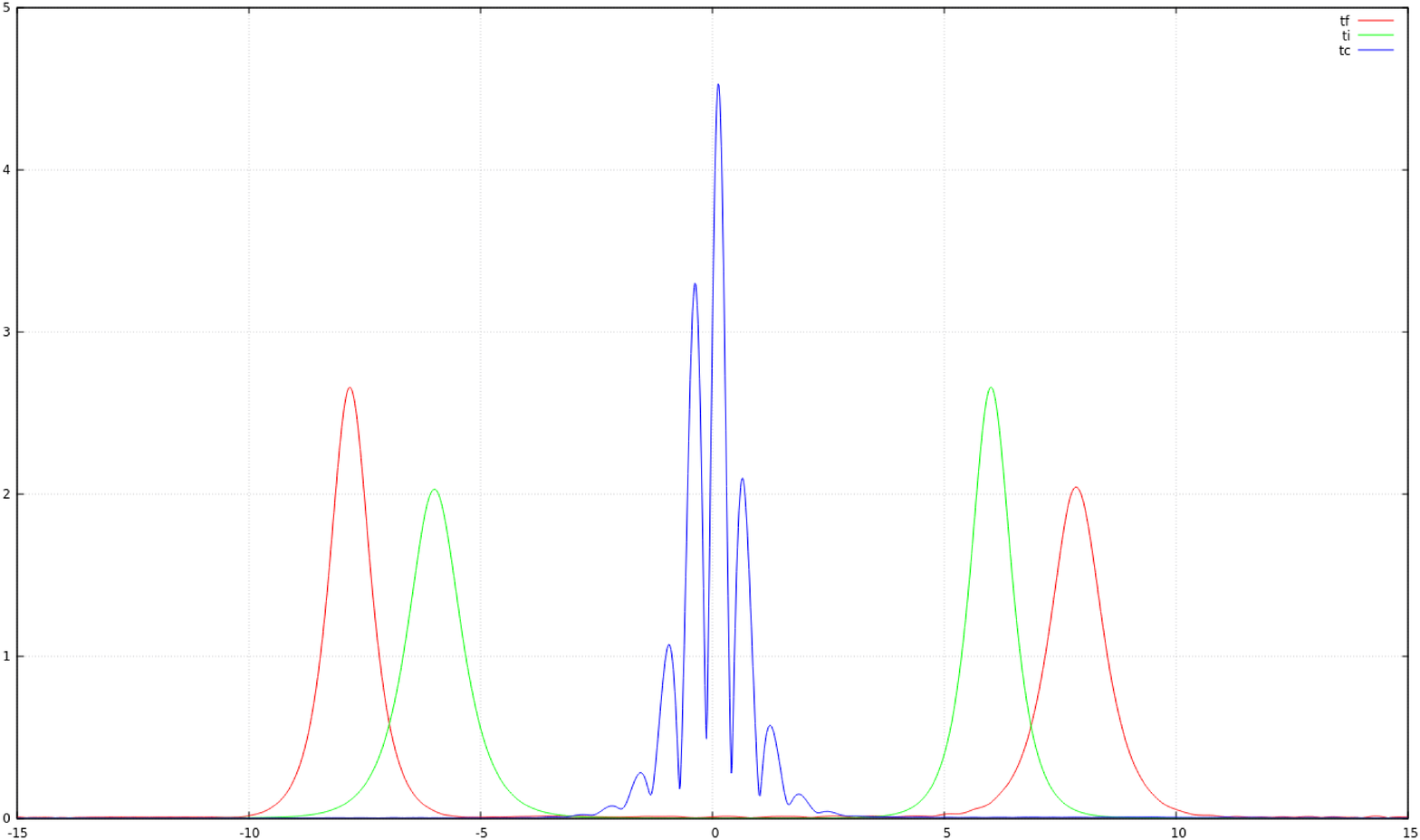} 
\includegraphics[width=2cm,scale=6, angle=0, height=4cm]{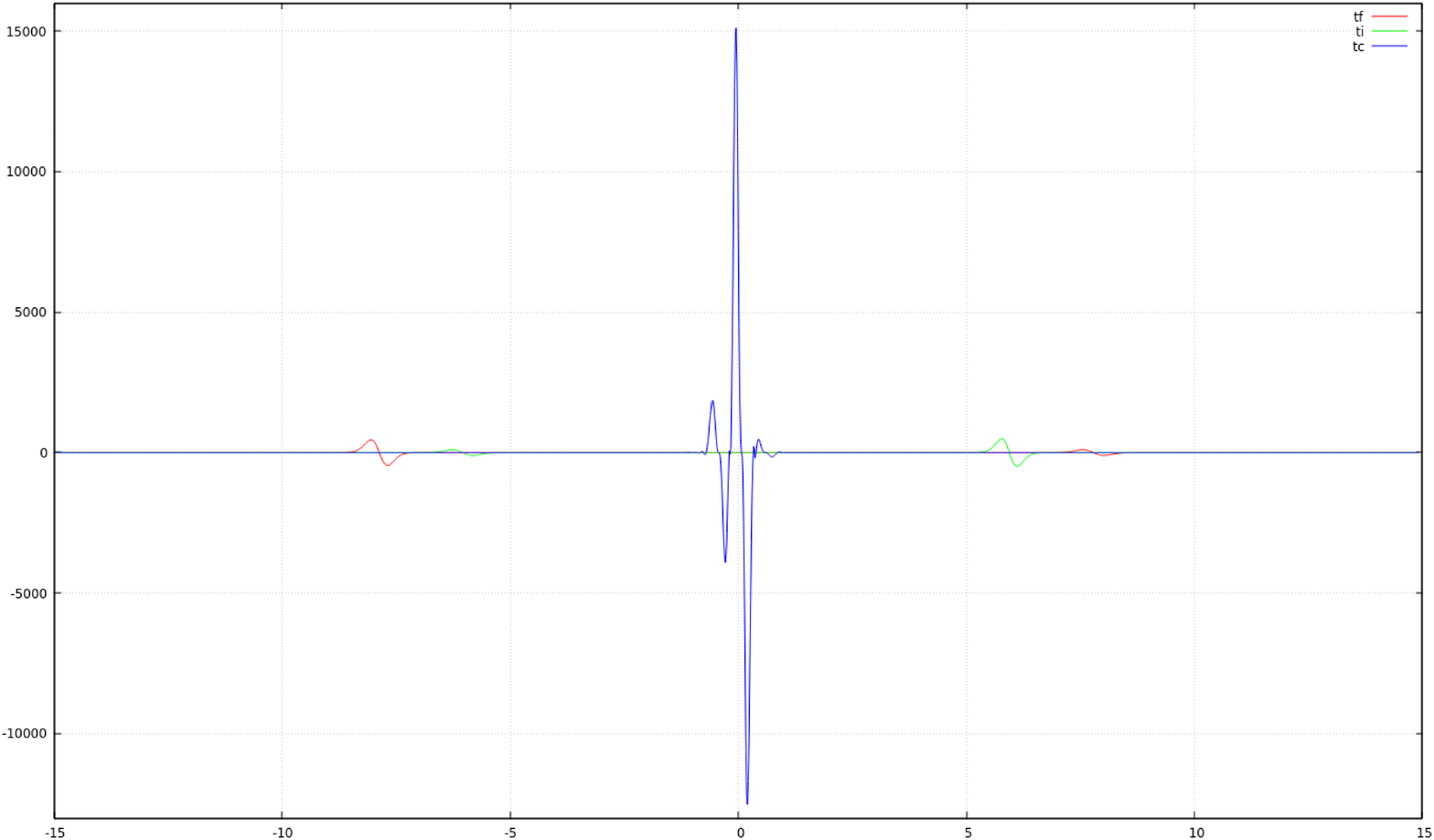}
\includegraphics[width=2cm,scale=6, angle=0, height=4cm]{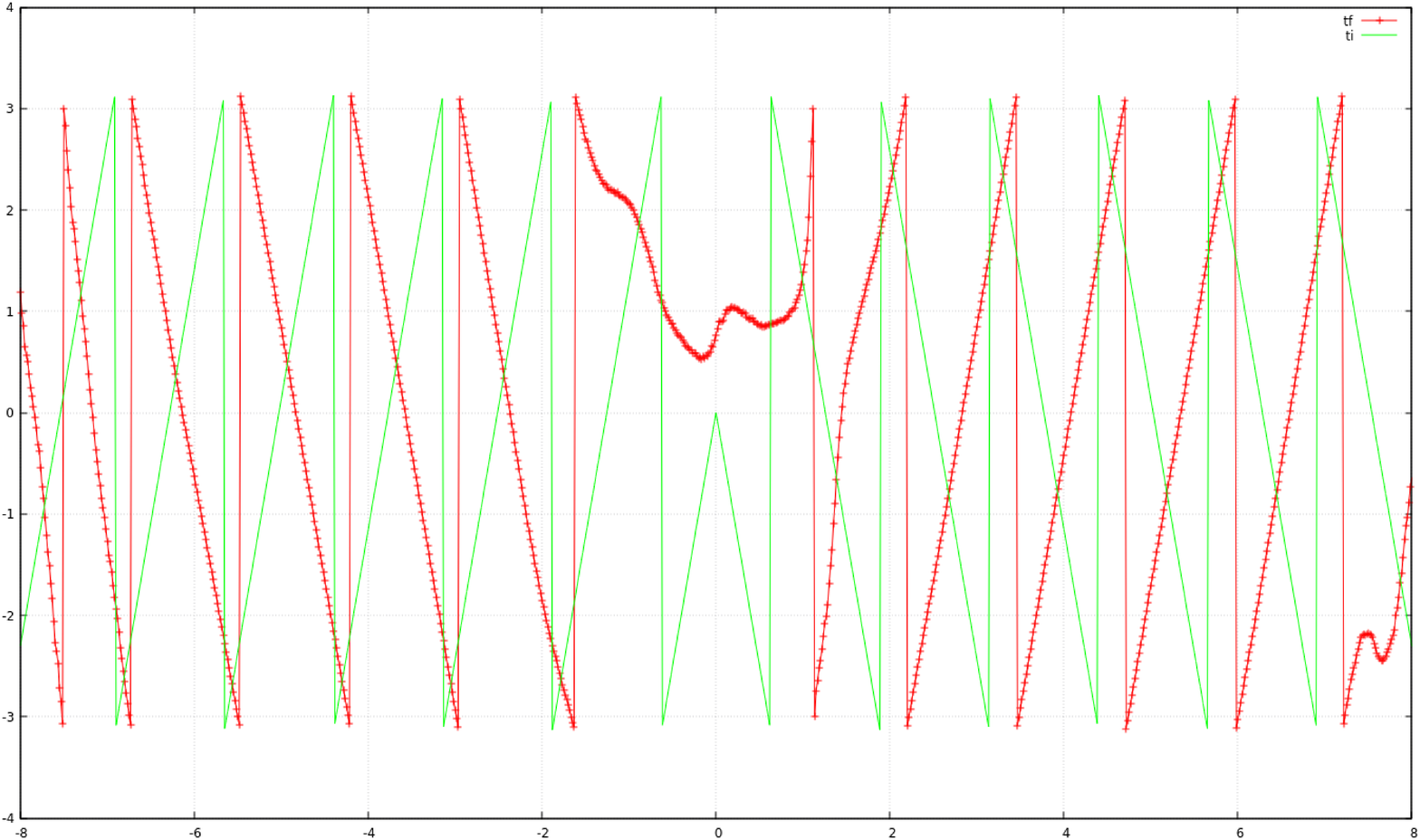} 
\includegraphics[width=2cm,scale=6, angle=0, height=4cm]{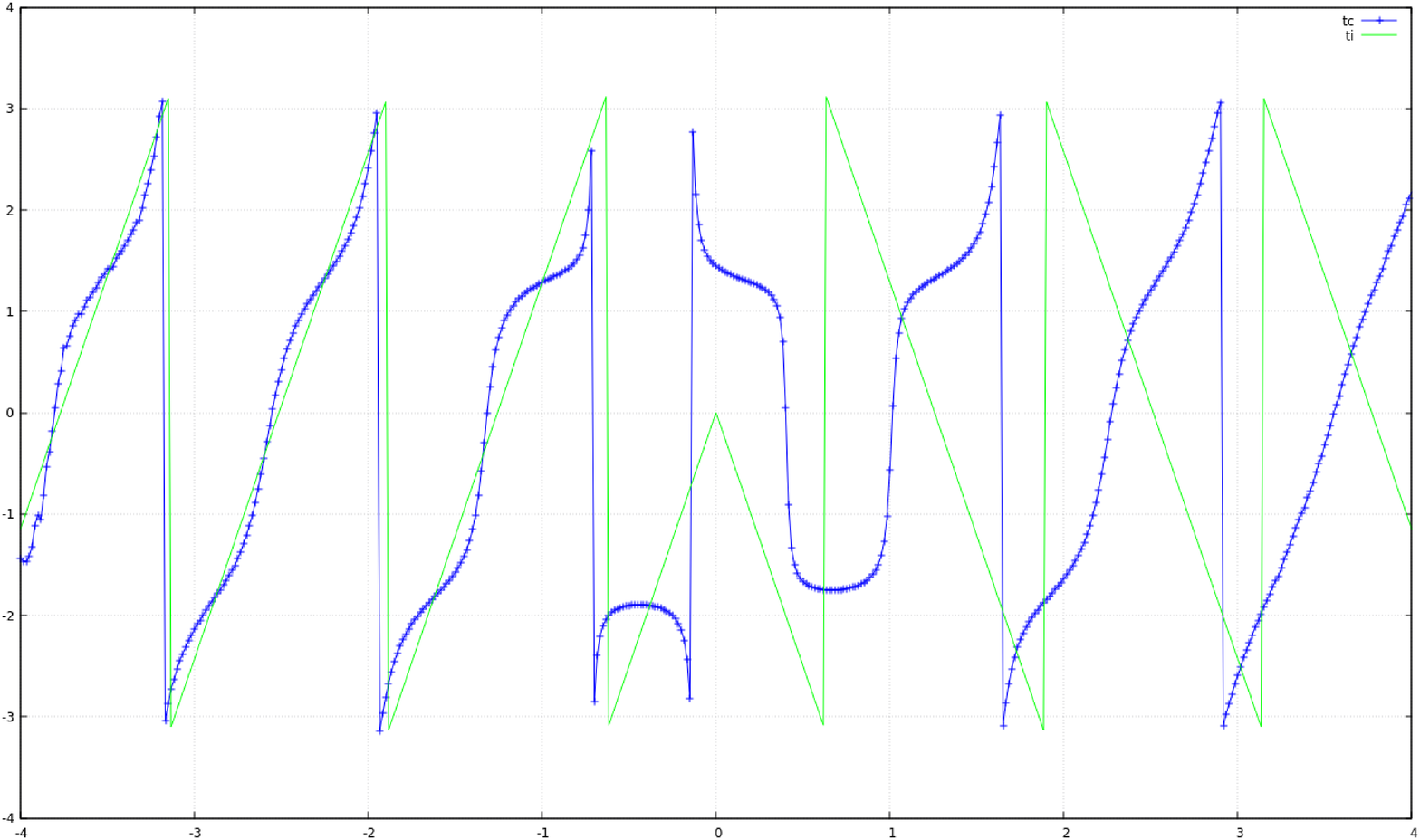}
\includegraphics[width=2cm,scale=6, angle=0, height=4cm]{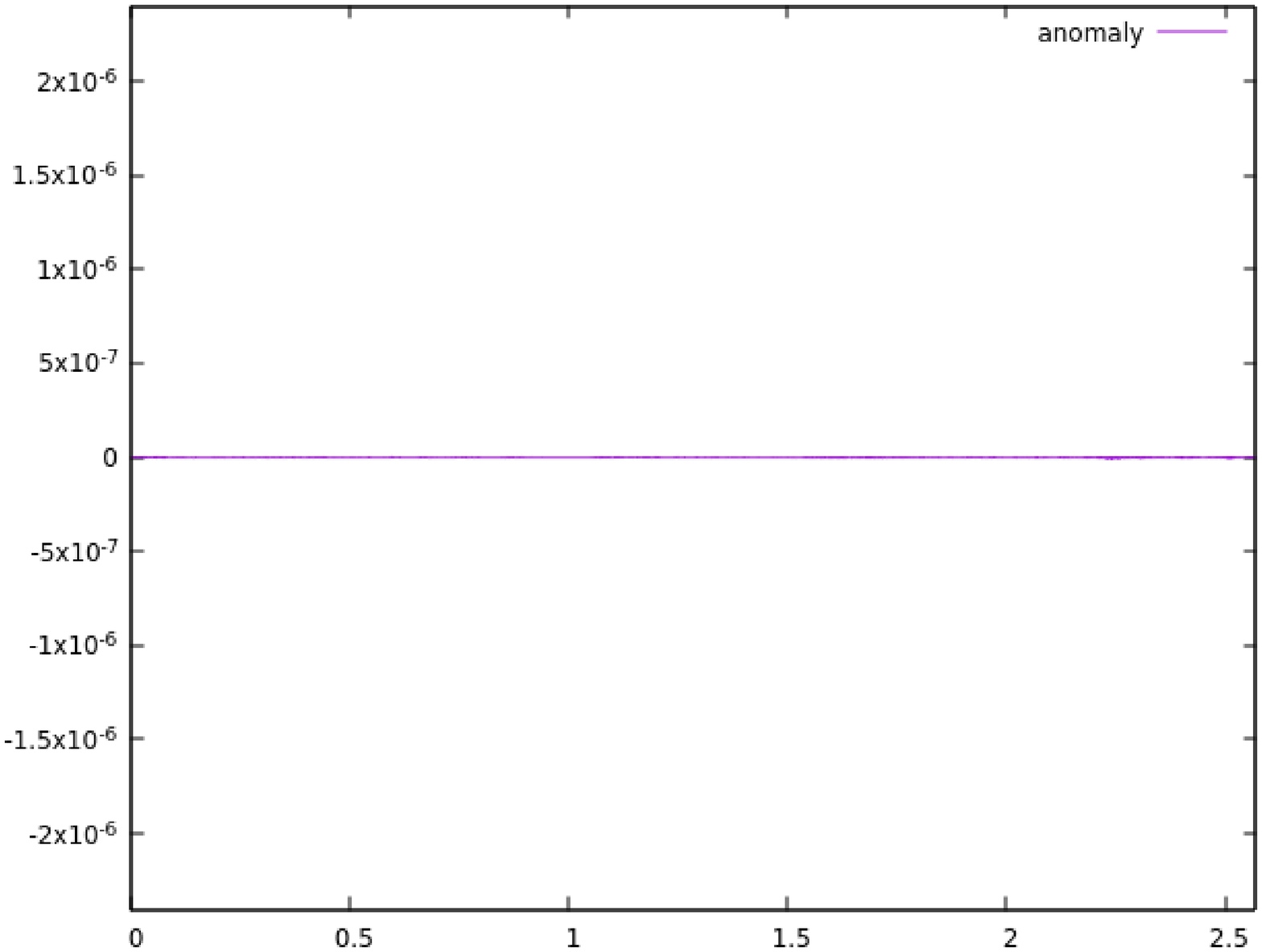}
\includegraphics[width=2cm,scale=6, angle=0, height=4cm]{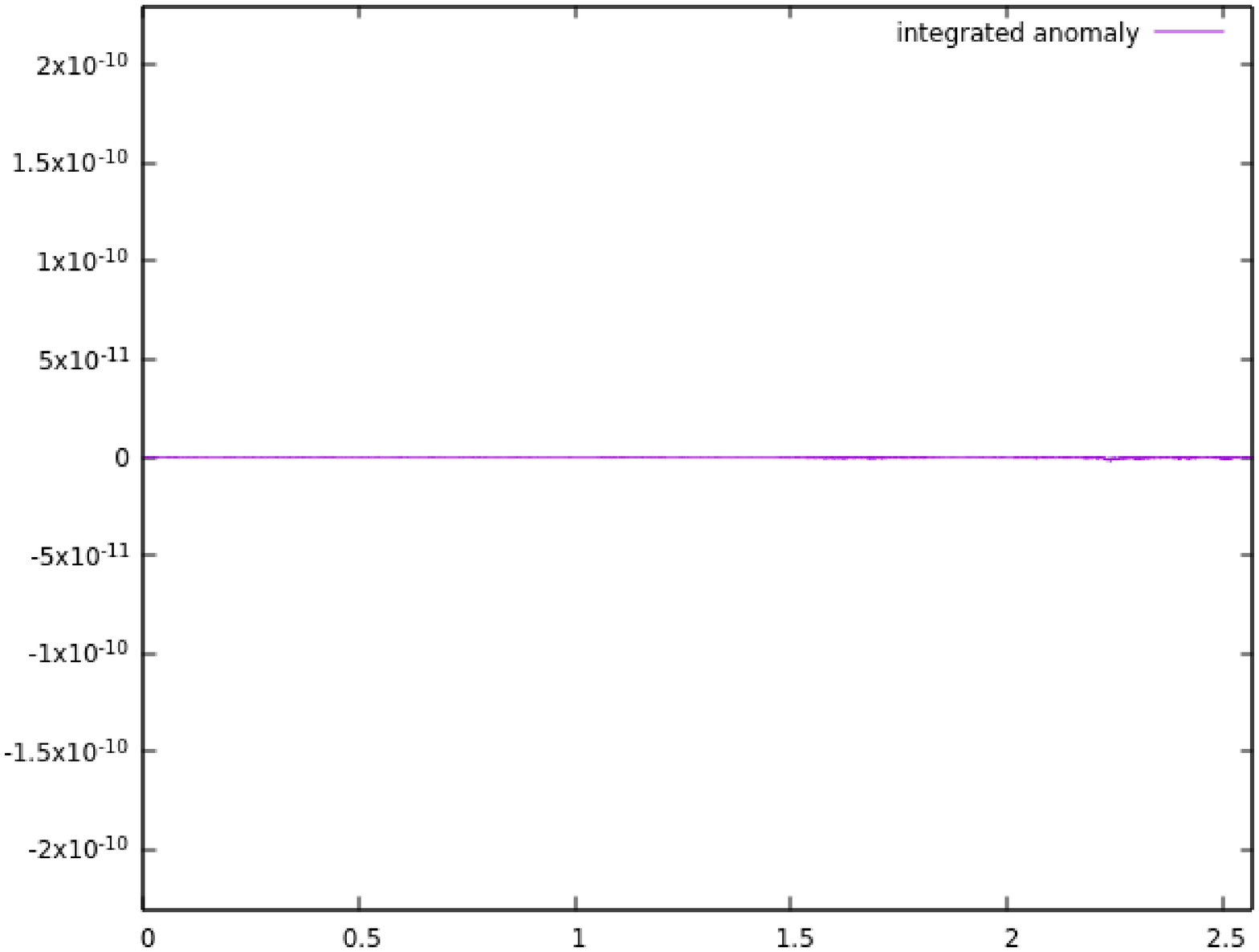}   
\parbox{6in}{\caption {(color online)Type IB.  Top left: the transmission of two bright solitons plotted for $\epsilon=-0.03$. The initial bright solitons $t_i$ (green) travel in opposite direction with velocity $v = 5$ and amplitudes $ 2.0$ and $2.6$  such that $n=-1.18$ in (\ref{phi12}). They form a collision pattern $t_c$ (blue) in their approximation and then transmit to each other. The solitons after collision are plotted as red line ($t_f$). Top right: the anomaly density $\g(x,t)$ plotted for three successive times ($t_i$, $t_c$ and $t_f$). Middle: the phase plotted for an initial time $t_i$ (green), collision time $t_c$ (blue)  and final time $t_f$ (red).  Bottom: the anomaly $\beta^{(4)}(t)$ and time integrated anomaly $\int^t dt' \beta^{(4)}(t')$, respectively.}}
\end{figure}

\begin{figure}
\centering
\label{fig6}
\includegraphics[width=2cm,scale=6, angle=0, height=4cm]{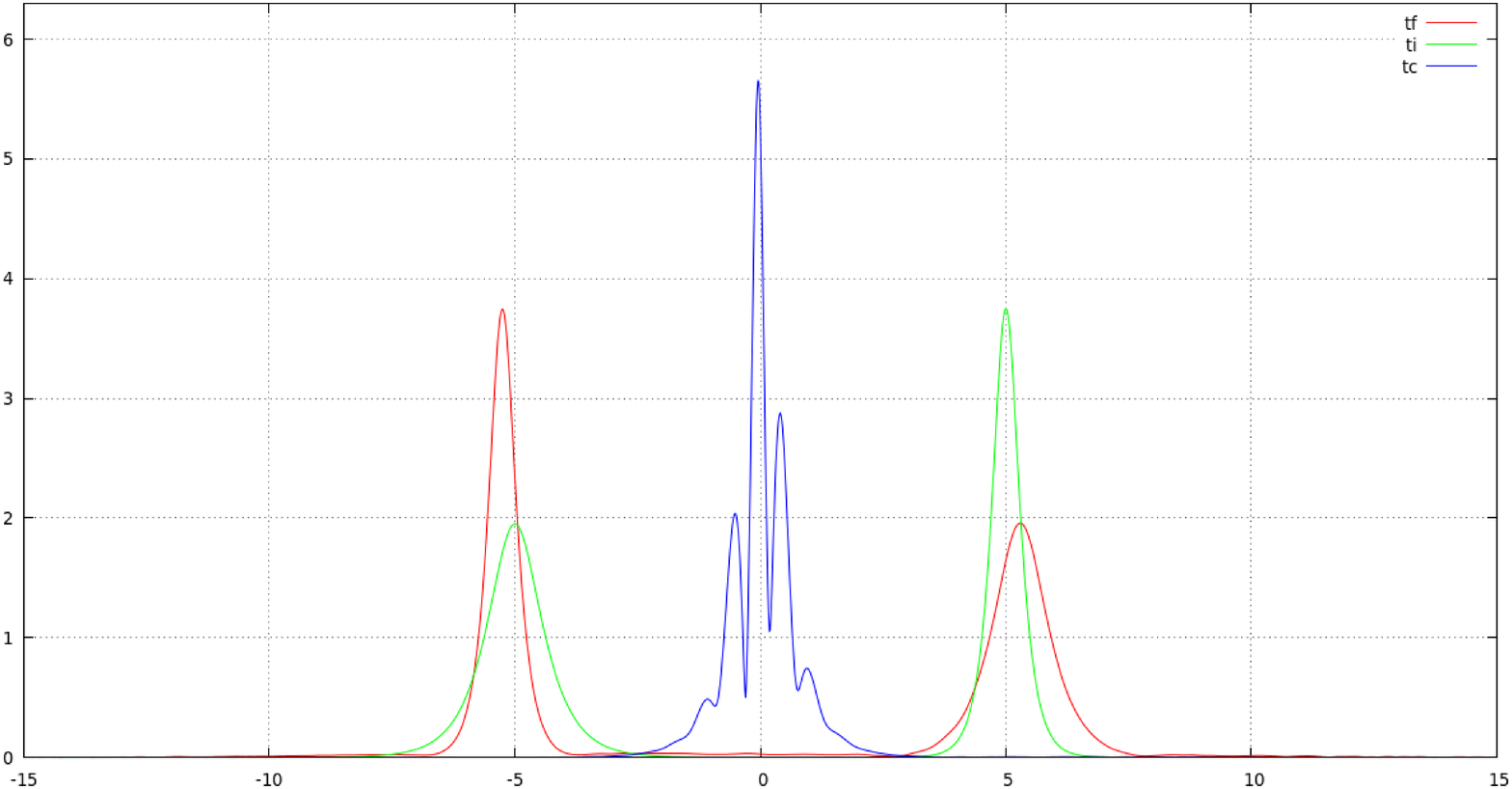} 
\includegraphics[width=2cm,scale=6, angle=0, height=4cm]{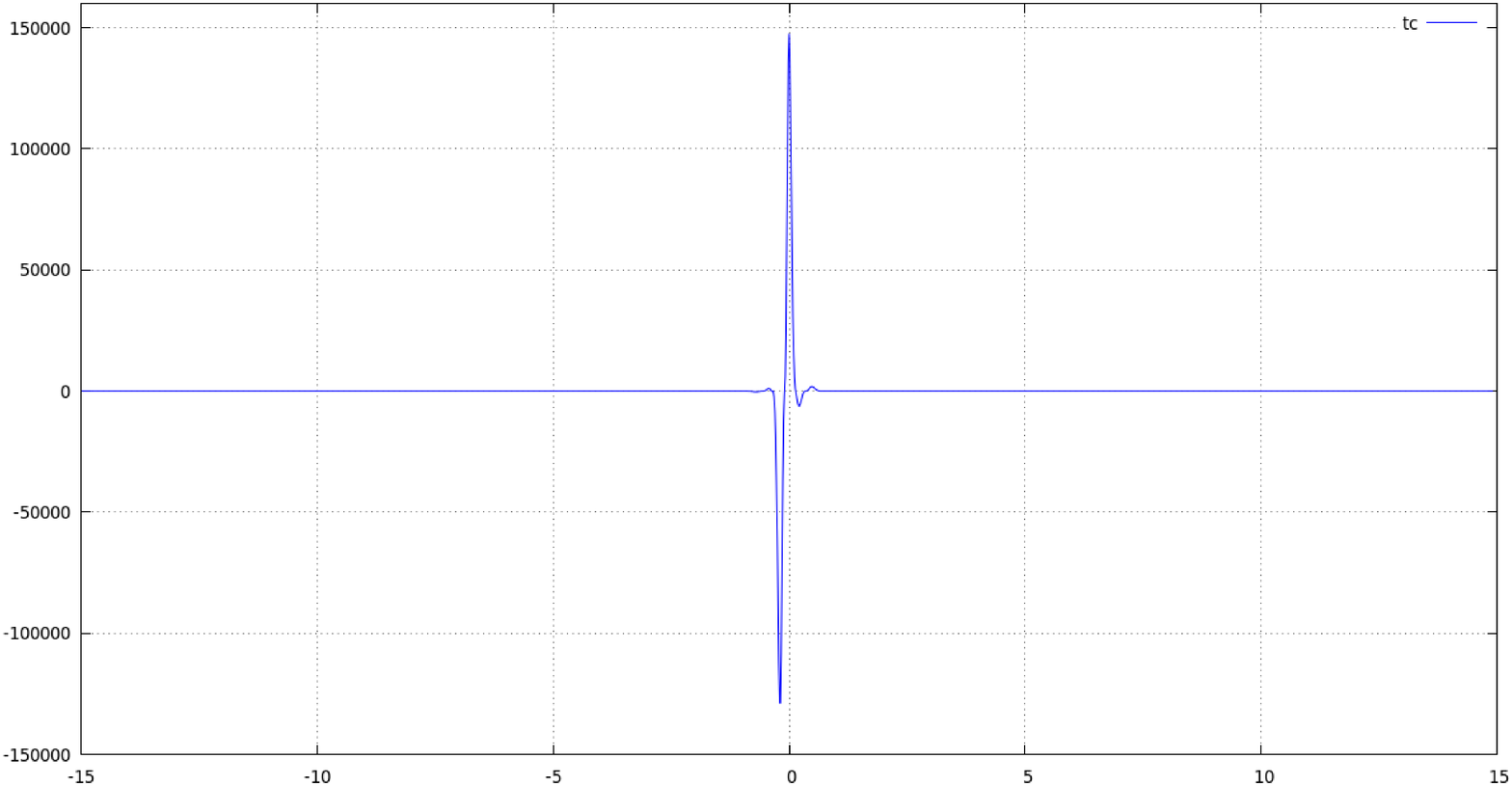} 
\includegraphics[width=2cm,scale=6, angle=0, height=4cm]{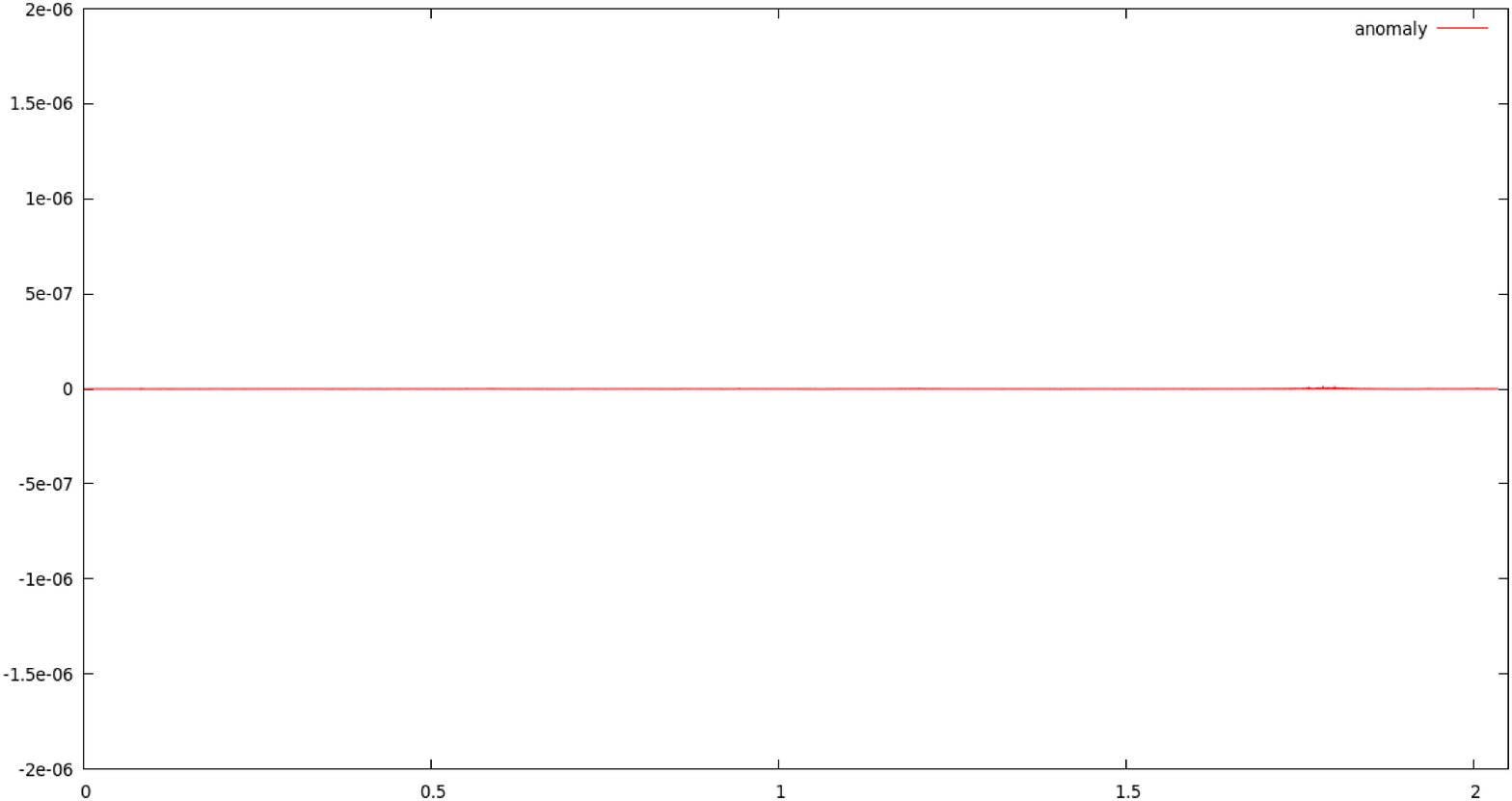}
\includegraphics[width=2cm,scale=6, angle=0, height=4cm]{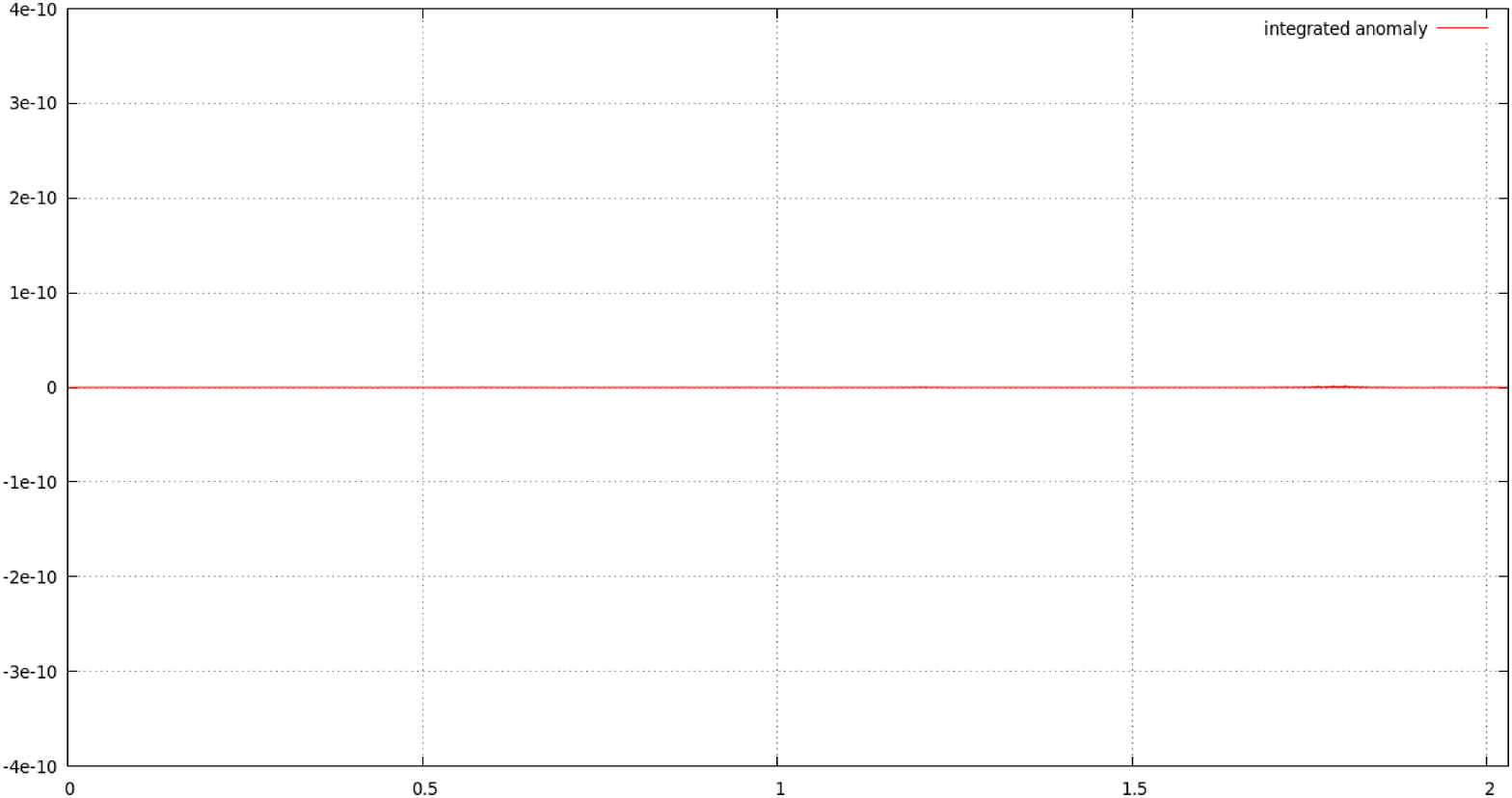}   
\parbox{6in}{\caption {(color online) Type IB. Top left: the transmission of two bright solitons  plotted for $\epsilon=0.06$. The initial bright solitons $t_i$ (green) travel in opposite direction with velocity $v = 5$ and amplitudes $ 1.98$ and $3.8$  such that $n=-3.734$ in (\ref{phi12}). They form a collision pattern $t_c$ (blue) in their approximation and then transmit to each other. The solitons after collision are plotted as red line ($t_f$). Top right: the anomaly density $\g(x,t)$ plotted at the collision time  $t_c$ . Bottom: the anomaly $\beta^{(4)}(t)$ and time integrated anomaly $\int^t dt' \beta^{(4)}(t')$, respectively.}}
\end{figure} 
 
\begin{figure}
\centering
\label{fig7}
\includegraphics[width=1cm,scale=6, angle=0, height=4cm]{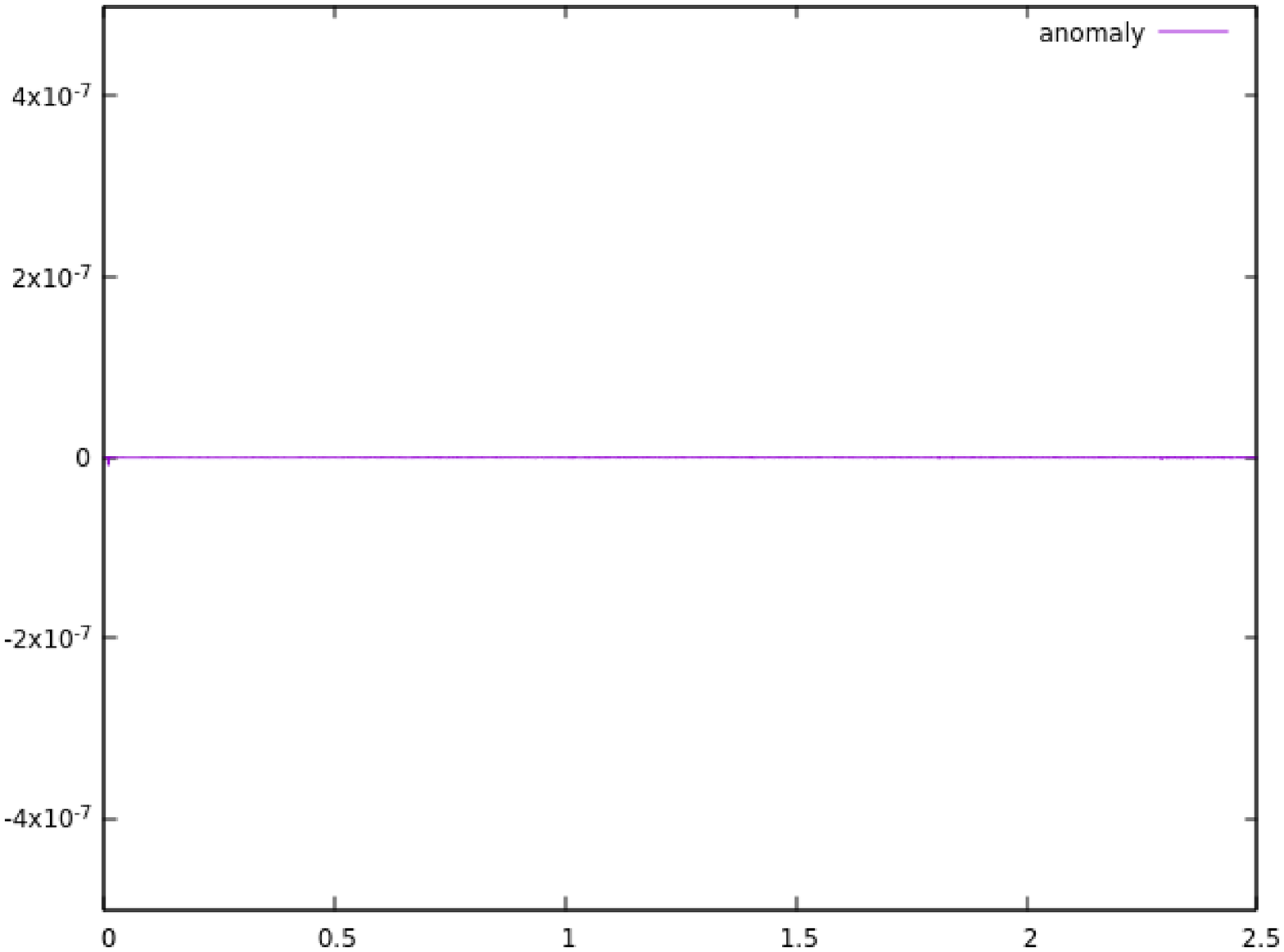}
\includegraphics[width=2cm,scale=6, angle=0, height=4cm]{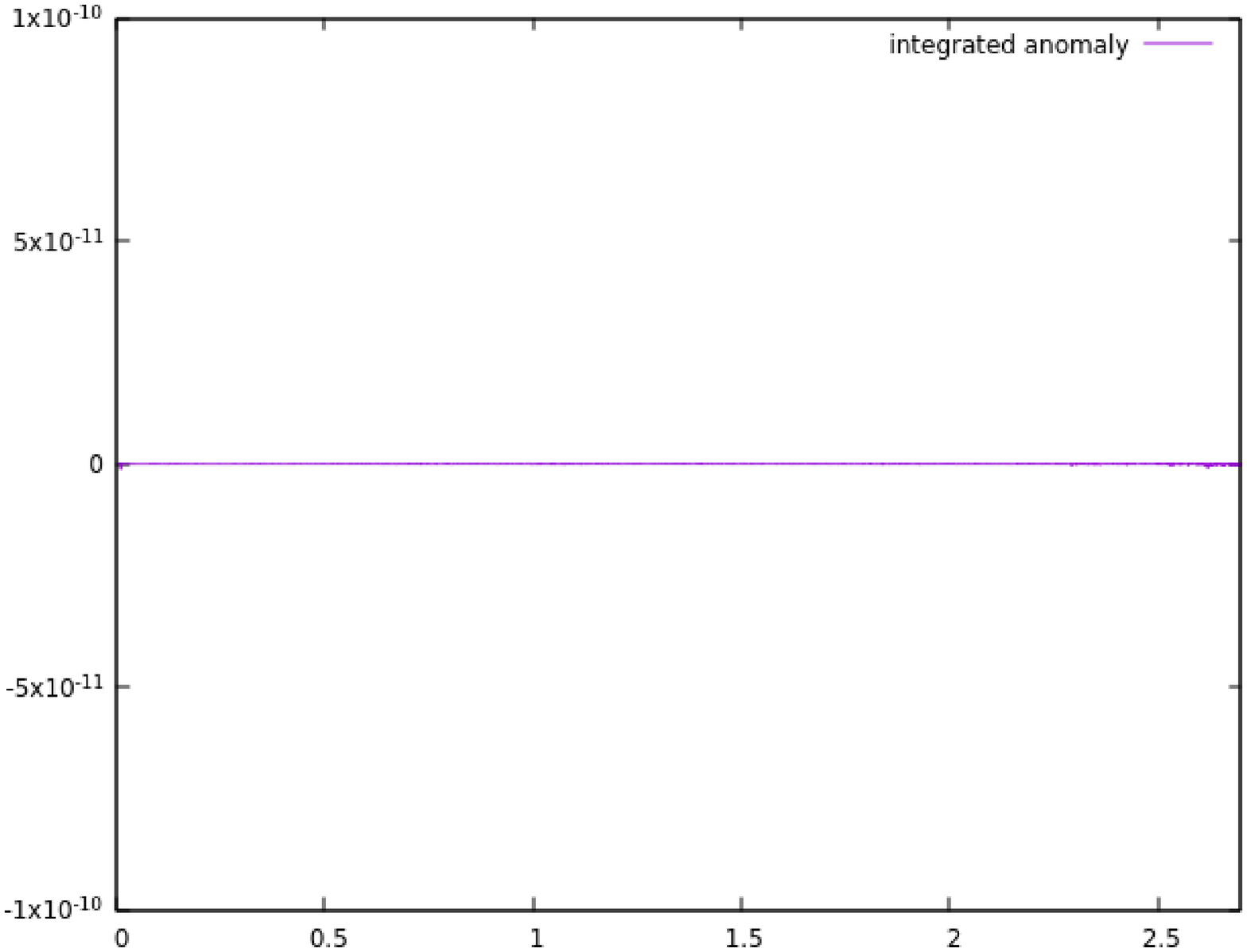}
\parbox{6in}{\caption {Type IB. Anomaly and integrated anomaly for  $\epsilon=-0.06,\,  v_1 = -5, v_2 = 4,\,\rho_1 = 2.5, \rho_2= 2, n=-1.074$.}}
\end{figure}

\begin{figure}
\centering
\label{fig8}
\includegraphics[width=2cm,scale=6, angle=0, height=4cm]{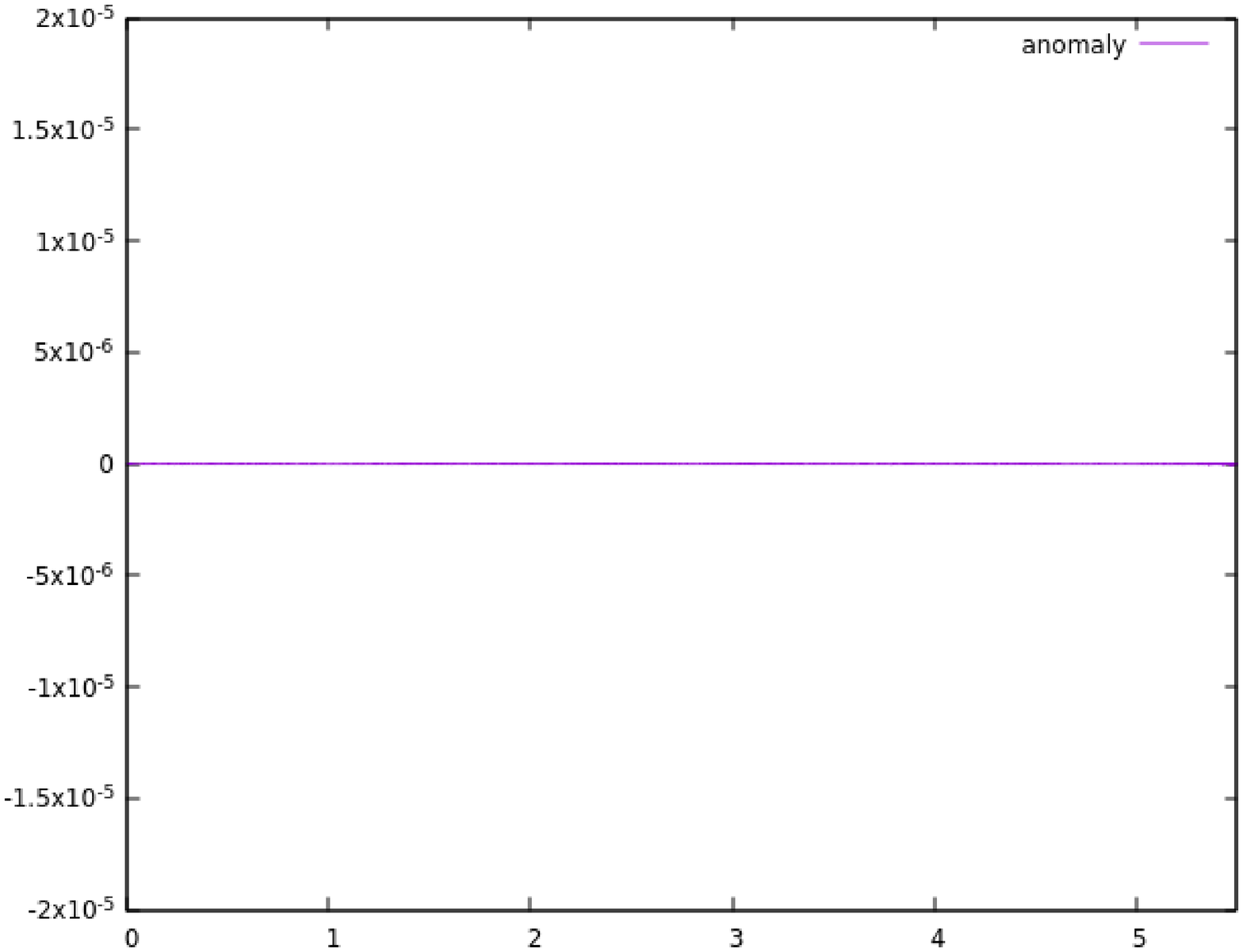}
\includegraphics[width=2cm,scale=6, angle=0, height=4cm]{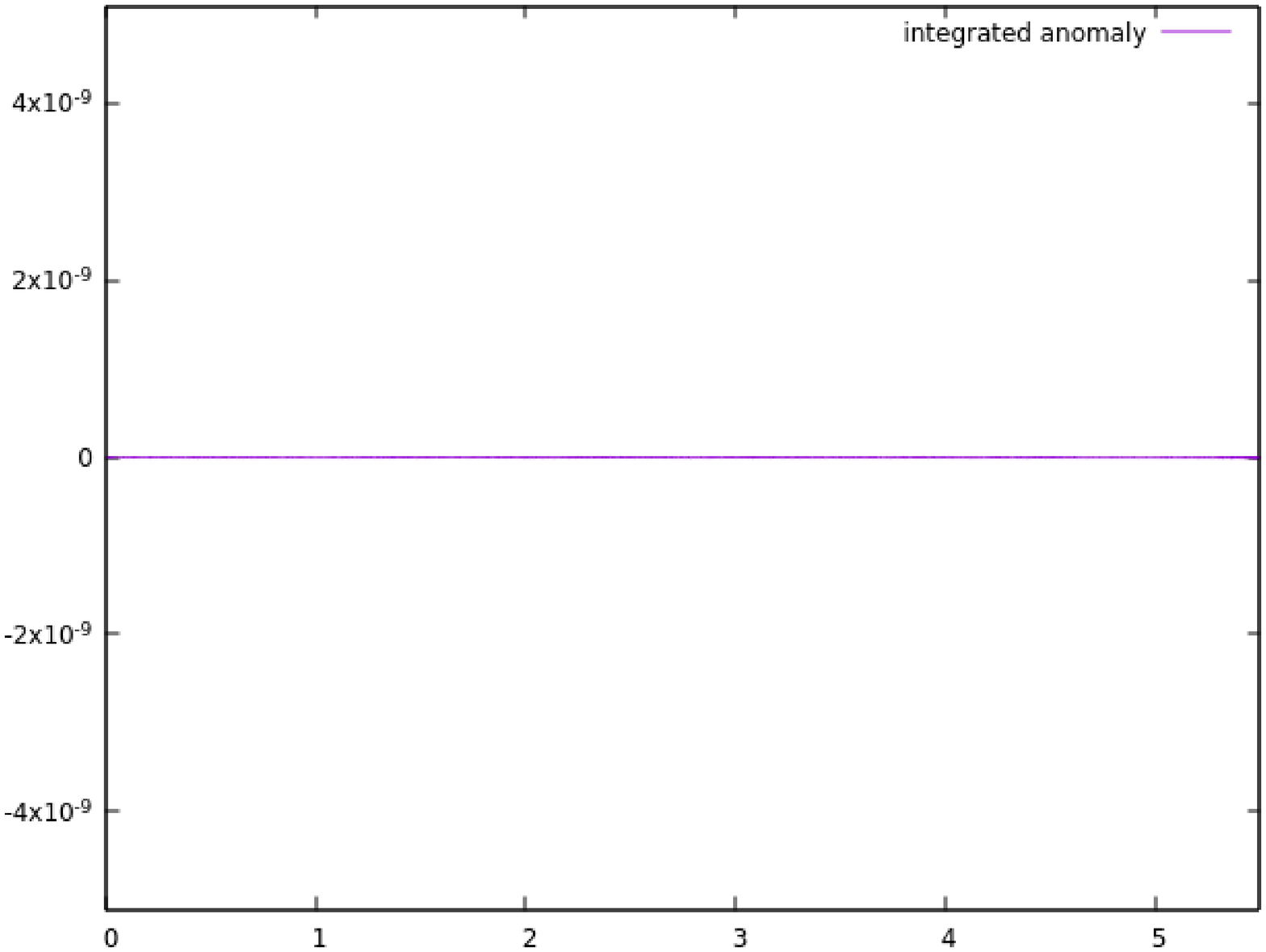}   
\parbox{6in}{\caption {Type IB. Anomaly and integrated anomaly for    $\epsilon=-0.06,\,  v_1 = -2.5, v_2 = 2,\,\rho_1 = 3.5, \rho_2= 3, n= -3.07$.}}
\end{figure}

\begin{figure}
\centering
\label{fig9}
\includegraphics[width=4cm,scale=12, angle=0, height=4cm]{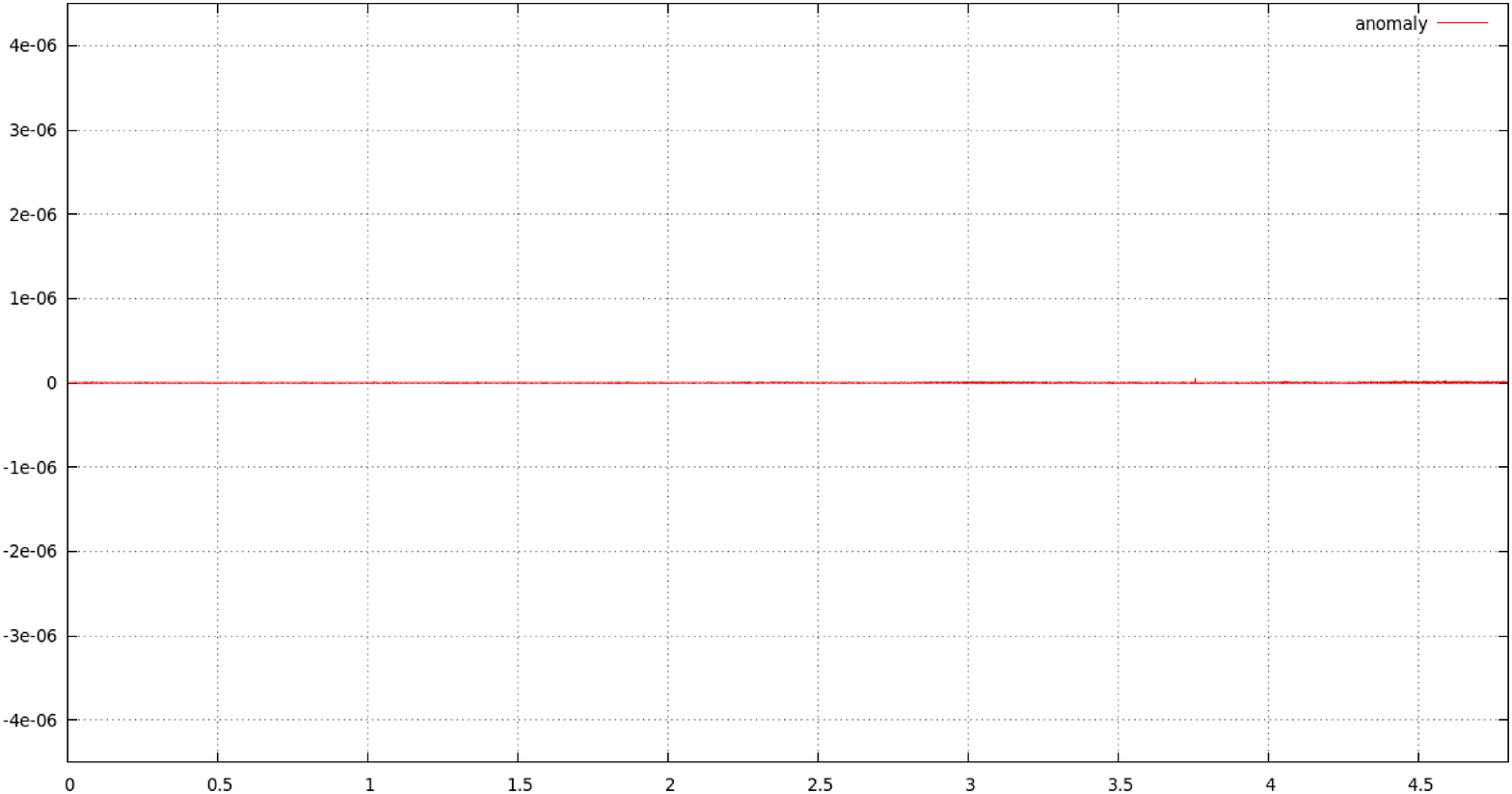}
\includegraphics[width=4cm,scale=12, angle=0, height=4cm]{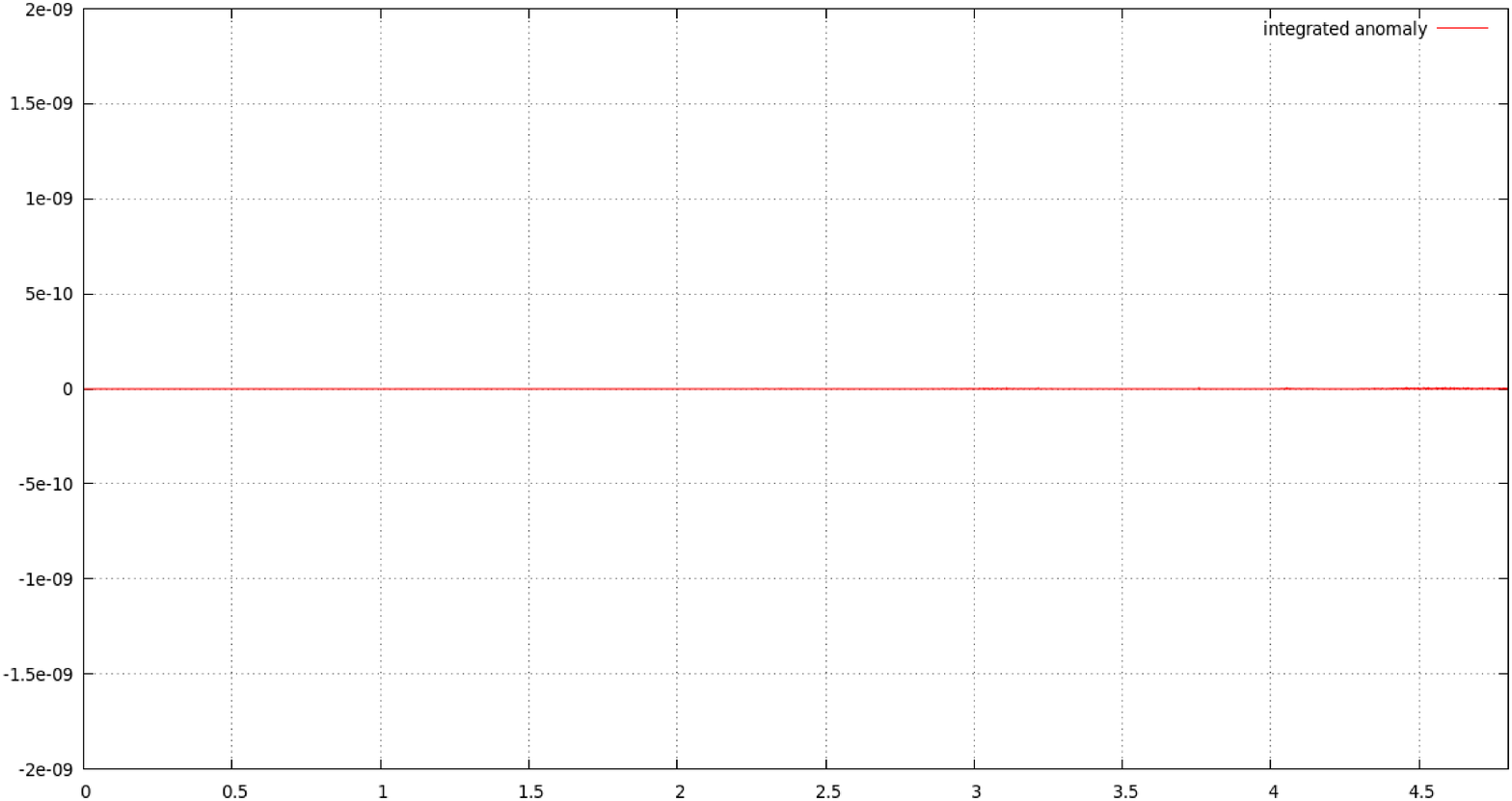}   
\parbox{6in}{\caption {Type IB. Anomaly and integrated anomaly for  $\epsilon=0.06,\,  v_1 = -3, v_2 = 2,\,\rho_1 = 2.5, \rho_2= 2, n= -1.96$.}}
\end{figure}

\begin{figure}
\centering
\label{fig10}
\includegraphics[width=4cm,scale=5, angle=0, height=4cm]{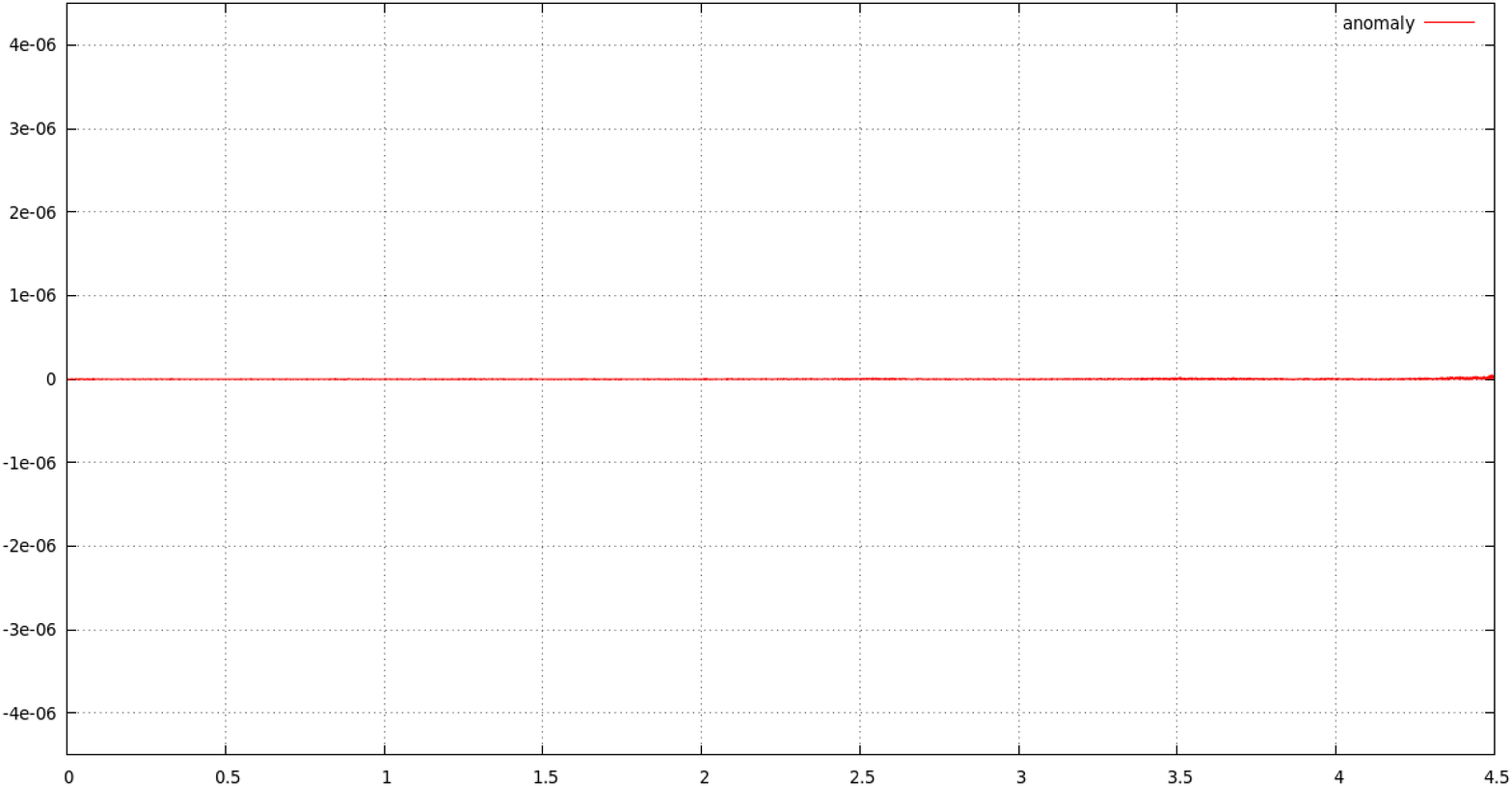}
\includegraphics[width=4cm,scale=5, angle=0, height=4cm]{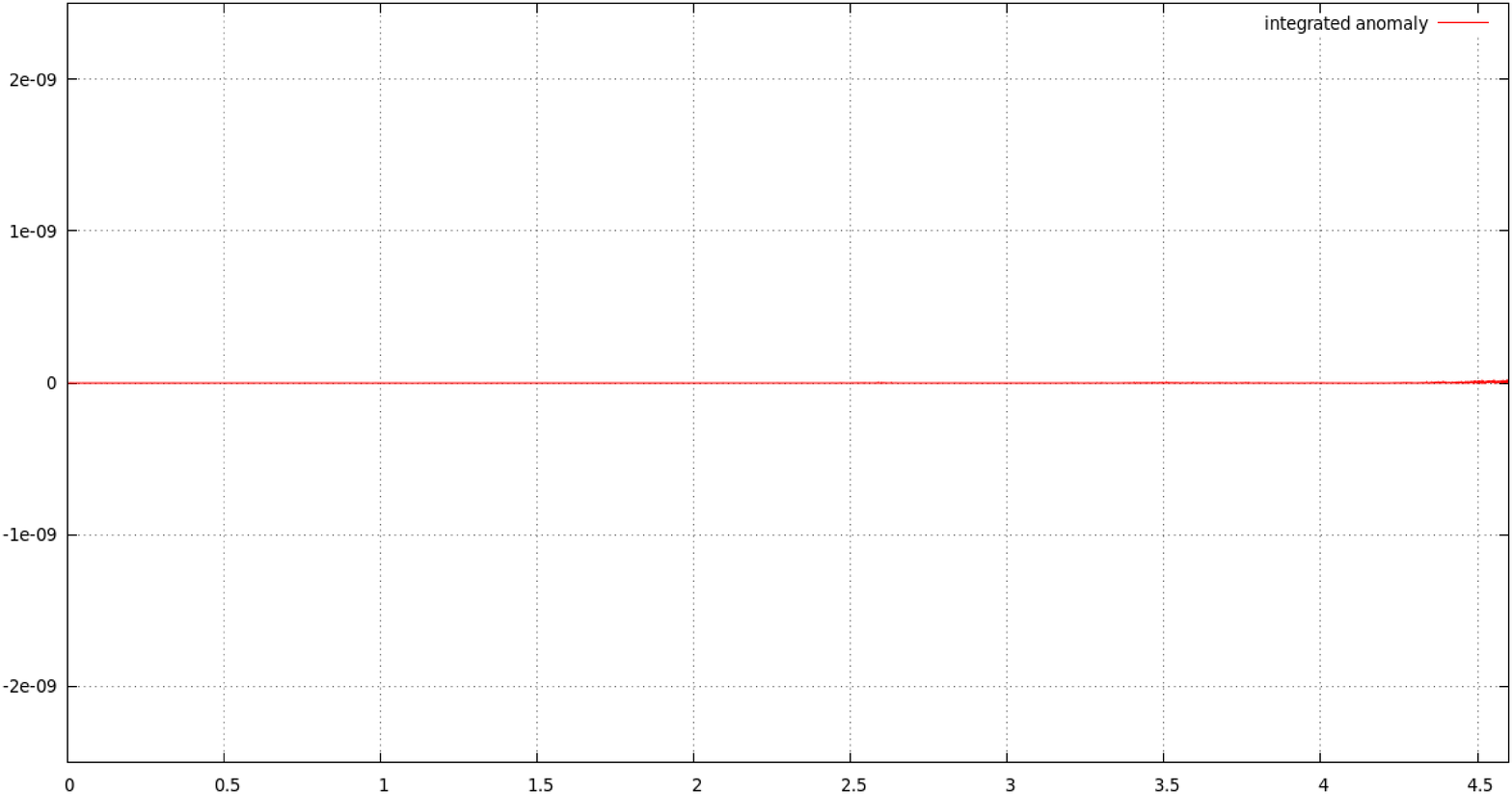}   
\parbox{6in}{\caption {Type IB. Anomaly and integrated anomaly for  $\epsilon=0.06,\,  v_1 = -3, v_2 = 2,\,\rho_1 = 3.5, \rho_2= 1.5, n=8.5$.}}
\end{figure}   
  
\newpage

\begin{figure}
\centering
\label{fig11} 
\includegraphics[width=2cm,scale=6, angle=0, height=5cm]{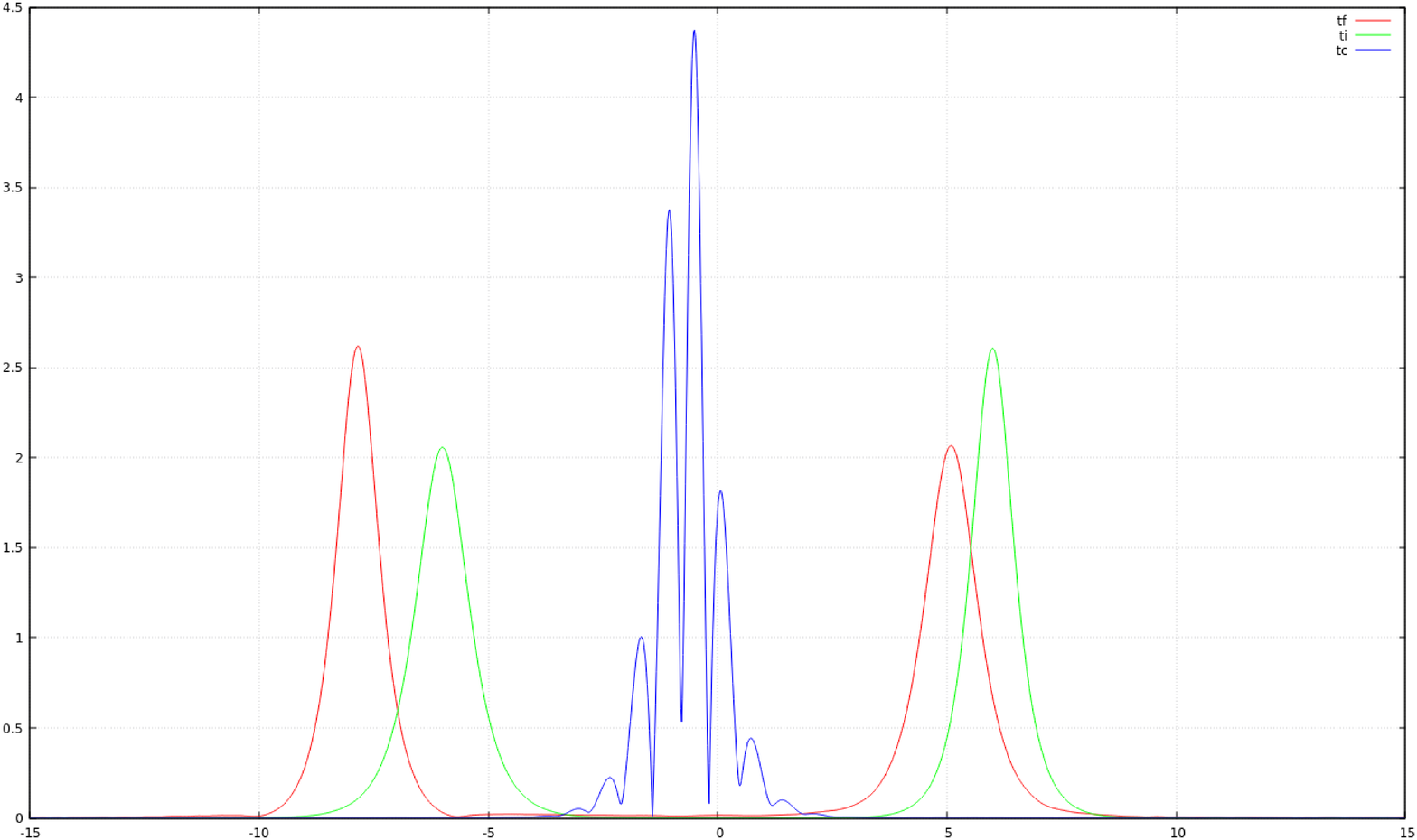} 
\includegraphics[width=2cm,scale=6, angle=0, height=5cm]{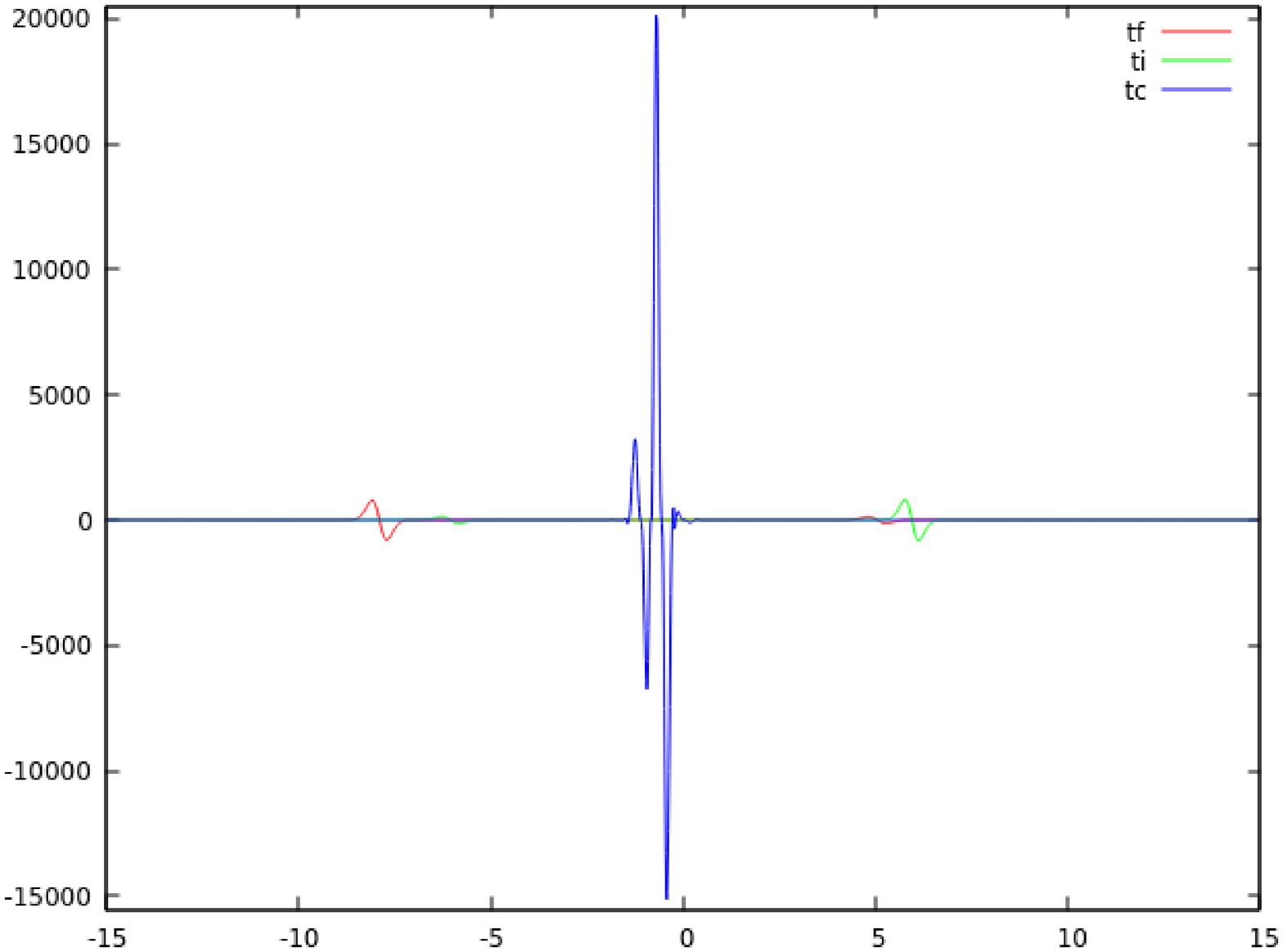} 
\includegraphics[width=2cm,scale=6, angle=0, height=5cm]{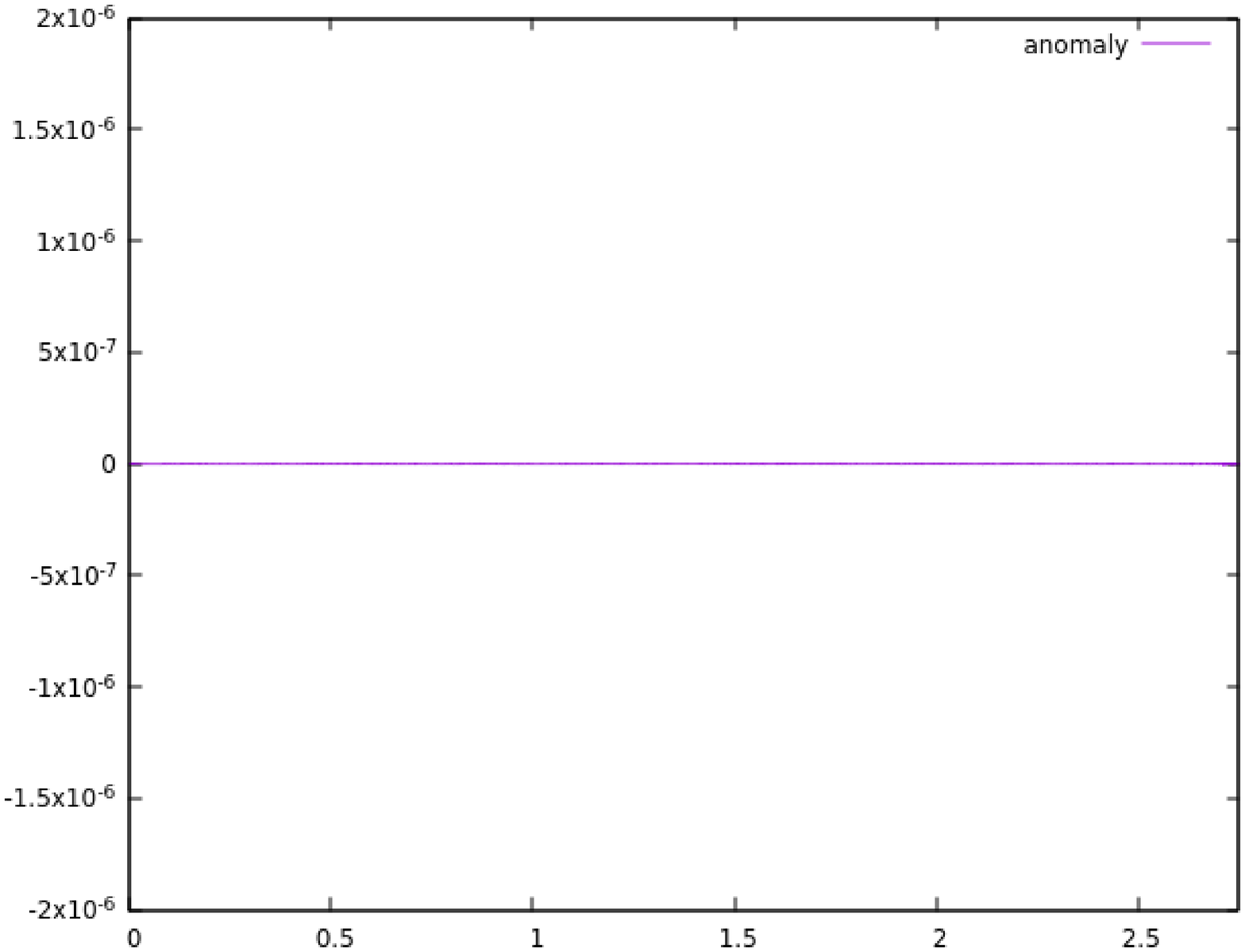}
\includegraphics[width=2cm,scale=6, angle=0, height=5cm]{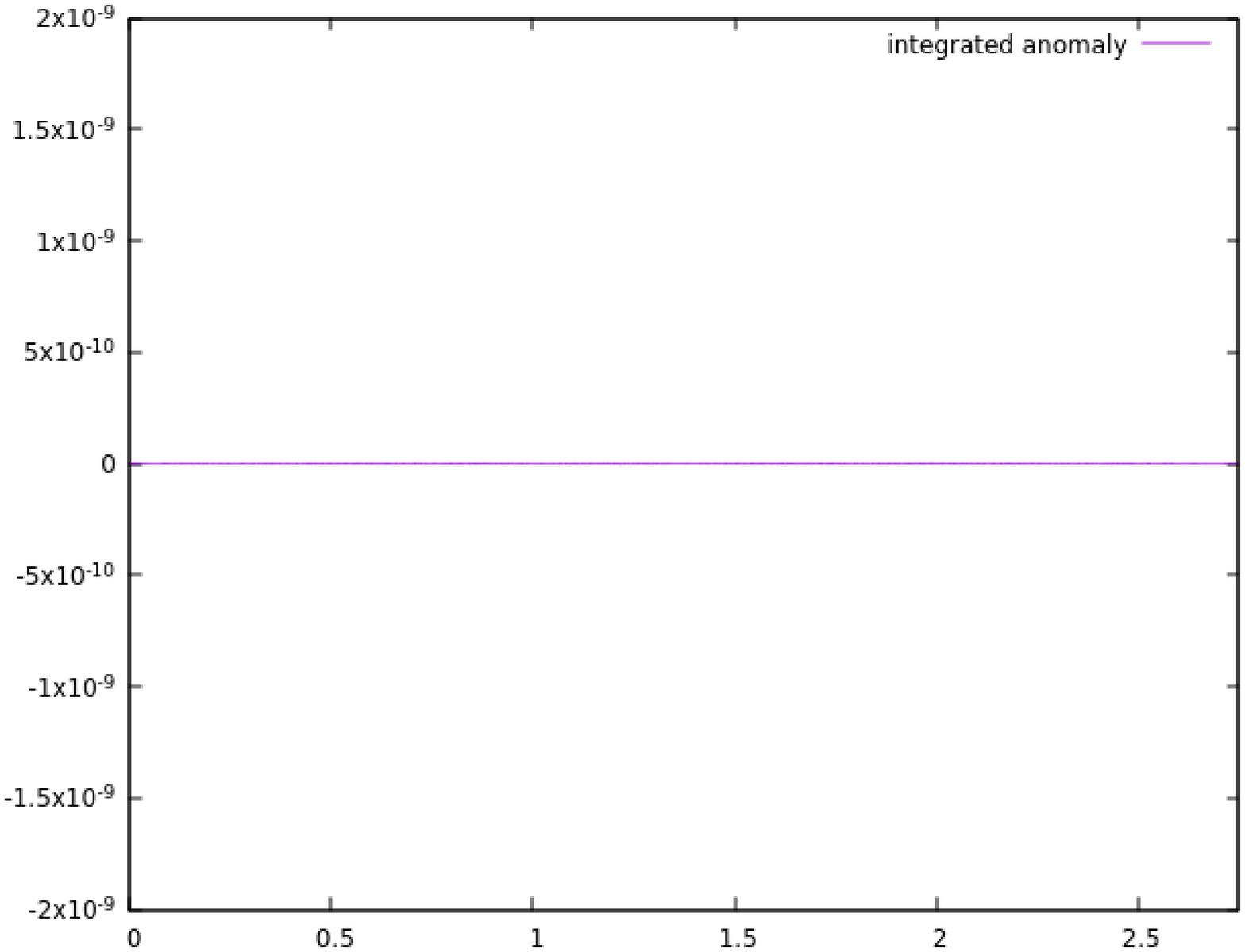} 
\parbox{6in}{\caption {(color online)Type IIB. Top left: the transmission of two bright solitons  plotted for $\epsilon=-0.06$. The initial bright solitons $t_i$ (green) travel in opposite direction with velocities $v_1 =-5, v_2 = 4$ and amplitudes $\rho_1 = 2.6, \rho_2 = 2.1$  such that $n=0.785, \phi_1=2.31, \phi_2=-3.53$  in (\ref{phi12}). They form a collision pattern $t_c$ (blue) in their approximation and then transmit to each other. The solitons after collision are plotted as red line ($t_f$). Top right: the anomaly density $\g(x,t)$ plotted at the collision time  $t_c$ . Bottom: the anomaly $\beta^{(4)}(t)$ and time integrated anomaly $\int^t dt' \beta^{(4)}(t')$, respectively.}}
\end{figure}

\begin{figure}
\centering
\label{fig12} 
\includegraphics[width=2cm,scale=6, angle=0, height=5cm]{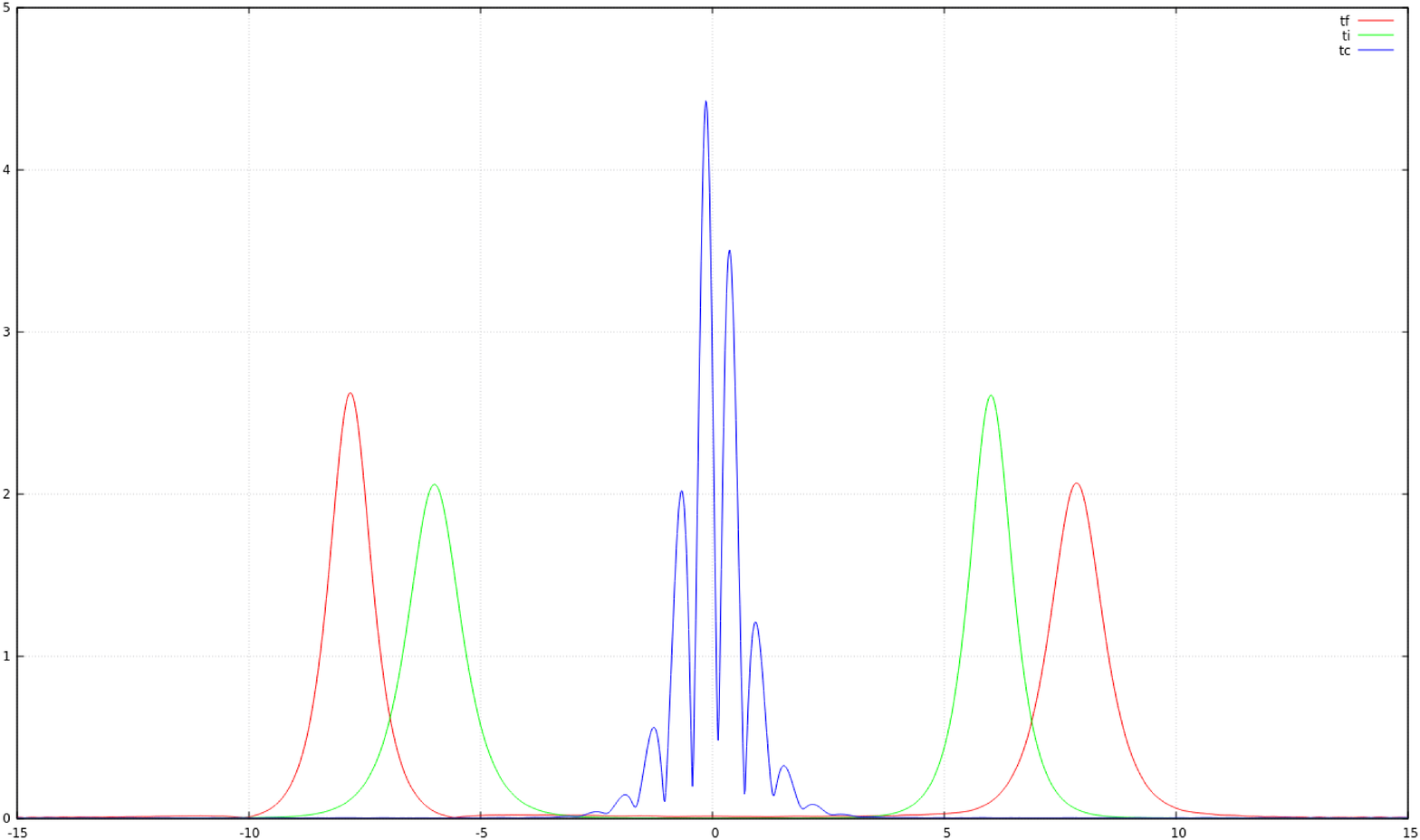} 
\includegraphics[width=2cm,scale=6, angle=0, height=5cm]{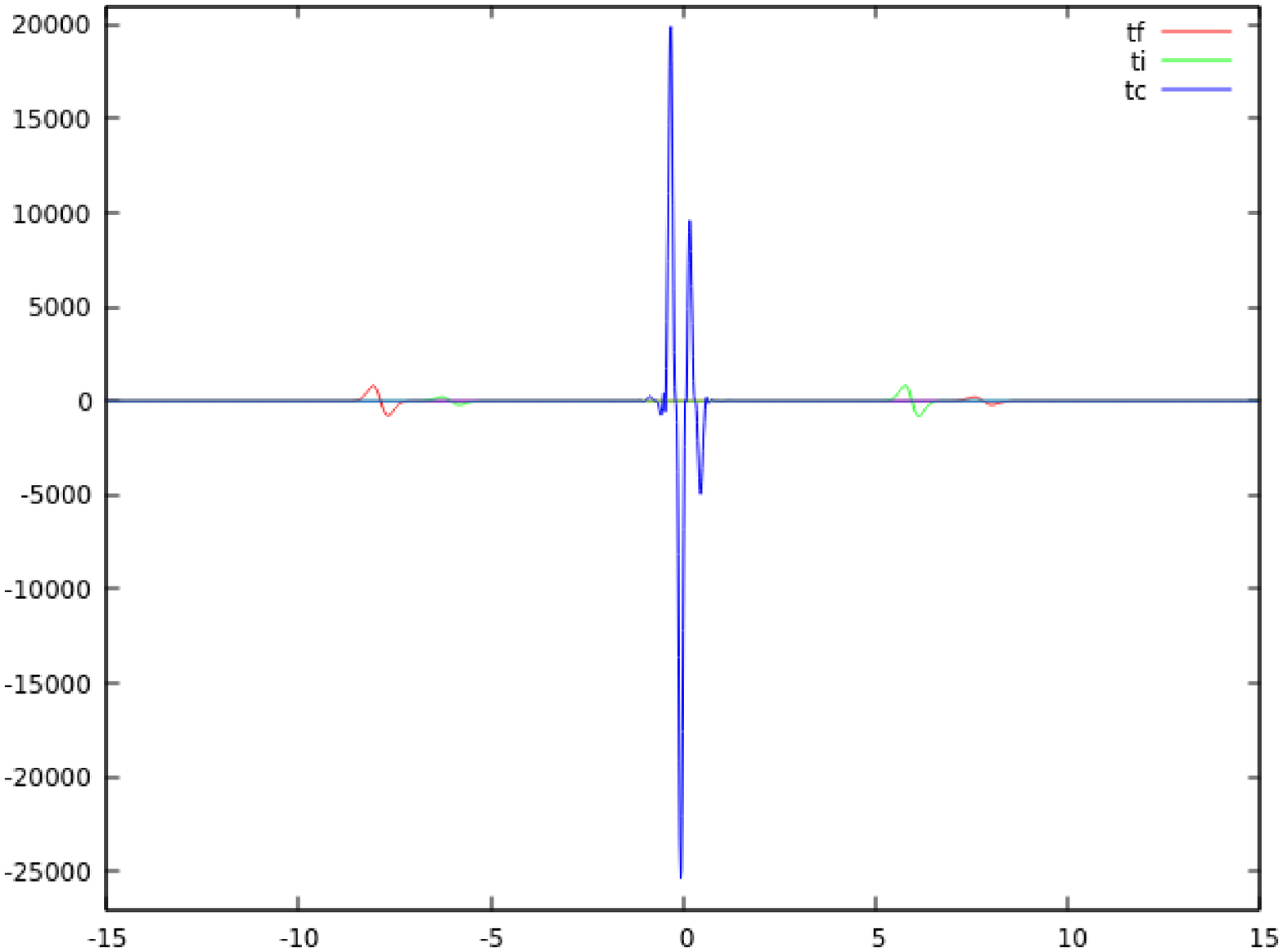} 
\includegraphics[width=2cm,scale=6, angle=0, height=5cm]{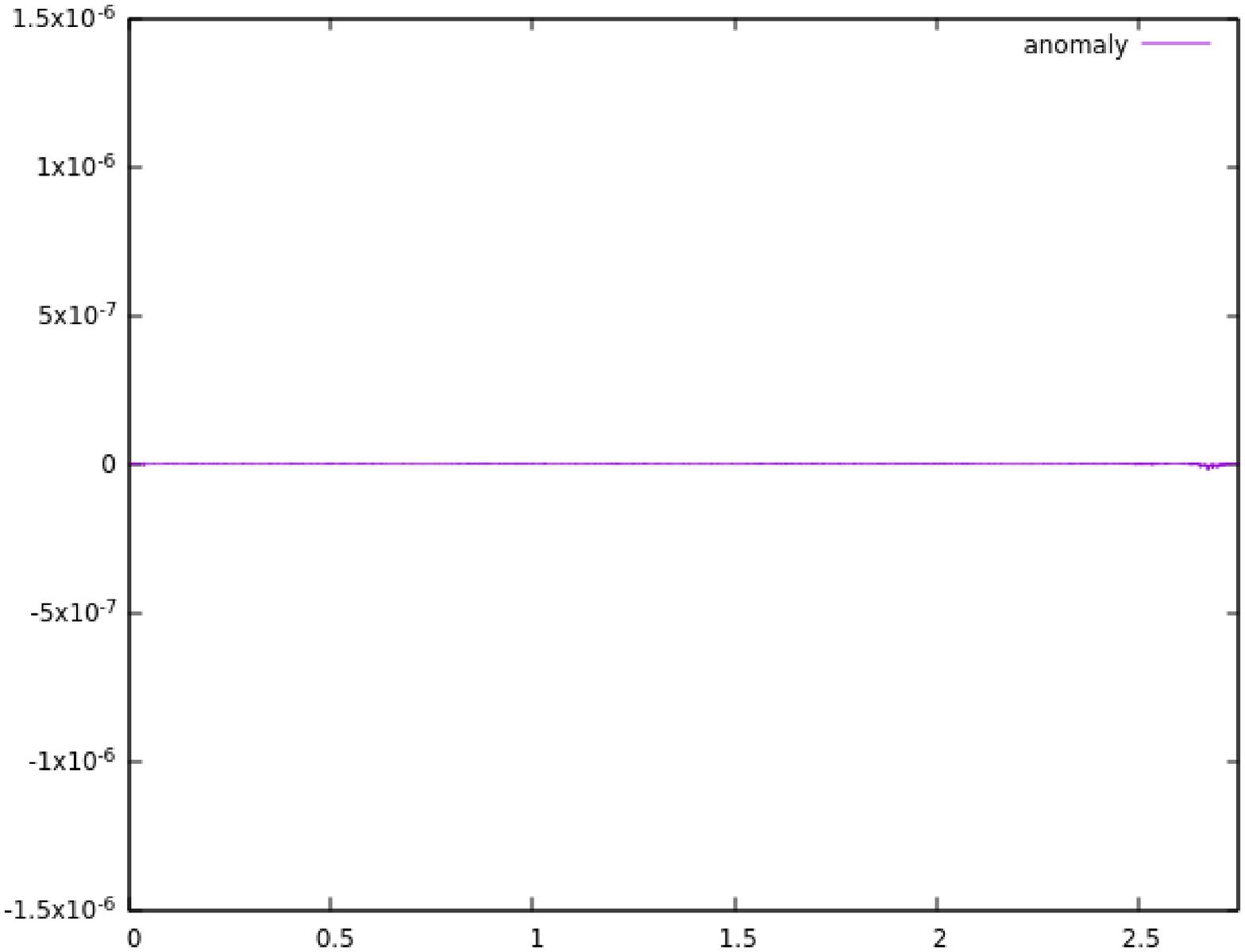}
\includegraphics[width=2cm,scale=6, angle=0, height=5cm]{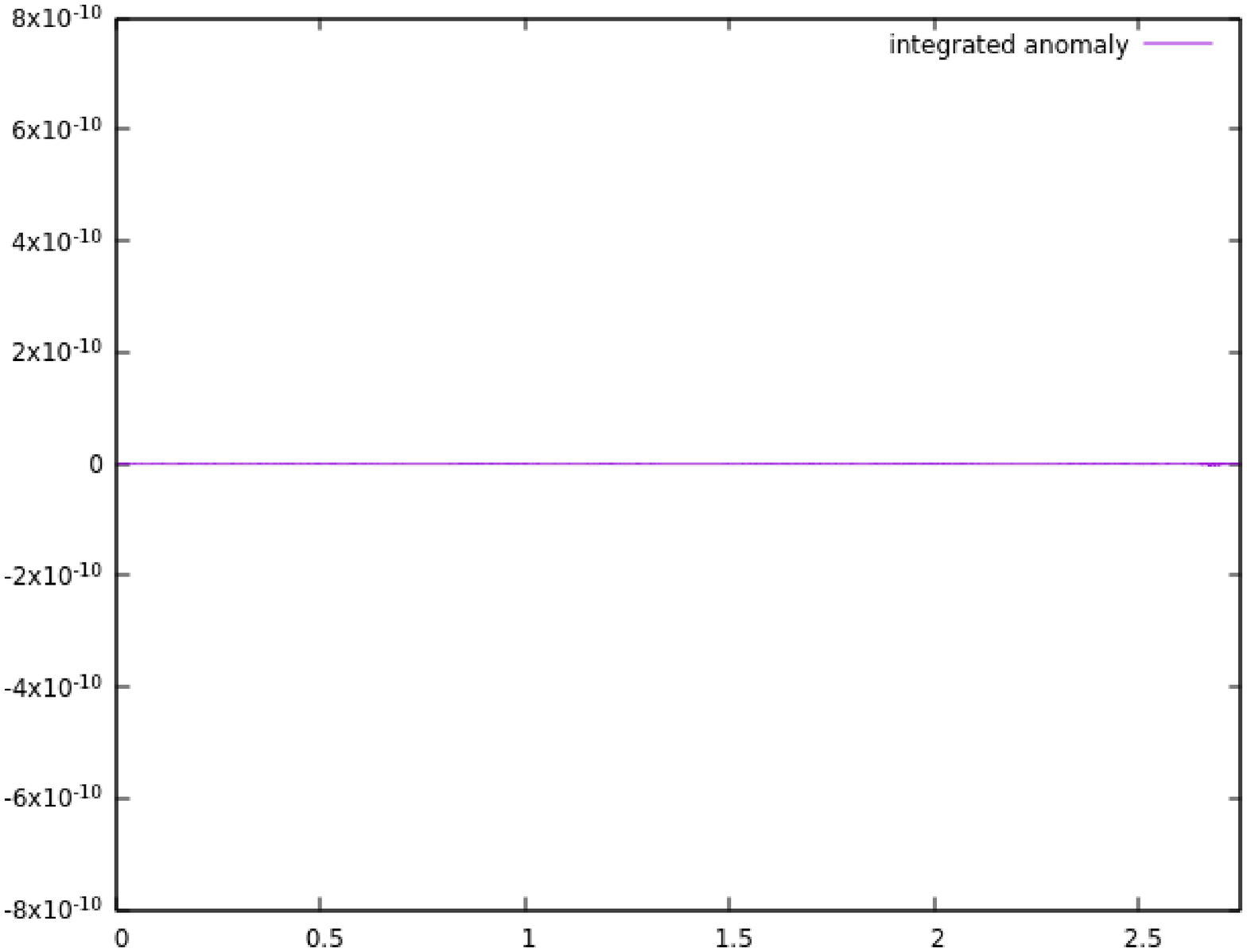}   
\parbox{6in}{\caption {(color online)Type IIA. Top left: the transmission of two bright solitons  plotted for $\epsilon=0.06$. The initial bright solitons $t_i$ (green) travel in opposite direction with velocities $v_1 =- v_2 = -5$ and amplitudes $\rho_1 = 2.6, \rho_2 = 2.1$  such that $n=2, \phi_1=2.545, \phi_2=-6.77$  in (\ref{phi12}). They form a collision pattern $t_c$ (blue) in their approximation and then transmit to each other. The solitons after collision are plotted as red line ($t_f$). Top right: the anomaly density $\g(x,t)$ plotted at the collision time  $t_c$ . Bottom: the anomaly $\beta^{(4)}(t)$ and time integrated anomaly $\int^t dt' \beta^{(4)}(t')$, respectively.}}
\end{figure}

\begin{figure}
\centering
\label{fig13} 
\includegraphics[width=2cm,scale=6, angle=0, height=4cm]{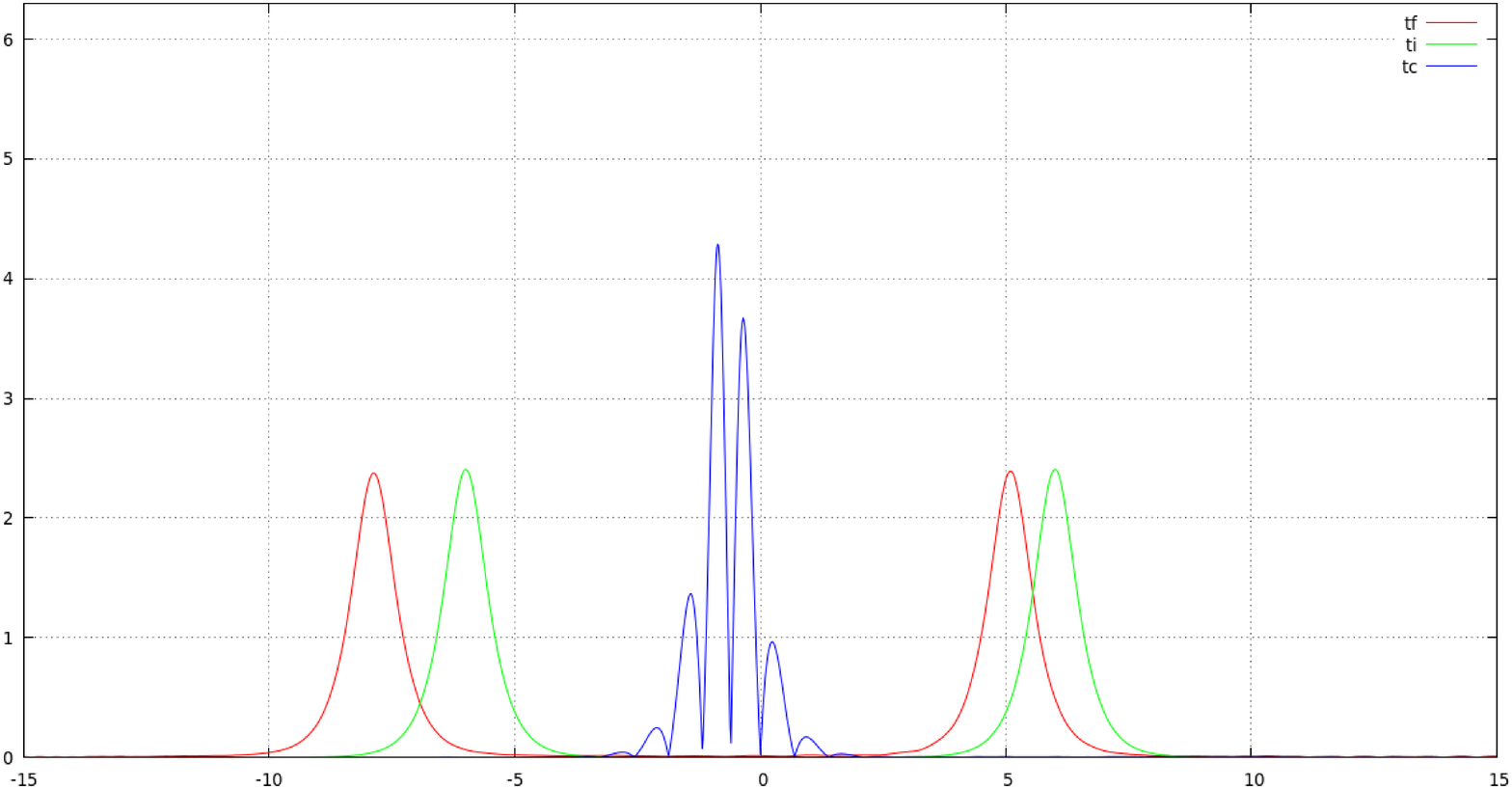} 
\includegraphics[width=2cm,scale=6, angle=0, height=4cm]{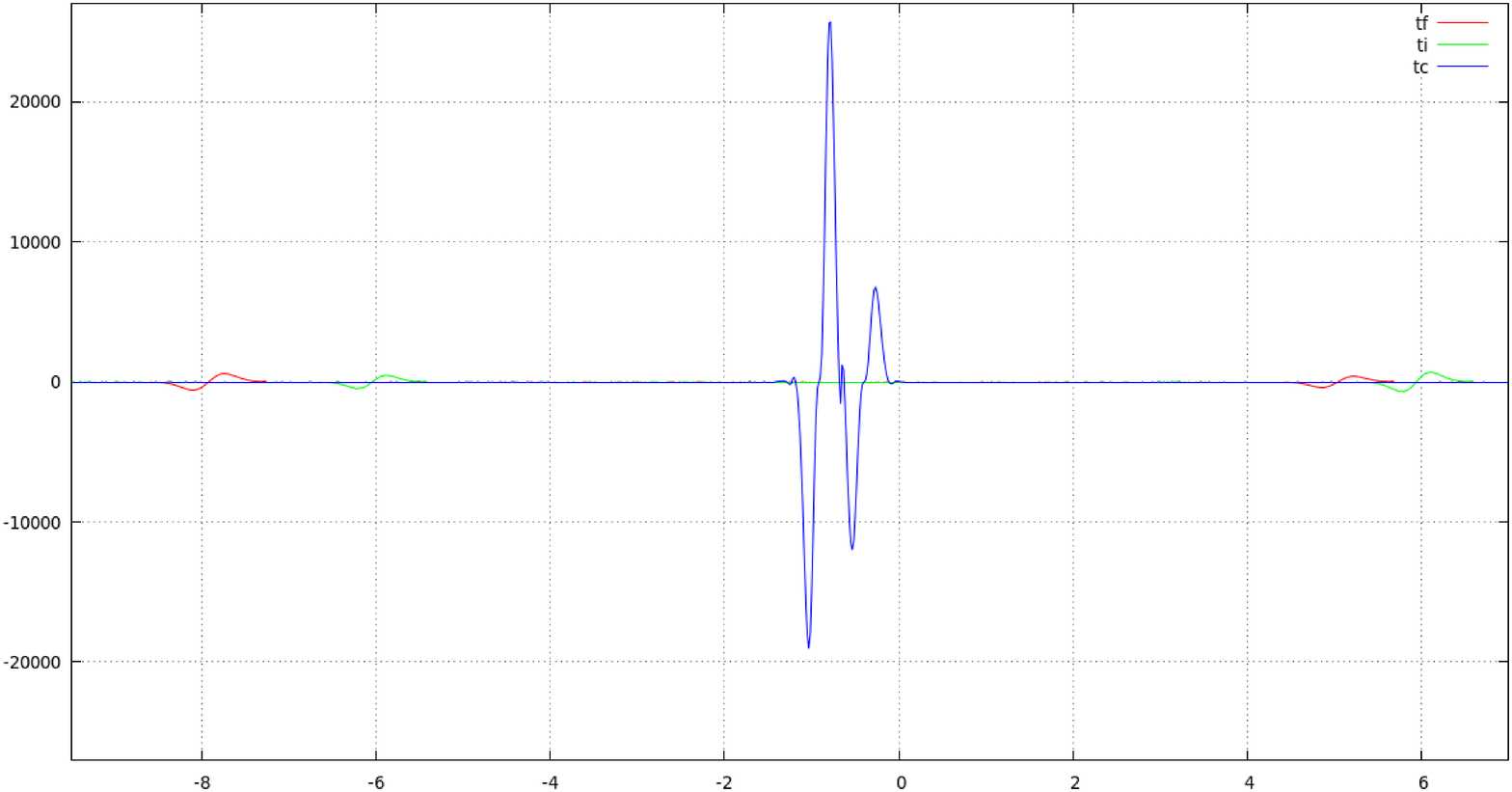} 
\includegraphics[width=2cm,scale=6, angle=0, height=4cm]{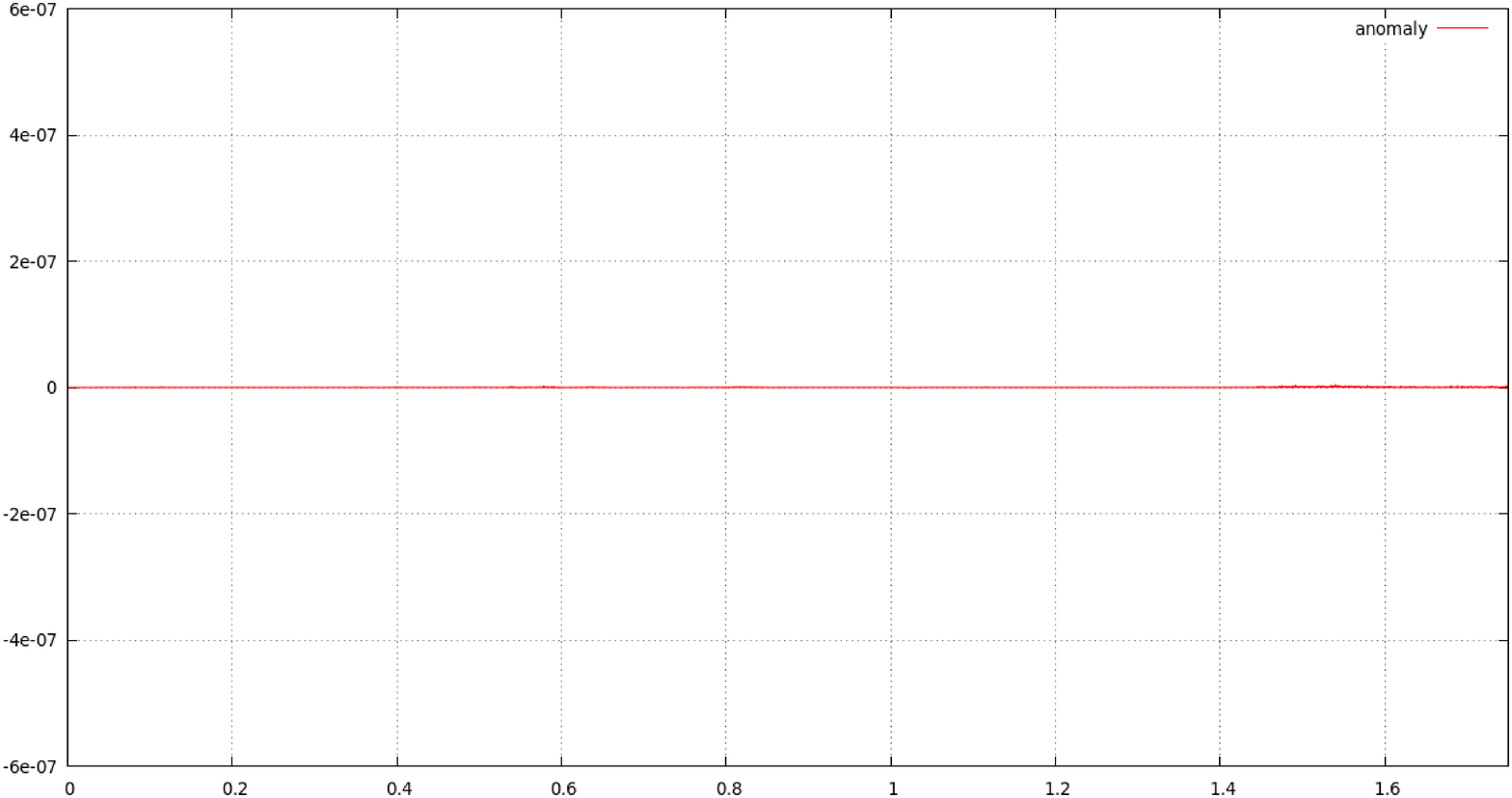}
\includegraphics[width=2cm,scale=6, angle=0, height=4cm]{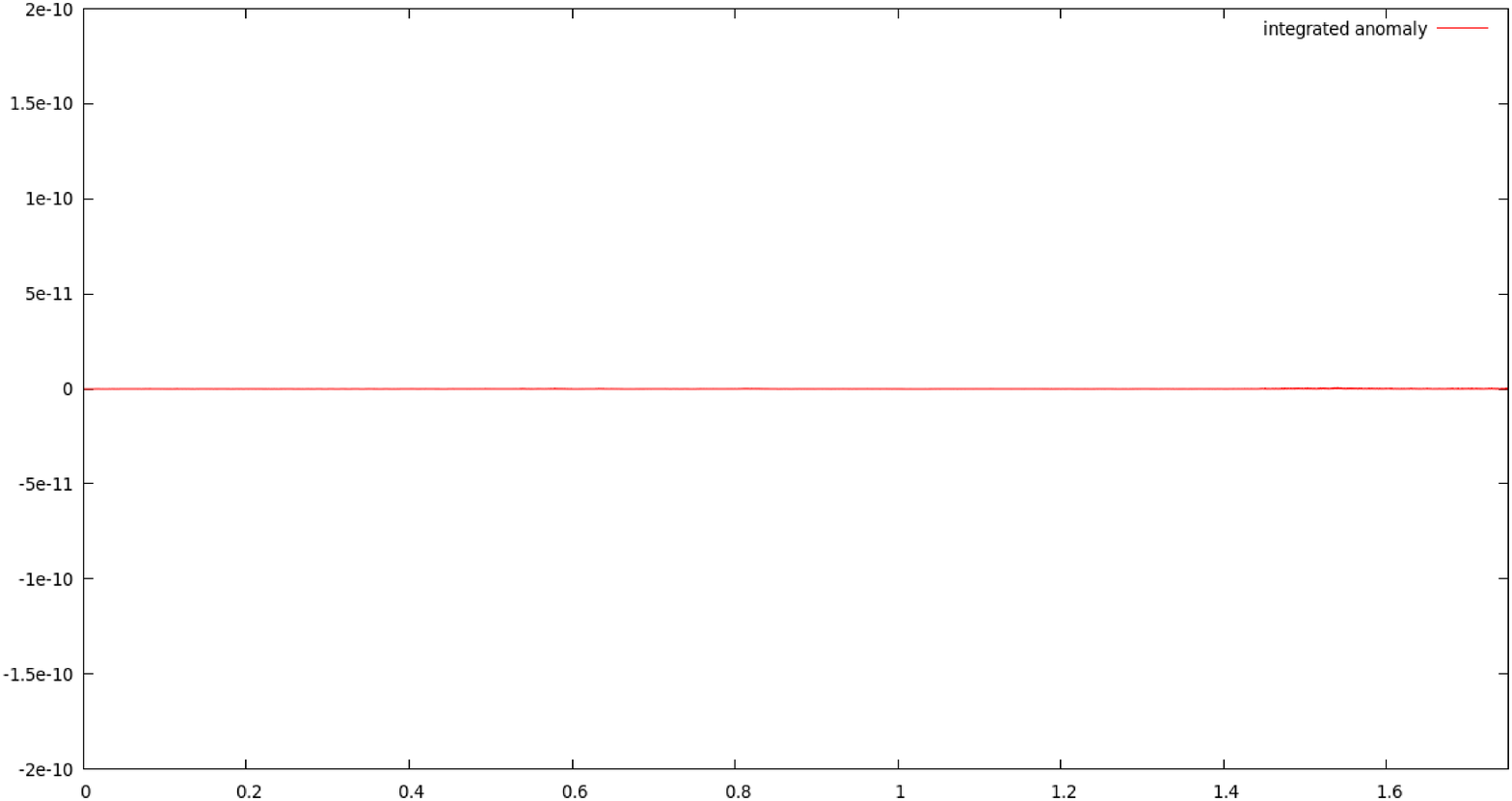}   
\parbox{6in}{\caption {(color online) Type IIB.  Top left: the transmission of two bright solitons  plotted for $\epsilon=0.06$. The initial bright solitons $t_i$ (green) travel in opposite direction with velocities $v_1 =- 5, v_2 = 4$ and amplitudes $\rho_1=\rho_2 = 2.4$  such that $n=0.7854, \phi_1=2.11, \phi_2=-0.36$  in (\ref{phi12}). They form a collision pattern $t_c$ (blue) in their approximation and then transmit to each other. The solitons after collision are plotted as red line ($t_f$). Top right: the anomaly density $\g(x,t)$ plotted at the collision time  $t_c$ . Bottom: the anomaly $\beta^{(4)}(t)$ and time integrated anomaly $\int^t dt' \beta^{(4)}(t')$, respectively.}}
\end{figure}

\begin{figure}
\centering
\label{fig14}
\includegraphics[width=4cm,scale=5, angle=0, height=4.5cm]{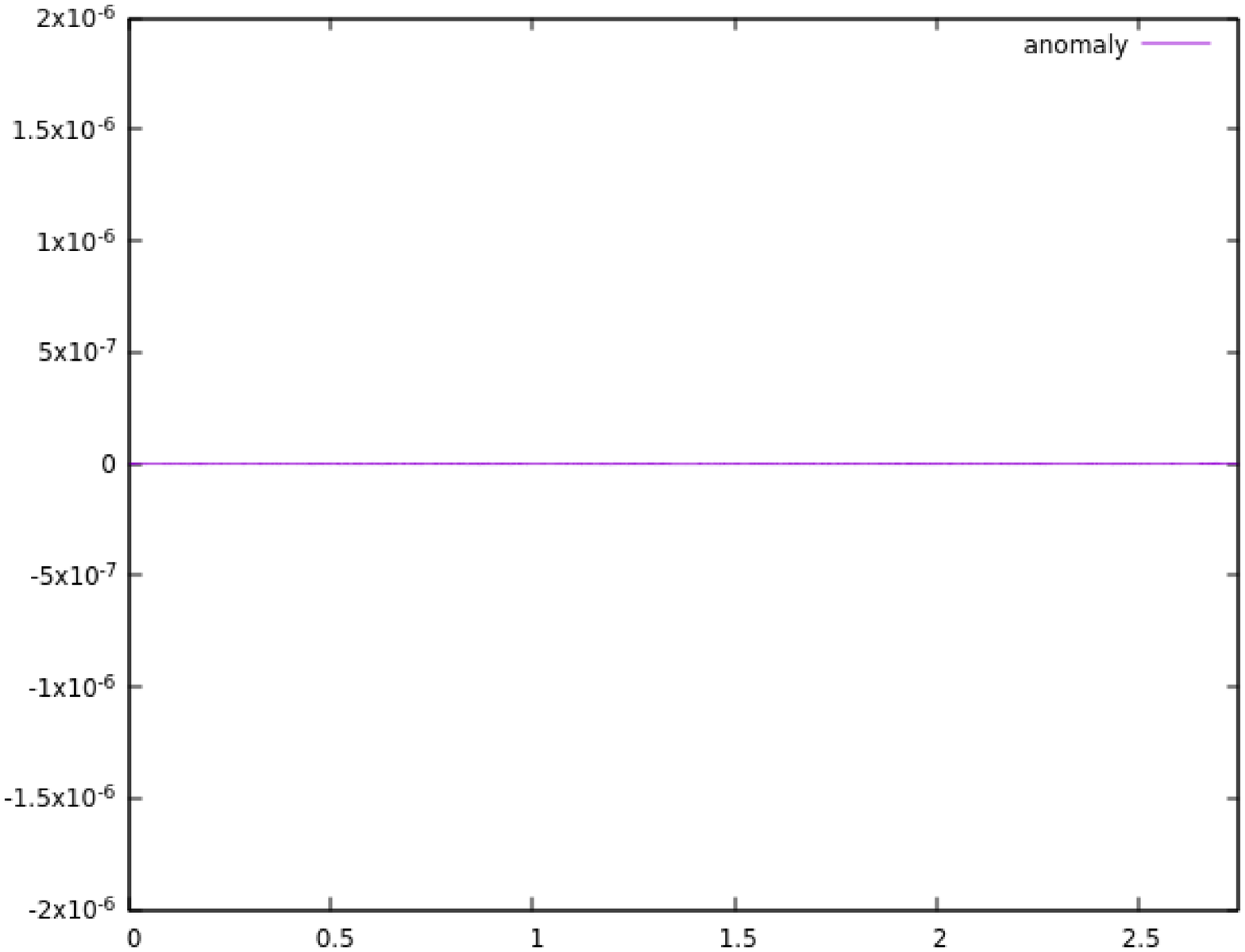}
\includegraphics[width=4cm,scale=5, angle=0, height=4.5cm]{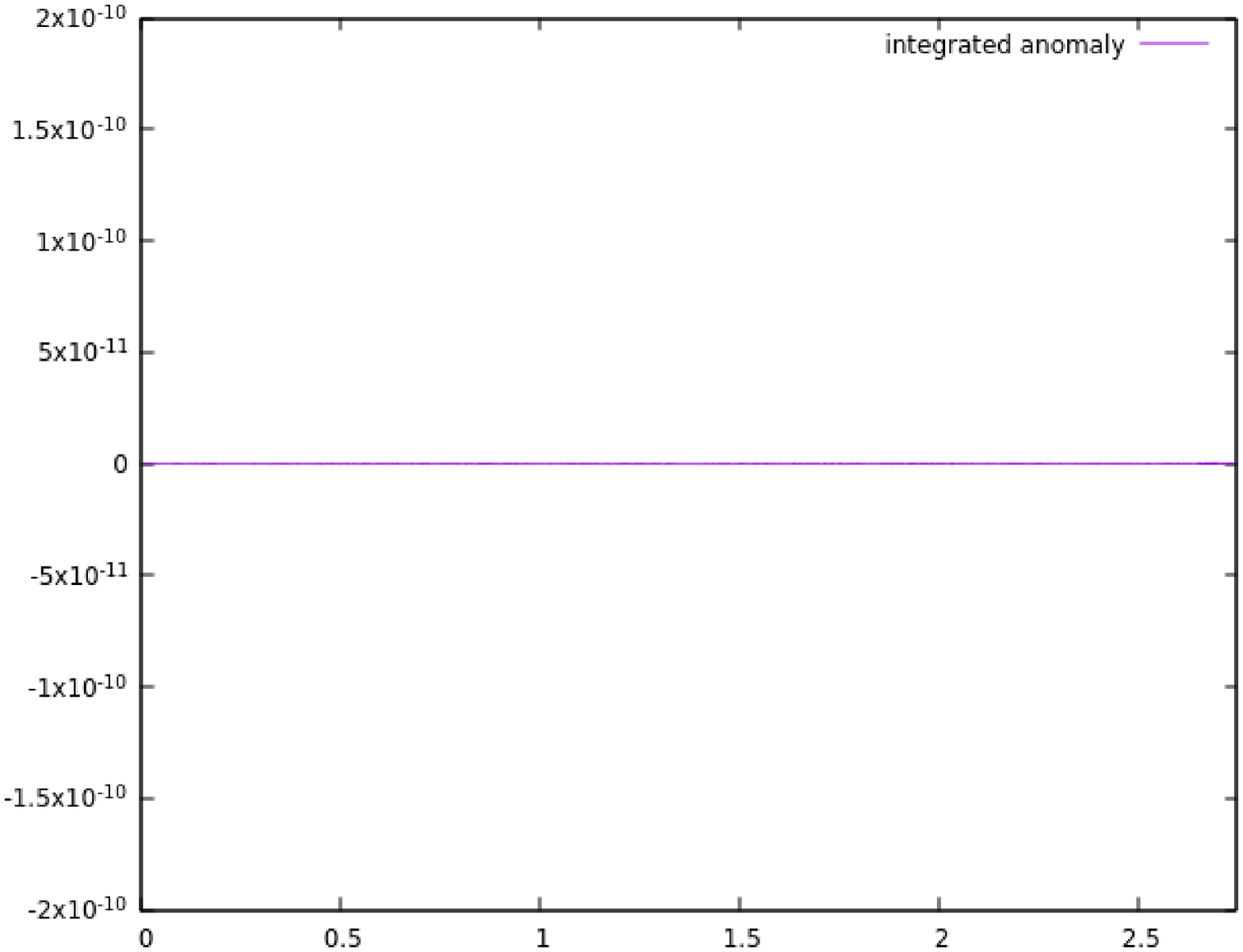}   
\parbox{6in}{\caption {Type IIB. Anomaly and integrated anomaly for  $\epsilon=0.06,\,  v_1 = -5, v_2 = 4,\,\rho_1 = 2.5, \rho_2= 2.0, n=0.7854, \phi_1= 2.31, \phi_2= -3.5312$.}}
\end{figure}   
 
 \begin{figure}
\centering
\label{fig15}
\includegraphics[width=4cm,scale=5, angle=0, height=4cm]{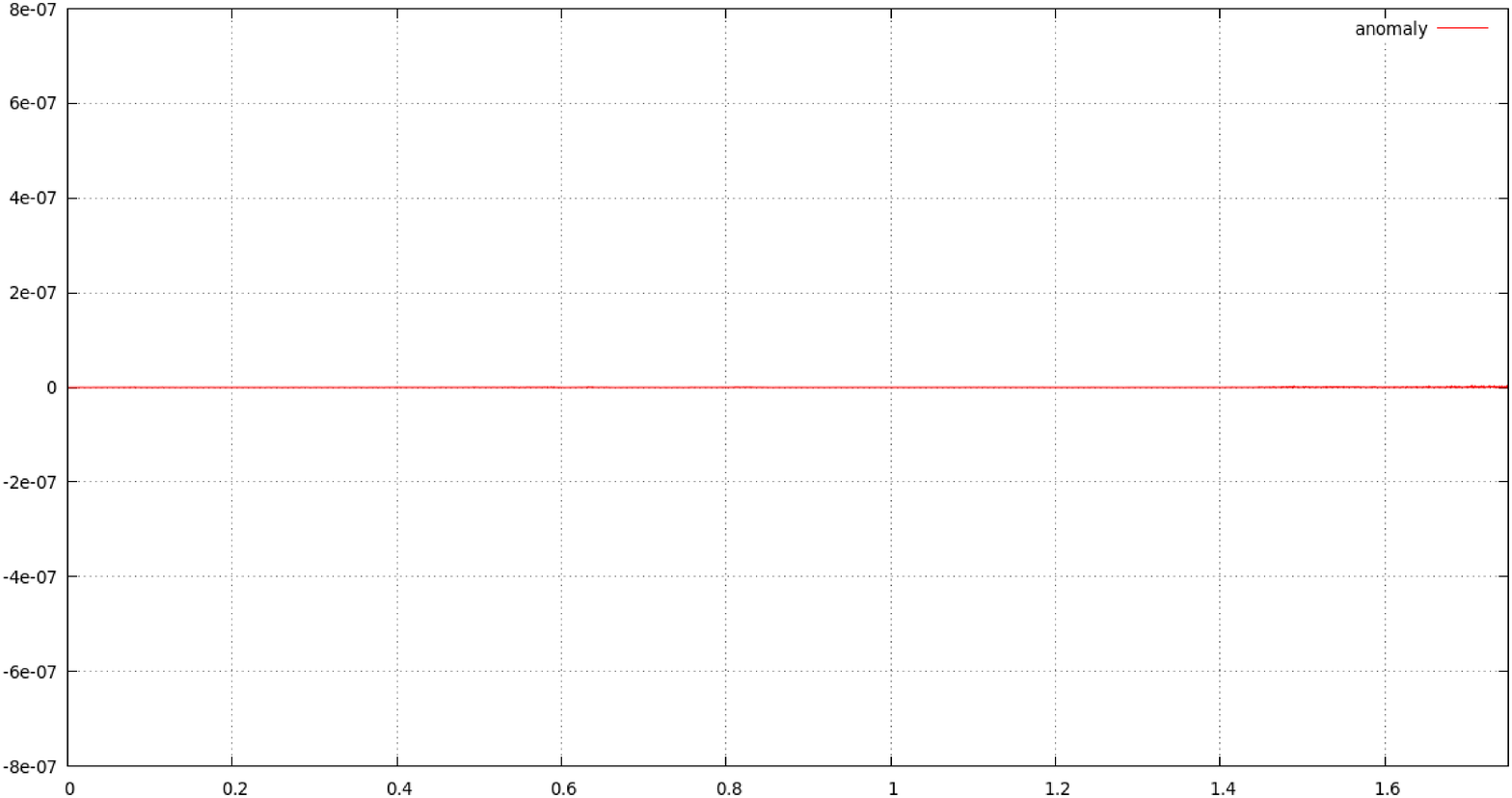}
\includegraphics[width=4cm,scale=5, angle=0, height=4cm]{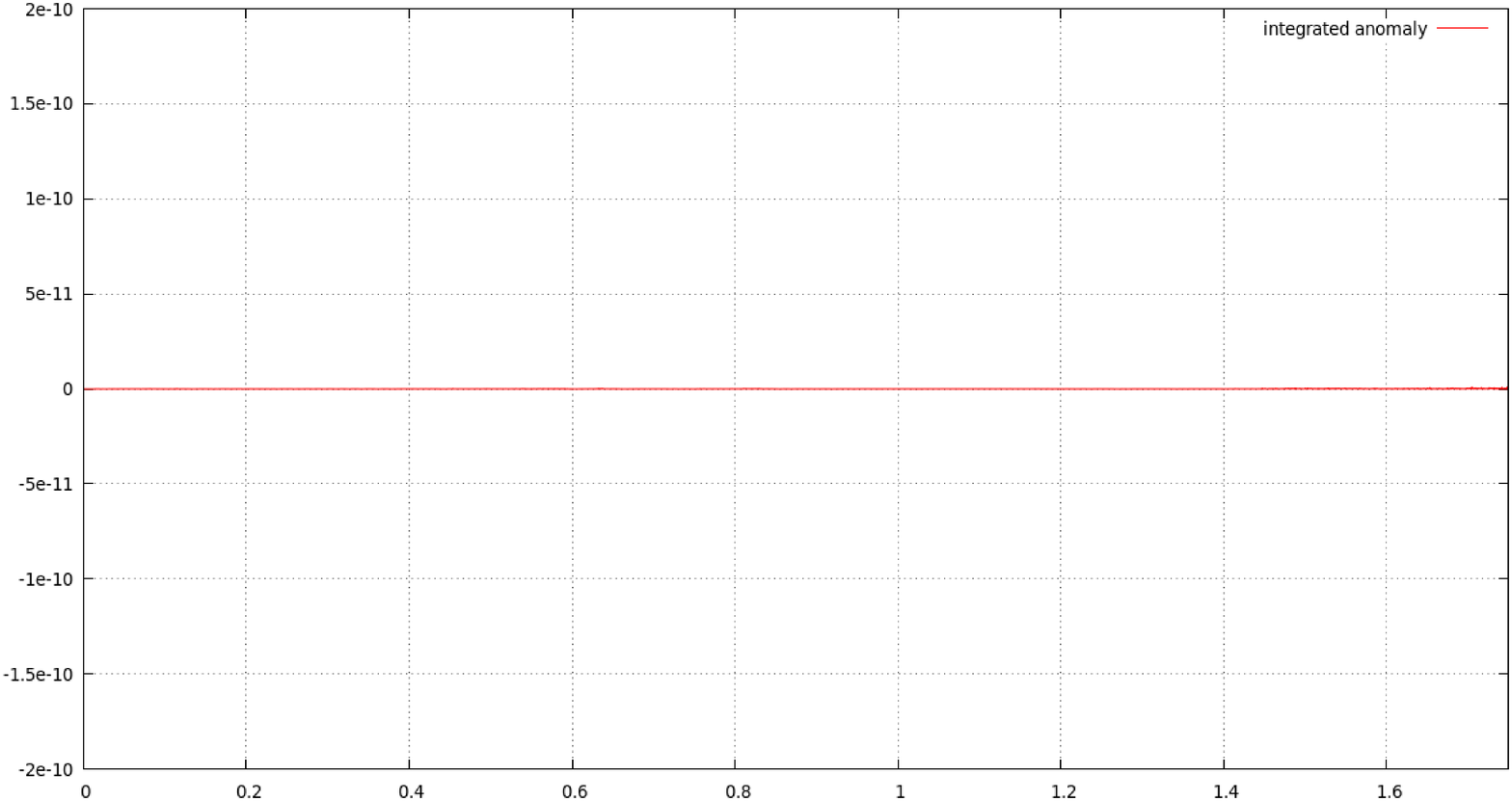}   
\parbox{6in}{\caption {Type IIA. Anomaly and integrated anomaly for  $\epsilon=0.06,\,  v_1 = -5, v_2 = 4,\,\rho_1 = 2.5, \rho_2= 2.5, n=1, \phi_1= 2.44, \phi_2= -0.697$.}}
\end{figure}    
 
\begin{figure}
\centering
\label{fig16}
\includegraphics[width=4cm,scale=5, angle=0, height=4.5cm]{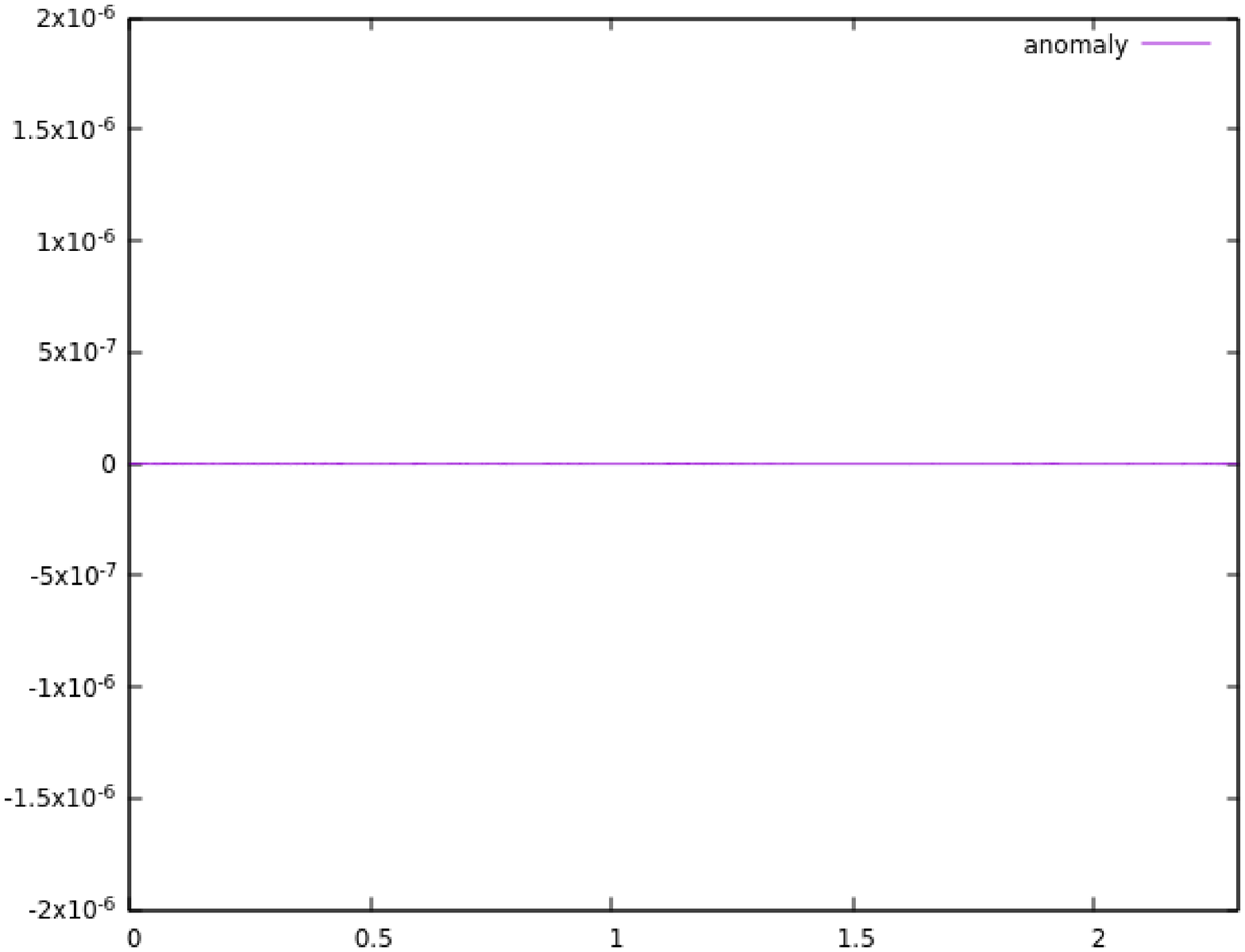}
\includegraphics[width=4cm,scale=5, angle=0, height=4.5cm]{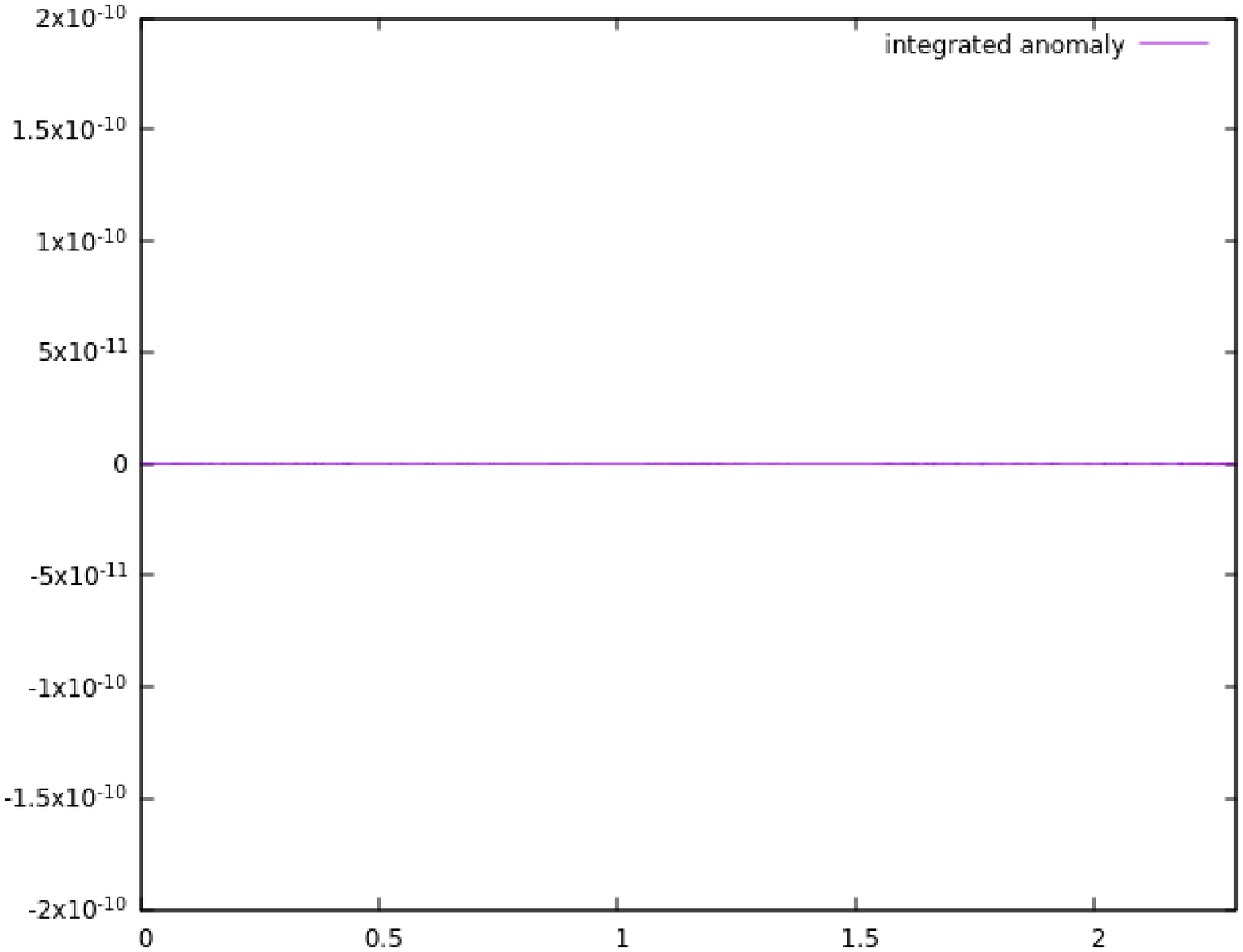}   
\parbox{6in}{\caption {Type IIB. Anomaly and integrated anomaly for  $\epsilon=- 0.01,\,  v_1 = - v_2 = -5,\,\rho_1 = 2.5, \rho_2= 2.0, n=0.7854, \phi_1= 0.637, \phi_2= - 4.862$.}}
\end{figure}   
 
\begin{figure}
\centering
\label{fig17}
\includegraphics[width=5cm,scale=5, angle=0, height=4.5cm]{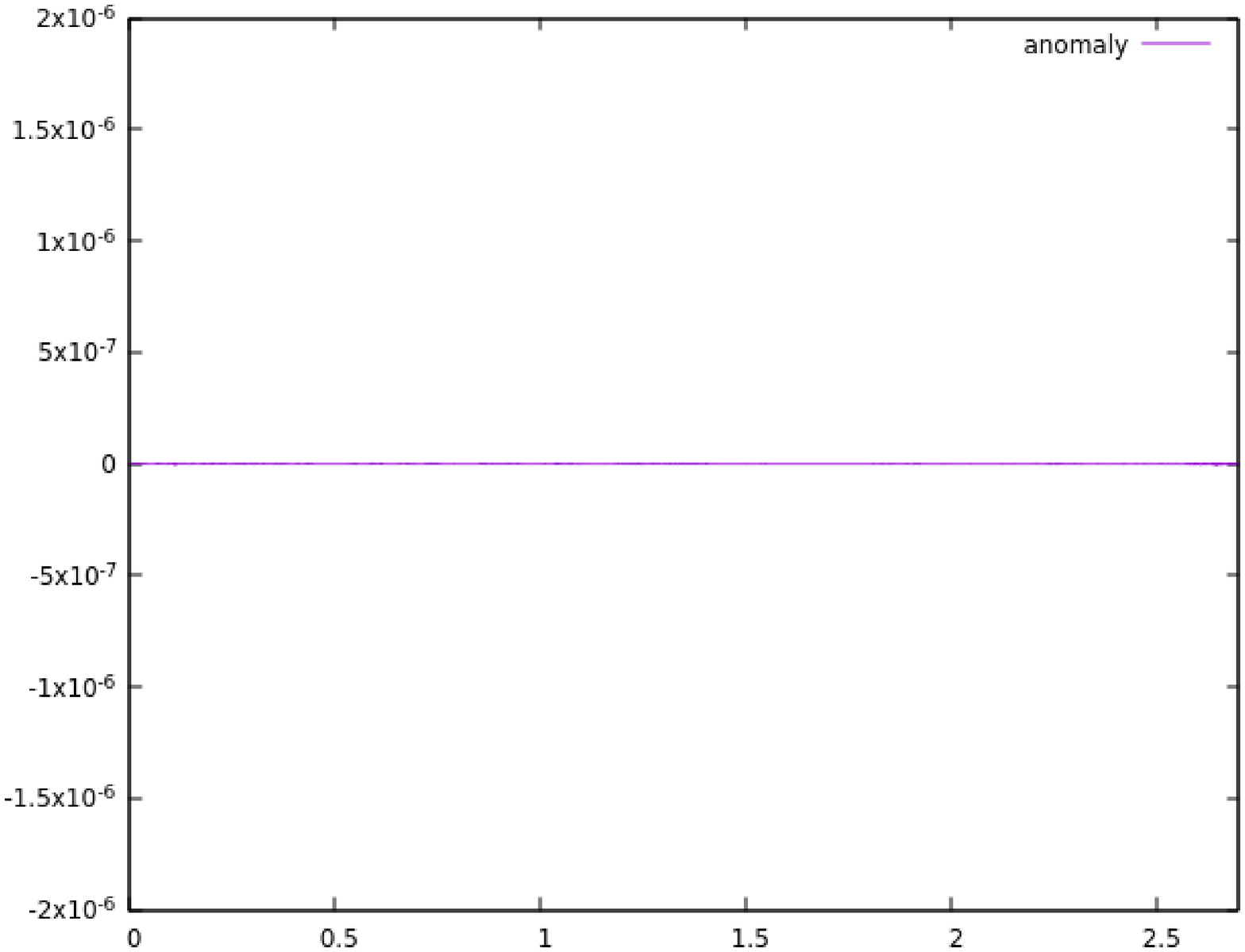}
\includegraphics[width=5cm,scale=5, angle=0, height=4.5cm]{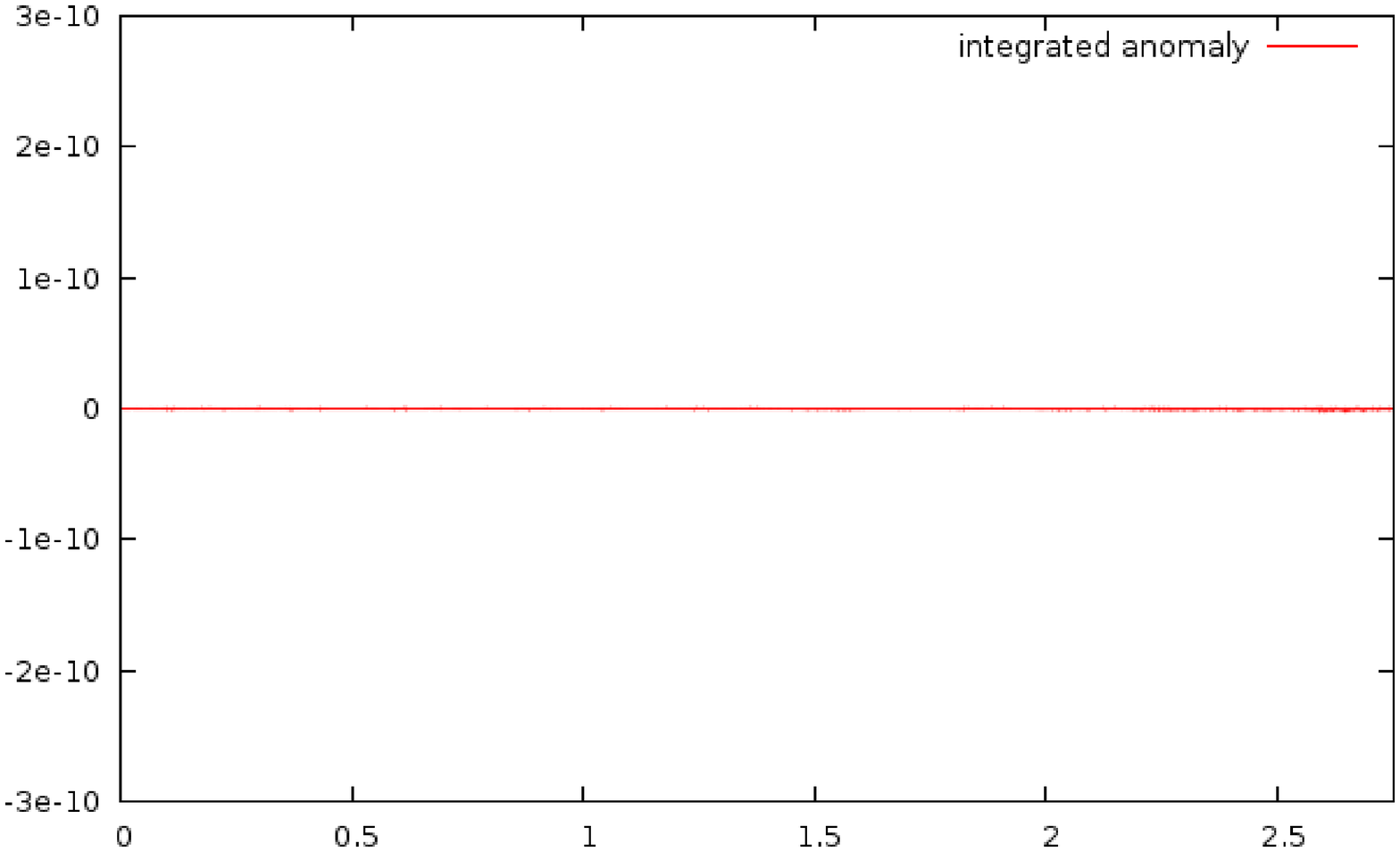}   
\parbox{6in}{\caption {Type IIA. Anomaly and integrated anomaly for  $\epsilon=- 0.01,\,  v_1 = - v_2 = -5,\,\rho_1 = 2.5, \rho_2= 2.0, n=2, \phi_1= 4.21, \phi_2= - 5.44$.}}
\end{figure}

\section{Discussions and some conclusions}
\label{conclu}

Let us summarize the main results of \cite{jhep2} for bright-solitons: a) the model (\ref{nlsd})-(\ref{pot1}) possesses an infinite number of exactly conserved quantities for one-bright solitary waves travelling with a constant speed. b) For two-bright solitons possessing a special space-time parity symmetry (\ref{pari1})-(\ref{pari0}), the charges are asymptotically conserved. This means that these quantities  vary in time during the collision process of two one-solitons but return, after the collision, to the values they had before collision.  

In this paper we have shown that the quasi-conserved charges of  the model (\ref{nlsd})-(\ref{pot1}) studied in \cite{jhep2} split into two subsets, with different conservation properties. We have obtained a subset comprising a new infinite tower of exactly conserved charges (\ref{charge11}), and a second subset containing the remaining asymptotically conserved ones (\ref{summ3}). The two-bright soliton solutions   possessing a special space-reflection parity symmetry (\ref{spari11})-(\ref{xsym}) give rise to a new tower of exactly conserved charges. In \cite{jhep3} it has been shown  the vanishing of the anomalies (\ref{qcon}) through some algebraic  techniques involving the interplay between the space-reflection and an order two $\IZ_2$ automorphism of the  $sl(2)$ loop algebra. We have computed the first non-trivial anomaly $\b^{(4)}$ of the $Q^{(4)}$ charges quasi-conservation law (\ref{qcon11}). We have verified that this anomaly vanishes, and consequently the exact conservation of the charge $Q^{(4)}$ holds for various two-soliton configurations, within numerical accuracy.  The only explanation we have found, so far, for the exact conservation  of the even order charges, is that the two-bright soliton solutions are eigenstates of the space-reflection parity transformation for a fixed time.

Remarkably, we have found that even for two-soliton solutions with different velocities and  amplitudes, which do not satisfy the space-reflection symmetry (\ref{spari11})-(\ref{xsym}),  the anomaly $\b^{(4)}$ vanishes (see Figs. 5-17), within the numerical accuracy. Then, we may argue that the parity property (\ref{spari11})-(\ref{xsym}) is not the cause of the exact conservation of the charges, but according to the  analytical results of \cite{jhep3} it is a sufficient condition for these phenomena to happen. So, the symmetries involved in the quasi-integrability phenomena in relativistic and non-relativistic models deserve further investigation and they may find interesting applications in many areas of non-linear science.

Earlier results in solitary wave collisions in deformed NLS models have been obtained through small perturbations of the usual NLS model \cite{keener, malomed, ablowitz}  allied to the inverse scattering transform (IST) method
\cite{kivshar}. In fact, most of the known results consider particular relationships between the parameters of
the colliding solitons, e.g. high relative velocity, different amplitudes, fast  and slow  solitons are among the cases considered in the
literature. In this context, the solitary wave interactions for the generalized KdV equation have been considered in \cite{martel}, in  a special regime. In \cite{galina} the bright solitary wave collision of the deformed NLS model, where one soliton is small with respect to the other, has been studied.  The asymptotic approaches to describe the evolution and collision of three waves of the generalized KdV
model were considered in \cite{omel}. The solitary wave collisions, beyond small
perturbations, have been considered in the recent literature; e.g. \cite{jpa2} provides the spatial shifts of fast dark-dark soliton collisions and \cite{theocharis} studies slow dark soliton collisions.

\section{Acknowledgements}

HB and AMV acknowledge FAPEMAT for partial financial support in the initial stage of the work. HB thanks Professor L. A. Ferreira, M. Zambrano and H. F. Callisaya for useful discussions. 

\appendix

\section{Expressions of the quasi-conservation laws (\ref{cons1})}

In this appendix we present some expressions corresponding to the gauge transformed curvature (\ref{cons1}) provided in \cite{jhep2}.  The $a_{x}^{(3,-n)}$ for $n=0,1,2,3, 4$  are 
\br
a_{x}^{(3,0)} &=& \frac{i}{2} \vp^{(1,0)}\\\
a_{x}^{(3,-1)} &=& 2 i \eta R,\\  
a_{x}^{(3,-2)} &=& i \eta \vp^{(1,0)} R,\\
a_{x}^{(3,-3)} &=& \frac{i \eta \( 4 \eta R^3 + (\vp^{(1,0)})^2  R^2 - 2 R^{(2,0)} R + (R^{(1,0)})^2 \)}{2 R}, \\
a_{x}^{(3,-4)} &=& \frac{i \eta}{4 R} \Big[ 12 \eta \vp^{(1,0)} R^3 - 6 R \(\vp^{(2,0)}  R^{(1,0)} +\vp^{(1,0)}  R^{(2,0)} \) + 3 \vp^{(1,0)}  (R^{(1,0)})^2 \\
&& + \( (\vp^{(1,0)})^3 - 4 \vp^{(3,0)} \) R ^2\Big],
\er
and the parameters $\a^{(j,-n)}$ become
\br
\a^{(3,0)} &=& 1,\\
\a^{(3,-1)} &=& 0,\\
\a^{(3,-2)} &=& 2 \eta R,\\
\a^{(3,-3)} &=& 2 \eta R \vp^{(1,0)},\\
\a^{(3,-4)} &=& 6 \eta^2 R^2 + \frac{3}{2} \eta  (\vp^{(1,0)})^2 R - 2 \eta R^{(2,0)}+ \frac{3 \eta (\vp^{(1,0)})^2 }{2 R},
\er


\begin{thebibliography}{**}
\bibitem{jhep1}
L.A. Ferreira and Wojtek J. Zakrzewski, \JHEP{05}{2011}{130}.\\
L.A. Ferreira and Wojtek J. Zakrzewski, \JHEP{01}{2014}{058}\\
V.H. Aurichio and L.A. Ferreira, \JHEP{03}{2015}{152}\\
 L. A. Ferreira and W. J. Zakrzewski, Breather-like structures in modified sine-Gordon models, \Nonl{29}{2016}{1622}.
\bibitem{jhep2}
L.A. Ferreira, G. Luchini  and Wojtek J. Zakrzewski, \JHEP{09}{2012}{103}.
\bibitem{jhep5}
H.E. Baron and W.J. Zakrzewski, \JHEP{06}{2016}{185}.
\bibitem{jhep4}
H. Blas and H. F. Callisaya, {\sl Quasi-integrability in deformed sine-Gordon models and infinite towers of conserved charges}; 	arXiv:1605.08957 [hep-th].
\bibitem{jhep3} 
H. Blas and M. Zambrano, \JHEP{03}{2016}{005}.
\bibitem{bao}
Weizhu Bao, \MAA{11}{2004}{001}.\\
Weizhu Bao, Qinglin Tang, Zhiguo Xu, Numerical methods and comparison for computing dark and bright
                                     solitons in the nonlinear Schrödinger equation, \JCP{235}{2013}{423}.
\bibitem{cowan}
S. Cowan, R. H. Enns, S. S. Rangnekar, S. S. Sanghera, Quasi-soliton and other behaviour of the nonlinear cubic-quintic Schrödinger equation, \CJP{64}{1986}{311}.
\bibitem{crosta}
M. Crosta, A. Fratalocchi and S. Trillo, Bistability and instability of dark-antidark solitons in the cubic-quintic nonlinear Schrödinger equation, \PRA{84}{2011}{063809}.
\bibitem{sombra}
A. Sergio Bezerra Sombra, Bistable pulse collisions of the cubic-quintic nonlinear Schrödinger equation, \OC{94}{1992}{92}.
\bibitem{kivshar}
Y. S. Kivshar, B. Luther-Davies, \PR{298}{1998}{81}.
\bibitem{kroli}
W. Krolikowski and B. Luther-Davies, \OL{18}{1993}{188}.
\bibitem{enns}
R. H. Enns, Bistable solitons and the Painlevé test, \PRA{36}{1987}{5441}. 
\bibitem{solvint} 
  H.~Blas,{\sl Vector NLS hierarchy solitons revisited: Dressing transformation and tau function approach,}
  solv-int/9912015. 
\bibitem{keener}
J.P. Keener and D.W. McLaughlin, Solitons under perturbations, \PRA{16}{1977}{777}.\\
J.P. Keener and D.W. McLaughlin, \JMP{18}{1977}{2008}.
\bibitem{malomed}
B. A. Malomed, Inelastic interactions of solitons in nearly integrable systems, \PHSD{15}{1985}{374}.
\bibitem{ablowitz}
M.J. Ablowitz, S.D. Nixon, T.P. Horikis and D.J. Frantzeskakis, Proceedings of the Royal Society A: Mathematical, Physical and Engineering Sciences, vol. 467, issue 2133, 2011, pp. 2597-2621.
\bibitem{martel}
Y. Martel and F. Merle, Inelastic interaction of nearly equal solitons for the quartic gKdV equation , Invent. Math. 183 (2011) 563.\\
Y. Martel and F. Merle, Description of two soliton collision for the quartic gKdV equation, Ann. of Math. 174 (2011) 757.
\bibitem{galina}
G. Perelman, Two soliton collision for nonlinear Schrödinger equations in dimension 1 , Ann. Inst. H. Poinc. Anal. Non Lin., 28 (2011) 357.
 
\bibitem{omel}
G. A. Omel'yanov,  {\sl Interaction of 3 solitons for the GKdV-4
equation}, arXiv:1504.02167v1 [math.AP].

\bibitem{jpa2}
H. Blas and M. Zambrano, Spatial shifts of colliding dark solitons in deformed non-linear Schrödinger models ,  \JPAMT{48}{2015}{275201}
\bibitem{theocharis}
G. Theocharis, A. Weller, J.P. Ronzheimer, C. Gross, M.K. Oberthaler, P.G. Kevrekidis and D.J. Frantzeskakis, Multiple atomic dark solitons in cigar-shaped Bose-Einstein condensates,\PRA{81}{2010}{063604}.
\end{thebibliography}
\end{document}